\documentclass[prx,amsmath,amssymb, notitlepage, twocolumn,
nofootinbib,
superscriptaddress,
longbibliography
]{revtex4-1}

\usepackage{amsmath}
\usepackage{tabularx,graphicx}
\usepackage{epstopdf}
\usepackage{graphicx}
\usepackage{latexsym}
\usepackage{amssymb}
\usepackage{amsmath}
\usepackage{color, colortbl}
\usepackage{psfrag}
\usepackage{bbm}
\usepackage{bm}
\usepackage{titlesec}
\usepackage{dsfont}
\usepackage{feynmp}
\usepackage{slashed}
\usepackage{multirow}
\usepackage{url}

\newcommand{\be}{\begin{equation}}
\newcommand{\ee}{\end{equation}}

\newcommand{\mb}[1]{\mathbf{#1}}
\newcommand{\mc}[1]{\mathcal{#1}}

\newcommand{\beq}{\begin{eqnarray}}
\newcommand{\eeq}{\end{eqnarray}}

\newcommand{\la}{\langle}
\newcommand{\ra}{\rangle}

\newcommand{\Tr}{{\rm Tr}}
\newcommand{\tr}{{\rm tr}}

\newcommand{\bsp}{\begin{split}}
	\newcommand{\esp}{\end{split}}

\newcommand{\eff}{{\rm eff}}

\newcommand{\half}{\frac{1}{2}}
\newcommand{\ie}{{i.e., }}
\newcommand{\eg}{{e.g., }}

\newcommand{\WZW}{{\rm WZW}}
\newcommand{\swzw}{S^{(N)}_\WZW}
\newcommand{\nn}{\nonumber \\}
\newcommand{\sq}{{\rm Sq}}

\usepackage{color}
\definecolor{darkblue}{rgb}{0.,0.,0.4}
\definecolor{darkred}{rgb}{0.5,0.,0.}
\definecolor{BlueViolet}{RGB}{138,43,226}
\definecolor{SkyBlue}{RGB}{30,144,255}
\definecolor{DarkGreen}{RGB}{0,100,0}
\usepackage[pdftex,colorlinks=true,linkcolor=darkblue,citecolor=blue,urlcolor=darkred]{hyperref}
\usepackage[normalem]{ulem}

\renewcommand{\vec}[1]{\bm{#1}}

\begin{document}
	\title{Stiefel Liquids: \\ Possible Non-Lagrangian Quantum Criticality from Intertwined Orders}
	\author{Liujun Zou}
	\author{Yin-Chen He}
	\author{Chong Wang}
	\affiliation{Perimeter Institute for Theoretical Physics, Waterloo, Ontario, Canada N2L 2Y5}

\begin{abstract}

We propose a new type of quantum liquids, dubbed \textit{Stiefel liquids}, based on $2+1$ dimensional nonlinear sigma models on target space $SO(N)/SO(4)$, supplemented with Wess-Zumino-Witten terms. We argue that the Stiefel liquids form a class of critical quantum liquids with extraordinary properties, such as large emergent symmetries, a cascade structure, and nontrivial quantum anomalies. We show that the well known deconfined quantum critical point and $U(1)$ Dirac spin liquid are unified as two special examples of Stiefel liquids, with $N=5$ and $N=6$, respectively. Furthermore, we conjecture that Stiefel liquids with $N>6$ are \textit{non-Lagrangian}, in the sense that under renormalization group they flow to infrared (conformally invariant) fixed points that cannot be described by any renormalizable continuum Lagrangian. Such non-Lagrangian states are beyond the paradigm of parton gauge mean-field theory familiar in the study of exotic quantum liquids in condensed matter physics. The intrinsic absence of (conventional or parton-like) mean-field construction also means that, within the traditional approaches, it will be difficult to decide whether a non-Lagrangian state can actually emerge from a specific UV system (such as a lattice spin system). For this purpose we hypothesize that a quantum state is \textit{emergible} from a lattice system if its quantum anomalies match with the constraints from the (generalized) Lieb-Schultz-Mattis theorems. Based on this hypothesis, we find that some of the non-Lagrangian Stiefel liquids can indeed be realized in frustrated quantum spin systems, for example, on triangular or Kagome lattice, through the intertwinement between non-coplanar magnetic orders and valence-bond-solid orders.

\end{abstract}

\maketitle

\tableofcontents

\section{Introduction}

The richness of quantum phases and phase transitions never ceases to surprise us. Over the years many interesting many-body states have been discovered or proposed in various systems, such as different symmetry-breaking orders, topological orders, and even exotic quantum criticality. One lesson \cite{Senthil2005} we have learnt is that the vicinity of several competing (or intertwining \cite{Fradkin2015}) orders may be a natural venue to look for exotic quantum criticality. For example, it is proposed that the deconfined quantum critical point (DQCP) may arise as a transition between a Neel antiferromagnet (AF) and a valance bond solid (VBS)~\cite{Senthil2003, Senthil2003a}, and a $U(1)$ Dirac spin liquid (DSL) may arise in the vicinity of various intertwined orders~\cite{Affleck1988, Wen1995, Hastings2000, Hermele2005, Hermele2008,Song2018,Song2018a}. The physics of DQCP and $U(1)$ DSL, which will be discussed in more detail later, have shed important light on the study of quantum matter. So a natural question is
\begin{itemize}
\item Can one find more intriguing quantum criticality, possibly again around the vicinity of some competing orders?
\end{itemize}

On the conceptual front, by now quantum states with low-energy excitations described by well-defined quasiparticles and/or quasistrings, are relatively well understood. These include Landau symmetry-breaking orders, various types of topological phases and conventional Fermi liquids. Such states are tractable because at sufficiently low energies they become weakly coupled and admit simple effective descriptions, even if the system may be complicated at the lattice scale. In contrast, understanding quantum states that remain strongly coupled even at the lowest energies -- and therefore do not admit descriptions in terms of quasiparticles --  remains a great challenge, especially in dimensions greater than $(1+1)$ due to the lack of exact analytic results. The widely held mentality, when dealing with such states, is to start from a non-interacting \textit{mean-field} theory, and introduce \textit{fluctuations} that are weak at some energy scale. The fluctuations may grow under renormalization group (RG) flow, in which case the low-energy theory will eventually become strongly coupled and describe the non-quasiparticle dynamics. Here the mean-field theory can be formulated in terms of the original physical degrees of freedom (DOFs), like spins, as is done for Landau symmetry-breaking orders. It can also be formulated in terms of more interesting objects called \textit{partons}, which are ``fractions'' of local DOFs -- examples include composite bosons/fermions in fractional quantum Hall effects and spinons in spin liquids. Fluctuations on top of a parton mean-field theory typically lead to a gauge theory, which forms the theoretical basis of a large number of exotic quantum phases in modern condensed matter physics \cite{Wen2004Book}. Most (if not all) states in condensed matter physics are understood within the mean-field mentality. In fact, this mentality is so deeply rooted in condensed matter physics that very often a state can be considered ``understood'' only if a mean-field picture is obtained. 

Although most (if not all) states theoretically studied in condensed matter physics can be described by a mean field plus some weak fluctuations at some scale, a priori, there is no reason to assume that all non-quasiparticle states admit some mean-field descriptions. One may therefore wonder if there is an approach that can complement the mean-field theory, and whether one can use this approach to study quantum phases or phase transitions that cannot be described by any mean-field theory plus weak fluctuations. This question can also be formulated in the realm of quantum field theories. The universal properties of a field theory are characterized by a fixed point under RG, and such a fixed point usually allows a description in terms of a weakly-coupled renormalizable continuum Lagrangian at certain energy scale. Such a renormalizable-Lagrangian description of a fixed point is essentially the field-theoretic version of the mean-field description of a quantum phase or phase transition. In this language, a mean field formulated in terms of partons corresponds to a \textit{gauge theory} that is renormalizable, i.e. weakly coupled at the UV scale\footnote{We should note that our identification of ``mean-field theory'' in condensed matter and ``renormalizable Lagrangian'' in field theory is sometimes loose. For example, the Sachdev-Ye-Kitaev model \cite{Sachdev1993,Kitaev2015} has a renormalizable Lagrangian, but the corresponding ``mean field'' in our definition would have a trivial zero Hamiltonian, which is not a useful mean field.}. So one may similarly wonder if there are interesting RG fixed points that are {\it intrinsically} non-renormalizable, \ie cannot be described by any weakly-coupled renormalizable continuum Lagrangian at any scale, a property sometimes refered to as ``{\it non-Lagrangian}". Some examples of such ``non-Lagrangian'' theories have been discussed in the string theory and supersymmetric field theory literature over the years (see, for example, Refs.~\cite{Garcia2015, Beem2016, Gukov2017} for some recent exploration and Ref. \cite{Heckman2018} for a review), but it is not clear whether those examples could be directly relevant in the context of condensed matter physics. In particular, we are interested in non-supersymmetric theories realizable in relatively low dimensions such as $(2+1)$. If such non-Lagrangian theories can be identified, they also enrich our understanding of the landscape of quantum field theories in an intriguing way.

So the following important questions arise:
\begin{itemize}

    \item In condensed matter systems, are there exotic quantum phases and phase transitions  beyond the paradigm of mean-field $+$ weak fluctuations? Equivalently, are there non-Lagrangian RG fixed points relevant in condensed matter physics?
    
    \item If the answers to the above questions are yes, how can we tell in which systems such RG fixed points can emerge? How can we predict the physical properties of these states?
    
\end{itemize}

Besides the conceptual importance, these questions may also be practically relevant. If the quantum phases and phase transitions envisioned above do exist, they should be included as candidate theories for many of the elusive experimental and numerical systems, for example in spin liquid physics \cite{Savary2017,Zhou2017, Broholm2019}.

In this paper we focus on critical quantum states (phases or phase transitions) that are effectively described by some conformal field theories (CFTs) at low energies. We also focus on bosonic systems such as spin models. We shall first look for inspirations from two well known exotic quantum critical states: the DQCP and the $U(1)$ DSL -- we review these two states in Sec.~\ref{sec: review and summary}. The effective theory of DQCP is usually formulated in terms of some gauge theories that flow to strong coupling in the IR. However, there is a non-renormalizable description of DQCP based purely on local (gauge invariant) DOFs, formulated as a non-linear sigma model (NLSM) supplemented with a topological Wess-Zumino-Witten (WZW) term \cite{Tanaka05,Senthil2005,Nahum2015a,Wang2017}. {\footnote{Known $(2+1)$-d non-supersymmetric Lorentz-invariant renormalizable Lagrangians all only involve critical bosons/soft spins, Dirac/Majorana fermions, gauge fields, and their combinations. In particular, a NLSM where the DOF lives in a manifold and bears some constraints is non-renormalizable.}} This description comes with some virtues such as the locality of all DOFs and a manifest emergent $SO(5)$ symmetry. The fact that this description is strongly coupled 
in both UV and IR is usually viewed as a drawback. However, since we are now aiming to study ``non-Lagrangian'' theories that do not have weak-coupling descriptions anyway, it seems natural to try to turn this ``bug'' into a ``feature'', by generalizing the NLSM construction to some ``non-Lagrangian'' critical states. To achieve this, it turns out to be useful to consider the $U(1)$ DSL, which is known to be closely related to the DQCP \cite{Wang2017,Song2018,Song2018a}. If we can extend the NLSM construction to the $U(1)$ DSL,  we may then ``extrapolate'' the two theories to obtain an entire series of theories, some of which could possibly go beyond any mean-field $+$ weak fluctuations description.

With these motivations, we study a special type of $(2+1)$-d quantum many-body states, each labeled by two integers $(N, k)$, with $N\geqslant 5$ and $k\neq 0$. Their effective theories are formulated purely in terms of local DOF, described by a NLSM defined on a target space $SO(N)/SO(4)$, supplemented with a WZW term at level $k$ \cite{Wess1971, Witten1983}. The manifold $SO(N)/SO(4)$ is known as a Stiefel manifold (\eg see Ref. \cite{NakaharaBook2003}), so we dub these states ``Stiefel liquids" (SLs), and we refer to an SL labeled by $(N, k)$ as SL$^{(N, k)}$, and SL$^{(N, k=1)}$ may also be simply written as SL$^{(N)}$. These SLs have many interesting properties, such as a large emergent symmetry. Furthermore, there is a cascade structure among them: for each $k$, an SL with a smaller $N$ can be obtained from an SL with a larger $N$ by appropriately perturbing the latter and focusing on the resulting low-energy sector.

We propose that these SLs form cascades of extraordinary critical quantum liquids. In fact, SL$^{(5)}$ is precisely the effective field theory for the DQCP mentioned above. Furthermore, we will argue that SL$^{(6)}$ describes the $U(1)$ DSL discussed above, and is thus a dual description of the latter purely based on local DOFs. Due to the cascade structure, SL$^{(N>6)}$ can be viewed as extrapolating theories of the DQCP and $U(1)$ DSL. We will argue that SL$^{(N>6)}$ can flow to conformally invariant RG fixed points in the IR. Furthermore, they appear to have no obvious(renormalizable) gauge-theoretic description, and we conjecture that they are in fact non-Lagrangian. We provide various reasonable arguments to support this conjecture, although a rigorous proof is currently lacking, and it is unclear if such a proof is possible at all. However, as we argue, even if this conjecture can be disproved, one has to necessarily invoke novel ingredients of renormalizable field theories that have not been appreciated so far, and in this way new general insights can still be gained.

SL$^{(5)}$ and SL$^{(6)}$, namely, the DQCP and the $U(1)$ DSL, can both emerge in some lattice spin systems. The standard way to establish the \textit{emergibility} of these states is to construct their corresponding mean-field theory on the lattice,  based on the parton trick. For their non-Lagrangian counterparts with $N>6$, we do not have any known mean-field construction, and some alternative approach has to be adopted. In Sec.~\ref{sec: N>6}, we propose an approach, which is complimentary to the traditional mean-field approach, to study in which systems they may emerge. This approach is based on the hypothesis that a quantum state described by some effective field theory is emergible from a lattice system if and only if the quantum anomalies of the field theory match that of the lattice system.

Quantum anomalies, in particular, 't Hooft anomalies, were originally introduced as an obstruction to consistently coupling a system with certain global symmetry to a gauge field corresponding to this symmetry \cite{Hooft1980}, and recently it has been realized that they are also related to whether this symmetry can be realized in an on-site fashion \cite{Chen2011}. That is, the anomaly detects the structure of locality and/or the interplay between symmetry and locality of a system. Furthermore, quantum anomaly is an RG invariant, and it is powerful in constraining the IR fate of a system based on its UV information, in that a theory with a nontrivial anomaly is forbidden to have a symmetric short-range entangled ground state \cite{Hooft1980}. For a lattice system, the 't Hooft anomaly is intimately related \cite{Cheng2015, Jian2017, Cho2017, Metlitski2017} to the Lieb-Schutz-Mattis-type theorems \cite{Lieb1961, Oshikawa1999, Hastings2003, Po2017} for quantum matters on lattice systems.

\begin{figure*}
    \centering
    \includegraphics[width=0.95\textwidth]{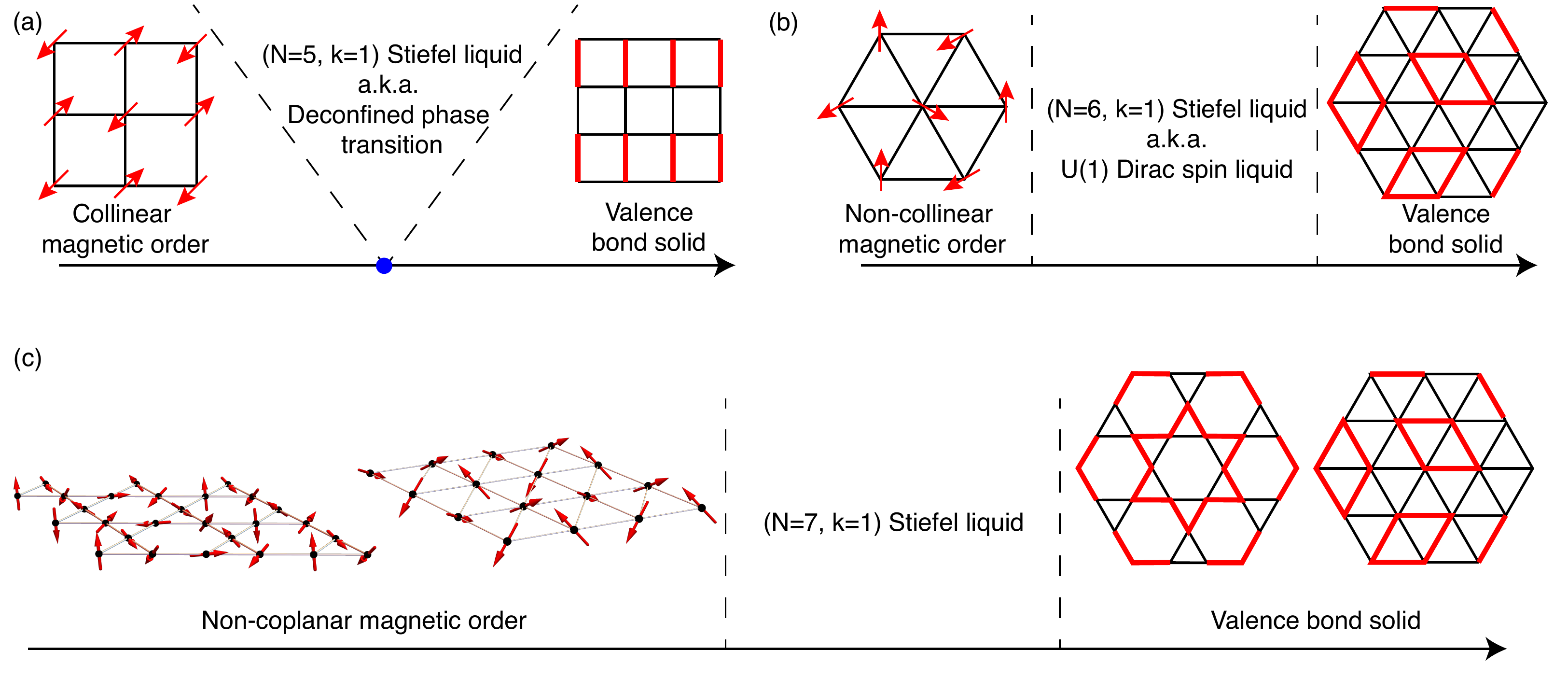}
    \caption{The Stiefel liquid out of intertwined orders in quantum magnets. (a) $(N=5, k=1)$ SL is the widely studied deconfined phase transition, which can arise from the intertwinement of  collinear magnetic order (e.g. Neel state) and valence bond solid. (b)  $(N=6, k=1)$ SL is the widely studied $U(1)$ Dirac spin liquid, which can arise from the intertwinement of  non-collinear magnetic order (but coplanar) and valence bond solid. (c) $(N=7, k=1)$ SL is a new critical quantum liquid, which can arise from the intertwinement of  non-coplanar magnetic order and valence bond solid. On triangular lattice the non-coplanar order is known as the tetrahedral order. On kagome lattice the non-coplanar order is known as the cuboctahedral order, in which the magnetizations on the three sublattices are $S_A=-\vec Q_3 \cos (n\pi)-\vec Q_1 \cos[(m+n)\pi]$, $S_B=\vec Q_3 \cos(n\pi)-\vec Q_2\cos(m\pi)$, $S_C=\vec Q_2\cos(m\pi)+\vec Q_1\cos[(m+n)\pi]$ with $Q_{1,2,3}$ being orthogonal to each other.}
    \label{fig:intertwined}
\end{figure*}

In this paper, matching the anomalies of two seemingly different theories motivates us to propose that these two theories can emerge in the same physical setup. In addition, anomaly-based considerations also enable us to make specific concrete predictions of a system, without referring to the details of its theory (such as its Hamiltonian). For example, we propose that SL$^{(7)}$ can be realized in spin-$1/2$ triangular and kagome lattice systems, and we are able to predict some of its detailed physical properties, such as the crystal momenta of gapless modes in this realization. One interesting observation is that SL$^{(7)}$ can naturally arise in the vicinity of competing non-coplanar magnetic order and VBS, which is a natural generalization of that SL$^{(5)}$ (DQCP) can naturally arise in the vicinity of competing collinear magnetic order and VBS, and SL$^{(6)}$ ($U(1)$ DSL) can naturally arise in the vicinity of competing non-collinear but coplanar magnetic order and VBS. We illustrate these in Fig.~\ref{fig:intertwined}.

Thinking more broadly, the absence of mean-field constructions or renormalizable continuum Lagrangians forces us to focus on more universal aspects of the critical states. The natural goal here is to obtain an   {\textbf {intrinsic characterization}} of universal many-body physics: a complete characterization of the universal properties of a many-body system, without explicitly referring to any Hamiltonian, Lagrangian, or wave function. This goal is also motivated by the observation that, although often useful, a Hamiltonian/Lagrangian/wave function is just a specific UV regularization of the universal physics of the underlying quantum phase or phase transition. Therefore, it is conceptually and aesthetically desirable to find such an intrinsic characterization. (Of course, to obtain non-universal details of a many-body system, its Hamiltonian/Lagrangian/wave function is needed.) In fact, such an intrinsic characterization has been (partly) achieved in various systems, such as CFTs in $(1+1)$-d \cite{CFTBook}, a large class of gapped phases in various dimensions \cite{Kitaev2006, Etingof2009, Chen2013, Bonderson2014, Lan2015, Lan2016, Lan2016a, Lan2017, Gaiotto2017, Lan2018, Gaiotto2019, Kong2020, Johnson-Freyd2020}, and symmetry-enriched $U(1)$ quantum spin liquids in $(3+1)$-d  \cite{Wang2013, Wang2015, Zou2017, Zou2017a, Hsin2019, Ning2019}. The present work can be viewed as a small step toward this ambitious goal for more complicated critical states of matter.

\section{Summary of results}

The main results of this paper are summarized below. This part also serves as a map for this paper.

\begin{enumerate}
    
    \item We propose a class of exotic $(2+1)$-d quantum many-body states dubbed Stiefel liquids (SLs), each indexed by two integers, $N\geqslant 5$ and $k\neq 0$. The effective theory of an SL with index $(N, k)$, SL$^{(N, k)}$, is formulated as a nonlinear sigma model (NLSM) on target manifold $SO(N)/SO(4)$, supplemented with a Wess-Zumino-Witten (WZW) term at level $k$. The target manifold is also known as Stifel manifold $V_{N,N-4}$ (or simply $V_N$ in this paper), hence the name Stiefel liquid. The Stiefel manifold can be parameterized using an $N\times (N-4)$ matrix $n_{ji}$ satisfying $n^Tn=I_{N-4}$, where $I_{N-4}$ is the $(N-4)$-dimensional identity matrix. The NLSM is defined using the action
    \beq
    \qquad S^{(N,k)}[n]=\frac{1}{2g}\int d^{2+1}x\Tr(\partial_\mu n^T\partial^\mu n)+k\cdot S^{(N)}_{\rm{WZW}}. \nonumber
    \eeq
    The first term is a standard kinetic term, and the WZW term is well defined since $\pi_4(V_N)=\mathbb{Z}$ and $\pi_i(V_N)$ are trivial for $i<4$. The detailed form of the WZW term will be discussed in Sec. \ref{subsec: SL begins}.
    As we will mainly be interested in $k=1$, we also use SL$^{(N)}$ to denote SL$^{(N, 1)}$.
    
    The above NLSM is non-renormalizable, so its dynamics at strong coupling ($g$ not small) is not clearly defined on face value. We can nevertheless argue, as we do in Sec.~\ref{subsec: IR fate}, that for each $k\neq0$ there is a critical $N_c(k)$, such that for $N>N_c(k)$ the theory can flow to a stable CFT fixed point at strong coupling. The stable fixed point is separated from the spontaneous symmetry breaking phase in the weak-coupling regime (small $g$) by a critical point. Those stable fixed points represent the critical Stiefel liquids which are the main focus of this paper.  We argue, based on existing numerical results,  that for $k=1$ it is likely that $5<N_c<6$.
    
    \item In Sec.~\ref{subsec: symmetries} we carefully discuss the symmetries of the Stiefel liquids. It turns out that the SL$^{(N, k)}$ theory has a rather nontrivial symmetry group. The continous symmetry is $SO(N)\times SO(N-4)$ if $N$ is odd, and for $N$ even it is $(SO(N)\times SO(N-4))/\mathbb{Z}_2$. An SL also has discrete $\mc{C}$, $\mc{R}$ and $\mc{T}$ symmetries. These discrete symmetries turn out to act nontrivially in the $SO(N)$ and $SO(N-4)$ internal spaces, so they form semi-direct products ($\rtimes$) with the continuous symmetries -- we discuss this carefully in Sec.~\ref{subsec: symmetries}. The SL theory also has the standard Poincar\'e symmetry group. The full symmetry is therefore the Poincar\'e plus
    \beq
    (SO(N)\times SO(N-4) )\rtimes(\mathbb{Z}_2^{\mc{C}}\times\mathbb{Z}_2^{\mc{T}}), &&\hspace{5pt} N=2n+1 \nonumber \\
    \left(\frac{SO(N)\times SO(N-4)}{\mathbb{Z}_2} \right)\rtimes(\mathbb{Z}_2^{\mc{C}}\times\mathbb{Z}_2^{\mc{T}}), &&\hspace{5pt} N=2n \nonumber
    \eeq

    \item In Sec.~\ref{subsec: cascade structure} we discuss a cascade structure of the SLs: for each $k$, by appropriately breaking the symmetry of SL$^{(N, k)}$ with a larger $N$, the low-energy sector of the resulting theory is another SL with the same $k$ but a smaller $N$. In particular, if we condense the first column of the $n_{ji}$ matrix to a fixed unit vector, say $n_{j1}=(1,0,0...)^T$, we break the $\sim SO(N)\times SO(N-4)$ symmetry of SL$^{(N,k)}$ down to $\sim SO(N-1)\times SO(N-5)$ and obtain the SL$^{(N-1,k)}$ theory.
    
    \item The SL$^{(5)}$ theory turns out to be nothing but the sigma-model description of the DQCP. In Appendix~\ref{app: gauge theory of (5, k)}, we extend this correspondence to general $k>0$ and propose that  SL$^{(5, k)}$ can be described by a $USp(2k)$ gauge theory with $N_f=2$ flavors of fermions -- the case with $k=1$ and $USp(2)=SU(2)$ is a familiar result. 
    
    \item In Sec.~\ref{sec: N=6: DSL}, we argue that SL$^{(6, k)}$ is equivalent to the $U(k)$ DSL, \ie 4 flavors of gapless Dirac fermions coupled to a $U(k)$ gauge field. The field $n_{ji}$ on the Stiefel manifold $SO(6)/SO(4)$ corresponds to the monopoles in the $U(k)$ gauge theory.  In particular, for $k=1$ this gives an SL description of the familiar $U(1)$ Dirac spin liquid. In Appendix ~\ref{app: DSL explicit}, we further support this correspondence by explicitly matching the anomalies of the $U(1)$ Dirac spin liquid with the SL$^{(6)}$ theory. In particular, we verify that various probe monopoles in the $U(1)$ DSL theory do have the nontrivial properties required by the anomalies.
    
    \item In Sec. \ref{subsec: SL begins}, we conjecture that SL$^{(N)}$ with $N>6$ is {\it non-Lagrangian}, \ie its conformally invariant fixed point has no description in terms of a weakly-coupled renormalizable continuum Lagrangian at any scale. Equivalently, SL$^{(N)}$ with $N>6$ cannot be accessed using any mean-field theory plus weak fluctuations. In particular, they are beyond the standard parton (mean-field) construction and gauge theory. We further elaborate on the arguments underlying this conjecture in Sec. \ref{sec: discussion}.
    
    \item In Sec. \ref{Sec: Anomaly Analysis}, we analyze the quantum anomalies of the SLs. One way to characterize the anomaly is to view our $(2+1)$-d system as the boundary of a $(3+1)$-d bulk, and the bulk has a nontrivial topological response when coupled to background gauge fields. We can couple the SL$^{(N,k)}$ theory to a background gauge field with gauge symmetry $(SO(N)\times SO(N-4) )\rtimes(\mathbb{Z}_2^{\mc{C}}\times\mathbb{Z}_2^{\mc{T}})$, which turns out (as we discuss carefully in Sec.~\ref{Sec: Anomaly Analysis}) to be equivalent to coupling to an $O(N)\times O(N-4)$ gauge field, with the following restriction on the gauge bundles:
    \be
    w_1^{O(N)}+w_1^{O(N-4)}+w_1^{TM}=0 \hspace{5pt} ({\rm{mod}}\hspace{2pt} 2). \nonumber
    \ee
    The three terms are the first Stiefel-Whitney (SW) classes of the $O(N)$, $O(N-4)$ and $(3+1)$-d spacetime tangent bundles, respectively. The anomaly is then given by the bulk topological response (see Sec.~\ref{Sec: Anomaly Analysis} for the derivation)
    \begin{widetext}
    \beq
    ik\pi\int_{X_4}\left\{w_4^{O(N)}+w_4^{O(N-4)}+\left[w_2^{O(N-4)}+\left(w_1^{O(N-4)}\right)^2\right]\left(w_2^{O(N)}+w_2^{O(N-4)}\right)+\left(w_1^{O(N-4)}\right)^4\right\}. \nonumber
    \eeq
    \end{widetext}
    Here $w_1$, $w_2$ and $w_4$ are the first, second and fourth SW classes of the corresponding bundles, respectively, and all the products involved are cup products. This anomaly is $\mathbb{Z}_2$-classified and only SLs with odd $k$ are nontrivial. The above bulk response gives the complete anomaly of SL$^{(N)}$ with an odd $N$. The case with an even $N$ is more complicated, and the above result is just a partial characterization in this case. To improve this result, in Sec. \ref{subsec: monopoles} we characterize the anomaly for the even-$N$ case by studying the monopoles corresponding to the global symmetry of the theory. This characterization, although still partial, contains physics beyond that in the above bulk action. In particular, SL$^{(N, k)}$ with an even $N$ and $k=2$ (mod $4$) turns out to be also anomalous, which cannot be detected by the above bulk action.
    
    The above anomaly of SL$^{(N, k)}$ with odd $k$ cannot be saturated by a gapped topological order -- the IR theory has to be either gapless or symmetry-breaking. In Sec. \ref{subsec: semion TO}, we show that if time-reversal and space reflection symmetries are broken, the anomalies of SL$^{(N)}$  can be realized by a semion (or anti-semion) topological order, for all $N\geqslant 5$.
    
    \item In Sec.~\ref{sec: N>6} we discuss possible lattice realizations of Stiefel liquids with $N>6$, which as we discussed are likely non-Lagrangian. The intrinsic absence of a mean-field construction for these states makes it challenging to decide whether a Stiefel liquid, say with $N=7$, can emerge out of a lattice system. We therefore propose an approach based on anomaly matching: we hypothesize that a state (like SL$^{(7)}$) is {\it emergible} from a lattice system if and only if the 't Hooft anomalies of the state match that of the lattice system. The anomalies of the lattice system come from the generalized Lieb-Schultz-Mattis theorems. The necessity of this condition is actually known, so we only hypothesize the sufficiency part.
    
    We then find that SL$^{(7)}$, the simplest non-Lagrangian SL, can indeed be realized on lattice spin systems if the microscopic physical symmetries are properly implemented in the low-energy theory. Here the ``microscopic symmetries'' include the $SO(3)$ rotation, time-reversal and lattice symmetries. Specifically, we identify two realizations of the SL$^{(7)}$ theory on triangular and Kagome lattices, respectively, both with one spin-$1/2$ moment sitting on each lattice site. Many observable properties of these specific realization of SL$^{(7)}$ are discussed. In particular, we find that this state can naturally arise in the vicinity of competing non-coplanar magnetic order and valance-bond solid (VBS). The corresponding non-coplaner magnetic orders are known as the tetrahedral order on triangular lattice and the cuboctahedral order on Kagome lattice, respectively. These realizations of the SL$^{(7)}$ state are very natural generalizations of the realizations of SL$^{(5)}$ (DQCP) and SL$^{(6)}$ ($U(1)$ DSL), since the DQCP  arises in the vicinity of competing collinear magnetic order and VBS, and the $U(1)$ DSL arises in the vicinity of competing  coplanar magnetic orders and VBS (as illustrated in Fig.~\ref{fig:intertwined}).
    
\end{enumerate}

We finish with some discussions in Sec. \ref{sec: discussion}. Various appendices contain additional details, as well as some general results.

Before proceeding, we will first briefly review the physics of the deconfined quantum critical point and the $U(1)$ Dirac spin liquid in Sec.~\ref{sec: review and summary}.

\section{Review of background} \label{sec: review and summary}

In this section, we first review some aspects of the DQCP and $U(1)$ DSL, which partly motivate the present work.

\subsection{Deconfined quantum critical point} \label{subsec: review of DQCP}

The classic DQCP was proposed as a critical theory for a quantum phase transition between a Neel AF and a VBS on a square lattice  \cite{Senthil2003, Senthil2003a}. Because the symmetry respected by either of these two phases is not a subgroup of the symmetry of the other, such a transition is considered to be beyond the Landau-Ginzburg-Wilson-Fisher paradiam if it is continuous. The original formulation of the DQCP is in terms of two flavors of bosons coupled to a dynamical $U(1)$ gauge field, and over the years many dual formulations have been proposed \cite{Senthil2005, Wang2017}. 

The formulation that is most relevant for our purpose is written in terms of a 5-component unit vector $\vec n$, whose first 3 and last 2 components can be thought of as the order parameters of the AF and the VBS, respectively. So this is a formulation directly based on local DOFs. The low-energy effective theory at the DQCP is a NLSM with a WZW term:
\beq \label{eq: DQCP effctive theory}
S^{(5, k)}=S_0+k\cdot S^{(5)}_{\rm WZW}
\eeq
The meaning of the superscripts ``$(5, k)$" will be clear later. For the DQCP, $k=1$, and this seemingly redundant factor is inserted for later convenience. The first term $S_0=\int d^3x\frac{1}{2g}(\partial_\mu\vec n)^2$ is the action of the usual NLSM. To define the WZW term, $S_\WZW^{(5)}$, one first needs to add one more dimension to the physical spacetime and extend the unit vector $\vec n$ into this extra dimension. We denote the coordinate of this extra dimension by $u$, and the extended unit vector by $\vec n^e$, such that $\vec n^e(x, y, t, u=0)=\vec n(x, y, t)$ and $\vec n^e(x, y, t, u=1)=\vec n^r$, with $n^r$ being an arbitrary fixed reference vector, which, for example, can be taken to be $\vec n^r=(1,0,0,0,0)^T$. For notational brevity, in the following we will drop the superscript ``e" in the extended vector and simply write it as $\vec n$, and the meaning of $\vec n$ should be clear from the context. In terms of the extended vector, the WZW term is
\beq
S_\WZW^{(5)}=\frac{2\pi}{\Omega_4}\int_0^1du\int d^3x\det(\tilde n)
\eeq
where $\Omega_4=8\pi^2/3$ is the volume of $S^4$ with unit radius, and $\tilde n$ is a 5-by-5 matrix defined as
$$
\tilde n\equiv(n, \partial_x n, \partial_y n, \partial_t n, \partial_u n)
$$
Namely, the first column of $\tilde n$ is $\vec n$, and its last 4 columns are derivatives of $\vec n$ arranged in the above way. More explicitly,
$$
\det(\tilde n)=\epsilon^{i_1i_2i_3i_4i_5}n_{i_1}\partial_xn_{i_2}\partial_yn_{i_3}\partial_tn_{i_4}\partial_un_{i_5}
$$
Geometrically, the WZW term (apart from the factor of $2\pi$) is the ratio of the volume swept by $\vec n$ (as its coordinates vary) and $\Omega_4$, the total volume of $S^4$ with unit radius. Physically, the WZW term intertwines the Neel and VBS orders \cite{Senthil2005, Wang2017} (see also earlier related works \cite{Read1989, Read1990}).

The theory Eq. \eqref{eq: DQCP effctive theory} enjoys an $I^{(5)}=SO(5)$ symmetry, under which $\vec n$ transforms in its vector representation. The purpose for the notation $I^{(5)}$ will be clear later. It is useful to imagine enlarging the $SO(5)$ symmetry group into $O(5)$. Due to the WZW term, the improper $Z_2$ rotation of this $O(5)$ group is {\em not} a symmetry of Eq. \eqref{eq: DQCP effctive theory}, but when it combines with the reversal of a space or time coordinate, it becomes the reflection or time reversal symmetry. We denote this symmetry as $O(5)^T$.

The Neel-VBS transition is driven by a rank-$2$ anisotropy term $\lambda(n_1^2+n_2^2+n_3^2-n_4^2-n_5^2)$, with $\lambda<0$ favoring the Neel order and $\lambda>0$ favoring the VBS order. At weak coupling the sigma model orders spontaneously and the Neel-VBS transition driven by the anisotropy will be first order. The DQCP, as a continuous Neel-VBS transition, then requires a nontrivial fixed point at strong coupling. The strong coupling dynamics, strictly speaking, is not well defined just from the sigma model Lagrangian since the theory is not renormalizable. Nevertheless, if such a strong-coupling fixed point exists, several nontrivial properties of this fixed point can be readily inferred: 

\begin{enumerate}
    \item The theory has the full  $O(5)^T$ symmetry.
    \item Local operators that transform trivially under $O(5)^T$ must be RG irrelevant.
    \item The theory has a 't Hooft anomaly, characterized by the topological action of a $(3+1)$-d symmetry-protected topological phase (SPT) whose boundary can host the DQCP:
    \beq \label{eq: DQCP complete anomaly}
    \mc{Z}^{(5)}_{\rm bulk}=\exp\left(i\pi\int_{X_4}w_4^{O(5)}\right)
    \eeq
    where $X_4$ is the $4$ dimensional spacetime manifold that the bulk SPT lives in, and $w_4^{O(5)}$ is the fourth Stiefel-Whitney (SW) class of the probe $O(5)^T$ gauge bundle that couples to the SPT. This topological response theory is subject to a constraint, $w_1^{O(5)}=w_1^{TM}\ ({\rm mod\ }2)$, with $w_1^{TM}$ the first SW class of the tangent bundle of $X_4$. This constraint guarantees that the orientation-reversal symmetry (\ie reflection and time reversal) is accompanied by an improper $Z_2$ rotation of the $O(5)$. The $SO(5)$ anomaly has been derived in Ref.~\cite{Wang2017}, and in Appendix~\ref{app: anomaly of DQCP} we extend it to the full $O(5)^T$.
\end{enumerate}

Below we collect some further results on the anomaly that are relevant to the present paper.

\begin{enumerate}

    \item The anomaly is $\mathbb{Z}_2$ classified, \ie they disappear if two copies of this theory are stacked together. This means that the anomalies of the theory described by Eq. \eqref{eq: DQCP effctive theory} remains the same if $k$ is changed by any even integer.
    
    \item In many cases, the anomaly of a $(2+1)$-d theory can be realized by a symmetric gapped topological order, but the anomaly of the DQCP cannot, if the system preserves both the $SO(5)$ and an orientation-reversal symmetry. In other words, due to this anomaly, as long as the system preserves the $SO(5)$ symmetry together with any orientation-reversal symmetry, it has to be gapless. This is an example of symmetry-enforced gaplessness \cite{Wang2014}.
    
    \item A physical way to characterize the anomaly of this theory is to gauge the $SO(5)$ symmetry, and examine the structure of the monopoles of the resulting $SO(5)$ gauge field. An $SO(5)$ monopole can be obtained by embedding a $U(1)$ monopole into one of the generators of the $SO(5)$ gauge group \cite{Wu1975}, and the field configuration of this monopole breaks the $SO(5)$ into $SO(3)\times SO(2)$. Then it is meaningful to ask what representation this $SO(5)$ monopole carries under the remaining $SO(3)\times SO(2)$. It turns out that it carries a spinor representation under $SO(3)$ and no charge under $SO(2)$ (up to binding the original matter fields in the fundamental representation of the $SO(5)$).
    
\end{enumerate}

Finally, we comment on the current status of the numerical studies on the actual IR dynamics of the DQCP. The emergence of the $SO(5)$ symmetry that rotates between Neel and VBS orders has been observed numerically \cite{Nahum2015a}. On the other hand,
the seemingly continuous transition \cite{Sandvik2006,Jiang2007,Melko2008,Charrier2008,Motrunich2008,Kuklov2008,Chen2009,Lou2009,Banerjee2010,Charrier2010,Sandvik2010,Bartosch2013,Harada2013,Chena2013,Nahum2015,Sreejith2015,Shao2016,Liu2019,Li2019,Sandvik2020} shows some puzzling features, including unconventional finite-size behavior \cite{Sandvik2006,Nahum2015,Shao2016} and measured critical exponents that violate bounds from conformal bootstrap \cite{Nakayama2016,Poland2019}. One plausible explanation is that the DQCP is pseudo-critical, \ie it is not a truly continuous phase transition, but its correlation length is very large. A universal mechanism for such pseudo-criticality based on the notion of complex fixed points \cite{Wang2017, Gorbenko2018} have been proposed. In terms of the WZW sigma model Eq.~\eqref{eq: DQCP effctive theory}, this means that the hypothesized strong-coupling fixed point does not really exist and the theory flows all the way to the weakly coupled, first order transition regime. However, there is a region, around some nontrivial coupling strength $g^*$, in which the RG flow is slow (also known as ``walking''\cite{Kaplan2009}). As a result the system behaves almost like a critical point up to some large length scale. A theory for the walking behavior in the sigma model has been put forward in Refs.~\cite{Ma2019,Nahum2019}.

\subsection{U(1) Dirac spin liquid} \label{subsec: review of DSL}

The $U(1)$ DSL was introduced as a critical quantum liquid that can emerge in certain spin systems \cite{Affleck1988, Wen1995}, and its contemporary standard model is formulated in terms of 4 flavors of gapless Dirac fermions minimally coupled to a dynamical $U(1)$ gauge field, with the Lagrangian
\beq \label{eq: DSL standard model}
\mc{L}=\sum_{i=1}^4\bar\psi_i i\slashed D_a\psi_i+\frac{1}{4e^2}f_{\mu\nu}f^{\mu\nu}
\eeq
where $\slashed D_a$ is the covariant derivative of the Dirac fermions, $\psi$, which are coupled to the dynamical $U(1)$ gauge field, $a$, whose field strength is $f_{\mu\nu}=\partial_\mu a_\nu-\partial_\nu a_\mu$. The Dirac fermion $\psi$ is {\em not} a local (gauge invariant) excitation here. Naively the simplest local operators are fermion biliners like $\bar{\psi}_i\psi_j$. It turns out that the most important local operators are the monopole operators \cite{Borokhov2002,Dyer2013} -- these are operators that insert $U(1)$ gauge flux, in units of $2\pi$, into the system.

The symmetries of the DSL are discussed in detail in Refs.~\cite{Borokhov2002,Dyer2013,Song2018}. In particular, it has an $I^{(6)}=(SO(6)\times U(1)_{\rm top})/Z_2$ symmetry, and the purpose for the notation $I^{(6)}$ will be clear later. The Dirac fermions transform under a flavor $SU(4)$ which is the spinor group of the $SO(6)$. The fermion bilinears $\bar{\psi}_i\psi_j$ form a singlet $\oplus$ an adjoint representation under $SO(6)$. The $U(1)_{\rm top}$ corresponds to the conservation of gauge flux, with conserved current $j_\mu=\epsilon_{\mu\nu\lambda}\partial^\nu a^\lambda/(2\pi)$ (the subscript ``top" is due to the fact that this current conservation does not rely on the detailed equations of motion and is therefore ``topological''). By definition only monopole operators are charged under the $U(1)_{\rm top}$. It turns out \cite{Borokhov2002} that the most fundamental monopoles also transform as a vector under the $SO(6)$. More concretely, the monopole can be represented by a 6-component complex bosonic field $\Phi$, such that the $SO(6)$ rotates the components of $\Phi$, and the $U(1)_{\rm top}$ acts by multiplying $\Phi$ by a phase factor.

Besides $I^{(6)}$, this theory also enjoys discrete charge conjugation $\mc{C}$, reflection $\mc{R}$, and time reversal $\mc{T}$ symmetries. To describe the actions of these discrete symmetries, it is useful to imagine enlarging the $SO(6)$ and $U(1)_{\rm top}$ symmetries to $O(6)$ and $O(2)$, respectively. Then it turns out \cite{Song2018} that the improper $Z_2$ rotation of neither $O(6)$ nor $O(2)$ is a symmetry of the DSL, but the $\mc{C}$ symmetry can be viewed as the combination of these two improper $Z_2$ rotations. The $\mc{R}$ and $\mc{T}$ symmetries can be viewed as a combination of spacetime orientation reversal and the improper $Z_2$ rotation of either $O(6)$ or $O(2)$ (but not both).

The $\Phi$ operators are the most fundamental local operators in the theory, in the sense that any other local operator can be built up using the $\Phi$'s.  Let us look at some examples. The $SU(4)$-singlet mass operator $\bar{\psi}\psi$ is identified as
$$
\bar\psi_i\psi_i\sim
i\epsilon^{abcdef}(\Phi_a^\dag\Phi_b-\Phi_a\Phi_b^\dag)(\Phi_c^\dag\Phi_d-\Phi_c\Phi_d^\dag)(\Phi_e^\dag\Phi_f-\Phi_e\Phi_f^\dag)
$$
where $a,b,c,d,e,f=1,2,3,4,5,6$. The $SU(4)$-adjoint mass operator (i.e., $\bar\psi_i \psi_j-\bar\psi_k \psi_k\delta_{ij}/4$) is identified as the rank-2 antisymmetric tensor of $\Phi$ that is neutral under the $U(1)_{\rm top}$, \ie $i(\Phi_a^\dag\Phi_b-\Phi_b^\dag \Phi_a)$. One can construct more of such identifications of operators. Here two operators are identified if they transform identically under all global symmetries (including both continuous and discrete symmetries).

The quantum anomalies of the $U(1)$ DSL have been partly analyzed in Ref.~\cite{Song2018} and more recently in Ref.~\cite{Calvera2021}, and we will study them further in this paper. 

The following facts about the nearby phases of the $U(1)$ DSL will be extremely useful for our later developments (see Refs. \cite{Wang2017, Song2018, Song2018a} for details).

\begin{enumerate}
    
    \item By condensing one component of $\Phi$ in the $U(1)$ DSL, the resulting state has the same symmetries and anomalies as the DQCP. It may even be possible that the theory indeed flows to DQCP once the monopole perturbation is turned on.
    
    \item Because of the above, just like the DQCP, the $U(1)$ DSL also enjoys symmetry-enforced gaplessness. One way to gap it out is to turn on an $SU(4)$-singlet mass of the Dirac fermions, which will drive the system into a semion (or anti-semion) topological order that breaks the orientation-reversal symmetries.
    
    \item By turning on a proper $SU(4)$-adjoint mass of the Dirac fermions, a condensate of $\Phi$ will be automatically induced, such that the remaining continuous symmetry of the system is $(SO(4)\times SO(2))/Z_2$, where the $SO(4)\subset SO(6)\subset I^{(6)}$ acts only on 4 components of $\Phi$, and the $SO(2)$ is a combination of $U(1)_{\rm top}$ and the $SO(2)\subset SO(6)\subset I^{(6)}$ acting on the other 2 components of $\Phi$. There are also remaining discrete $\mc{C}$, $\mc{R}$ and $\mc{T}$ symmetries. The anomalies are completely removed in this case. For $U(1)$ DSLs realized on lattice spin systems, this ``chiral symmetry breaking'' is the mechanism for realizing conventional Landau symmetry-breaking orders from the DSL -- examples include the coplanar ($120^{\circ}$) magnetic orders on triangular and Kagome lattices and various VBS orders \cite{Song2018a}.
    
\end{enumerate}

The $U(1)$ DSL is likely to be realized in spin-1/2 Heisenberg magnets on kagome and triangular lattices \cite{ Ran2007,Iqbal2016,He2017, Hu2019}. Furthermore, lattice Monte Carlo simulations support that the gauge theory Eq. \eqref{eq: DSL standard model} indeed flows to a CFT \cite{Karthik2015, Karthik2016}.

\section{Stiefel liquids: generality} \label{sec: SL in general}

In this section we introduce the general theory of SLs and discuss some of their  interesting properties. Recall that each Stiefel liquid will be labeled by two integers $(N, k)$, with $N\geqslant 5$ and $k\neq 0$. We will denote a SL corresponding to $(N, k)$ by SL$^{(N, k)}$. Since we will mostly focus on the case with $k=1$, we will also use the shorthand SL$^{(N)}$ to denote SL$^{(N, k=1)}$.

\subsection{Wess-Zumino-Witten sigma model on Stiefel manifold $SO(N)/SO(4)$} \label{subsec: SL begins}

The DOF of SL$^{(N, k)}$ is characterized by an $N$-by-$(N-4)$ real matrix, denoted by $n$, such that the columns of $n$ are orthonormal, \ie $n^Tn=I_{N-4}$, with $I_{N-4}$ the $(N-4)$-dimensional identity matrix. In mathematical terms, this matrix $n$ defines an $(N-4)$-frame in the $N$-dimensional Euclidean space. These $(N-4)$-frames live in a manifold $V_N\equiv SO(N)/SO(4)$, known as a Stiefel manifold \cite{NakaharaBook2003}. Taking the Stiefel manifold as the target space, a NLSM with the following action can be defined in any dimension:
\beq \label{eq: bare NLSM}
S_0[n]=\frac{1}{2g}\int d^{d+1}x\Tr(\partial_\mu n^T\partial^\mu n)
\eeq where the $n$ in the square braket indicates the dependence of the action on the configuration of $n$.

It is known that the homotopy groups of $V_N$ with any $N\geqslant 5$ satisfy that $\pi_{n}(V_N)=0$ for $n<4$, and $\pi_4(V_N)=\mathbb{Z}$, so a WZW term based on a closed 4-form on $V_N$ can be defined for any $N\geqslant 5$ in three spacetime dimensions \cite{Wess1971, Witten1983, Hull1991}. To define this WZW term, we will first add one more dimension to the physical spacetime and extend the matrix $n$ into this extra dimension. Denote the coordinate of the extra dimension by $u$, and the extended matrix by $n^e$, such that $n^e(x, y, t, u=0)=n(x, y, t)$ and $n^e(x, y, t, u=1)=n^r$, with $n^r$ a fixed reference matrix with entries $(n^r)_{ji}=\delta_{ji}$, where $(\cdot)_{ji}$ represents the entry in the $j$th row and $i$th column of the relevant matrix. For notational brevity, in the following we will drop the superscript ``e" in the extended matrix and simply write it as $n$, and the meaning of the matrix $n$ should be clear from the context.

To the best of our knowledge, the expression for such a WZW term or closed 4-form on $V_N$ with $N\geqslant 6$ is unavailable in the previous literature. We propose that the WZW term on $V_N$ is given by the following (real-time) action:
\beq \label{eq: WZW begins}
S_{\WZW}^{(N)}[n]=\frac{2\pi}{\Omega_4}\int_0^1du\int d^3x\sum_{i,i'=1}^{N-4}\det\left(\tilde n_{(ii')}\right)
\eeq
where the $N$-by-$N$ matrix $\tilde n_{(ii')}$ is given by
\beq \label{eq: n tilde}
\tilde n_{(ii')}=(n, \partial_x n_i, \partial_y n_i, \partial_t n_{i'}, \partial_u n_{i'})
\eeq
where $n_i$ represents the $i$th column of $n$ (the repeated indices $i$ and $i'$ are not summed over in the right hand side of Eq. \eqref{eq: n tilde}). That is, the first $N-4$ columns of $\tilde n_{ii'}$ are just $n$, and its last 4 columns are derivatives of the columns of $n$ arranged in the above way. More explicitly,
\begin{widetext}
\beq
\det(\tilde n_{(ii')})=\frac{1}{(N-4)!}\epsilon^{i_1i_2\cdots i_{N-4}}\epsilon^{j_1j_2\cdots j_N}n_{j_1i_1}n_{j_2i_2}\cdots n_{j_{N-4}i_{N-4}}\partial_xn_{j_{N-3}i}\partial_yn_{j_{N-2}i}\partial_tn_{j_{N-1}i'}\partial_un_{j_Ni'}
\eeq
\end{widetext}
where the $\epsilon$'s are the fully anti-symmetric symbols with rank $N-4$ and $N$, respectively. More comments on the mathematical aspects of this action are given in Appendix \ref{app: more on WZW}.

Taken together, the effective action of SL$^{(N)}$ is given by
\beq \label{eq: total action}
S^{(N)}[n]=S_0[n]+S_{\WZW}^{(N)}[n]
\eeq
The effective action of SL$^{(N, k)}$ is the level-$k$ generalization of Eq. \eqref{eq: total action}:
\beq \label{eq: level-k generalization}
    S^{(N, k)}=S_0+k\cdot\swzw .
\eeq

We remark that the $(2+1)$-d WZW-NLSM is non-renormalizable at strong couplings, so these theories should be defined with an explicit UV regularization. However, a symmetry-preserving local UV regularization should not affect the quantum anomalies of the theory. As for the IR dynamics of the theory, strictly speaking, it depends on the specific UV regularization, which is similar to the situation where a quantum phase or phase transition is described by a lattice Hamiltonian. In Sec. \ref{subsec: IR fate}, we will argue that there should exist UV regularizations under which $S^{(N, k)}$ flows to a conformally invariant fixed point under RG if $N$ is larger than a $k$-dependent critical value, $N_c(k)$, and thus describes a critical quantum liquid. If $N<N_c(k)$, $S^{(N, k)}$ does not flow to a nontrivial CFT; instead, its most likely fate is to flow to a Goldstone phase. In general, $N_c(k)$ increases with $k$, and we will argue that $N_c(1)<6$.

Notice when $N=5$, $n$ is just a column vector that can be identified as the vector $\vec n$ in Sec. \ref{subsec: review of DQCP}, and Eq. \eqref{eq: total action} is precisely Eq. \eqref{eq: DQCP effctive theory}. So SL$^{(5)}$ is precisely the DQCP. Since the DQCP, or SL$^{(5)}$, can be described by (renormalizable) gauge theories \cite{Wang2017}, it is natural to ask if other SLs can also be reformulated in terms of a renormalizable field theory, such as a gauge theory. In appendix \ref{app: gauge theory of (5, k)}, we provide an alternative description of SL$^{(5, k)}$ in terms of a $USp(2k)$ gauge theory\footnote{Refs. \cite{Lee2014} and \cite{Ippoliti2018} argue that SL$^{(5, k)}$ with $k=1,2$ can also arise in certain fermionic systems, such as monolayer and bilayer graphenes. We note that the $USp(2k)$ gauge theories in Appendix. \ref{app: gauge theory of (5, k)} are purely bosonic theories, and all Stiefel liquids are also fundamentally bosonic theories and do not need to involve fermions in any intrinsic way.}, and in Sec. \ref{sec: N=6: DSL} we provide a $U(k)$-gauge-theoretic description of SL$^{(6, k)}$.
    
However, we cannot find any gauge-theoretic formulation for SL$^{(N)}$ with $N>6$. In fact, due to their delicate symmetry structure (discussed below) and intricate anomaly properties (discussed in Sec. \ref{Sec: Anomaly Analysis}), we conjecture that the conformally invariant fixed points corresponding to SL$^{(N>6)}$ are {\it non-Lagrangian}, \ie they have no description in terms of a weakly-coupled renormalizable continuum Lagrangian at any scale. In Sec. \ref{sec: discussion}, we present more detailed arguments supporting this conjecture. As for SL$^{(N, k)}$ with $N>6$ and $k>1$, it may also be non-Lagrangian if it flows to a CFT.

Below we discuss the basic physical properties of the SLs in more detail.

\subsection{Symmetries} \label{subsec: symmetries}

In addition to the Poincar\'e symmetry, the actions in Eqs. \eqref{eq: total action} and \eqref{eq: level-k generalization} are invariant under an $SO(N)$ transformation, which acts as:
\beq \label{eq: SO(N) action}
n\rightarrow Ln, L\in SO(N),
\eeq
and another $SO(N-4)$ transformation, which acts as
\beq \label{eq: SO(N-4) action}
n\rightarrow nR, R\in SO(N-4).
\eeq
Notice that for even $N$, the two $\mathbb{Z}_2$ centers $L=-I_{N}$ and $R=-I_{N-4}$ act identically. So $S^{(N)}$ and $S^{(N, k)}$ have a continuous symmetry group $I^{(N)}$, where $I^{(N)}=(SO(N)\times SO(N-4))/Z_2$ for even $N$ and $I^{(N)}=SO(N)\times SO(N-4)$ for odd $N$. Recall that $I^{(5)}$ and $I^{(6)}$ were already introduced in Sec. \ref{sec: review and summary}.
    
Besides this continuous symmetry,  $S^{(N)}$ and $S^{(N, k)}$ also have discrete charge conjugation, reflection, and time reversal symmetries, \ie $\mc{C}$, $\mc{R}$ and $\mc{T}$. These symmetries can be combined with elements of $I^{(N)}$ to be redefined, and we will utilize this freedom to redefine these symmetries whenever useful later in the paper.
    
A particular implementation of these discrete symmetries for $N\geqslant 6$ is
\beq \label{eq: canonical CRT actions}
\begin{split}
&\mc{C}: n_{ji}\rightarrow(-1)^{f_{ji}}n_{ji}\\
&\mc{R}:
n_{ji}\rightarrow
\left\{
\begin{array}{lr}
n_{ji}, & j\leqslant N-1\\
-n_{ji}, & j=N
\end{array}
\right.\\
&\mc{T}:
n_{ji}\rightarrow
\left\{
\begin{array}{lr}
n_{ji}, & j\leqslant N-1\\
-n_{ji}, & j=N
\end{array}
\right.
\end{split}
\eeq
with $f_{ji}=1$ if $(j=N\ \&\  i< N-4)$ or $(j< N\ \&\ i=N-4)$, and $f_{ji}=0$ otherwise.
Notice that $\mc{R}$ and $\mc{T}$ also need to flip a spatial or temporal coordinate, respectively.
    
Another useful way to characterize these symmetries for $N\geqslant 6$ is to imagine enlarging the $SO(N)$ and $SO(N-4)$ in $I^{(N)}$ to $O(N)$ and $O(N-4)$, respectively. Then the improper rotation of neither the $O(N)$ nor the $O(N-4)$ is a symmetry due to the WZW term, but the combination of these two improper rotations is the $\mc{C}$ symmetry. Also, $\mc{R}$ and $\mc{T}$ can be viewed as an improper rotation of either $O(N)$ or $O(N-4)$ combined with a reversal of the appropriate spacetime coordinate.
    
When combined with the continuous symmetries, the full symmetry group (besides the Poincar\'e) can be written as 
\beq
(SO(N)\times SO(N-4) )\rtimes(\mathbb{Z}_2^{\mc{C}}\times\mathbb{Z}_2^{\mc{T}}), &&\hspace{5pt} N=2n+1, \nonumber \\
\left(\frac{SO(N)\times SO(N-4)}{\mathbb{Z}_2} \right)\rtimes(\mathbb{Z}_2^{\mc{C}}\times\mathbb{Z}_2^{\mc{T}}), &&\hspace{5pt} N=2n,
\eeq
where $\mathbb{Z}_2^{\mc{C}}$ and $\mathbb{Z}_2^{\mc{T}}$ are generated by $\mathcal{C}$ and $\mathcal{T}$, respectively. The semi-direct product $\rtimes$ comes from nontrivial relations like Eq.~\eqref{eq: canonical CRT actions}. We do not need to list $\mathbb{Z}_2^{\mc{R}}$ above since it is related to $\mathbb{Z}_2^{\mc{T}}$ through a Lorentz rotation.
    
For $N=5$, the matrix $n$ contains only a single column, and we can suppress its column index and denote it by $n_j$, with $j=1,2,\cdots, 5$. In this case, the above $\mc{C}$ symmetry does not exist{\footnote{However, some elements of the $I^{(5)}=SO(5)$ group may be interpreted as a charge conjugation symmetry in certain gauge-theoretic formulation of this theory \cite{Wang2017}.}}, and we take the actions of the $\mc{R}$ and $\mc{T}$ symmetries to be
\beq \label{eq: CRT for N=5}
\begin{split}
&\mc{R}: n_{1,2,3,4}\rightarrow n_{1,2,3,4},
n_5\rightarrow-n_5\\
&\mc{T}:
n_{1,2,3,4}\rightarrow n_{1,2,3,4},
n_5\rightarrow-n_5
\end{split}
\eeq
which is analogous to the case with $N\geqslant 6$.

\subsection{Cascade structure of the SLs} \label{subsec: cascade structure}

Now we discuss the relations between different SLs.

As is common for WZW theories, SL$^{(N, k)}$ with $k>1$ can be viewed as $k$ copies of SL$^{(N)}$ that have strong ``ferromagnetic" couplings between them. More precisely, the action of $k$ copies of decoupled SL$^{(N)}$ is $\sum_{i=1}^kS^{(N)}[n_i]$. A strong ferromagnetic coupling, $-\sum_{i\neq j}g_{ij}\int d^3x\Tr(n_i^Tn_j)$ with $g_{ij}>0$, has the tendency of identifying different $n_i$'s as a single matrix, $n$. Focusing on the low-energy sector that contains only $n$, the total action takes the form of Eq. \eqref{eq: level-k generalization}. That is, we can get SL$^{(N, k)}$ by appropriately coupling $k$ copies of SL$^{(N)}$. Notice that this coupling does not break the symmetries discussed in Sec. \ref{subsec: symmetries}, as expected.
    
SL$^{(N, k)}$ and SL$^{(N, -k)}$ almost have identical properties, because one of them can be obtained from the other by an improper $Z_2$ operation of either the $O(N)$ or $O(N-4)$. But they have opposite quantum anomalies, since a composed system of SL$^{(N, k)}$ and SL$^{(N, -k)}$ can be turned into a trivial state by a strong ferromagnetic coupling. Because of the similarity between SL$^{(N, \pm k)}$, in this paper we will mostly focus on the case with $k>0$.
    
Next, we remark on an interesting and important specific property of the SLs: the cascade structure.
    
Suppose we start with SL$^{(N, k)}$ described by $S^{(N, k)}$ with $N\geqslant 6$, and fix, say, $n_{11}=n_{22}=\cdots=n_{mm}=1$ with $m\leqslant N-5$, while allowing the other components of $n$ to fluctuate under the orthonormal constraint, $n^Tn=I_{N-4}$. Now fluctuations of the entries in the first $m$ rows and the first $m$ columns of $n$ are frozen, while fluctuations of the other entries remain at low energies. The $I^{(N)}$ symmetry of $S^{(N, k)}$ is {\em explicitly} broken to $I^{(N,m)}$, where $I^{(N, m)}=[SO(N-m)\times SO(N-4-m)\times SO(m)]/Z_2$ if both $N$ and $m$ are even, and $I^{(N, m)}=SO(N-m)\times SO(N-4-m)\times SO(m)$ if at least one of $N$ and $m$ is odd. Here the $SO(N-m)\subset SO(N)\subseteq I^{(N)}$ acts on the last $N-m$ rows of $n$, $SO(N-4-m)\subset SO(N-4)\subset I^{(N)}$ acts on the last $N-4-m$ columns of $n$, and $SO(m)$ is a combination of the $SO(m)\subset SO(N)\subseteq I^{(N)}$ acting on the first $m$ rows of $n$ and the $SO(m)\subset SO(N-4)\subset I^{(N)}$ acting on the first $m$ columns of $n$.
    
\begin{figure}
    \centering
    \includegraphics[width=0.45\textwidth]{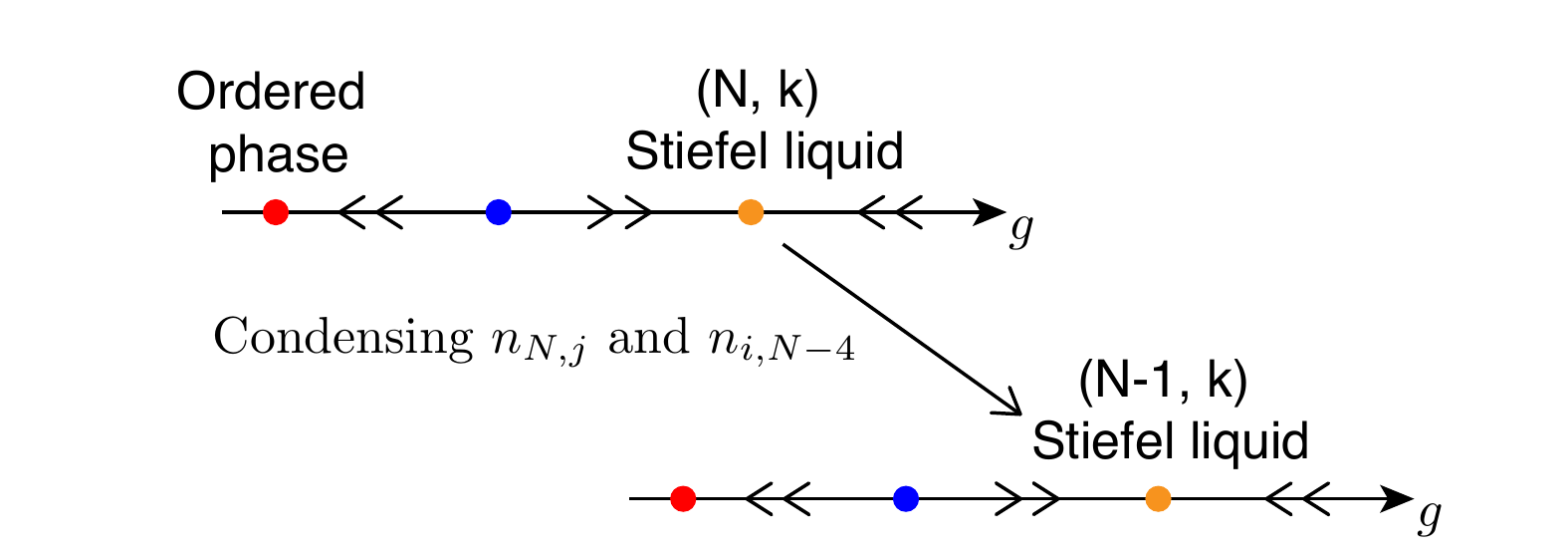}
    \caption{The cascade structure of Stiefel liquids.}
    \label{fig:cascade}
\end{figure}

To focus on the low-energy fluctuations, we can define a reduced $(N-m)$-by-$(N-4-m)$ matrix $n_{\rm red}$, by removing the first $m$ rows and first $m$ columns of $n$. This reduced matrix still satisfies an orthonormal condition, $n_{\rm red}^T n_{\rm red}=I_{N-4-m}$. In addition, $S^{(N, k)}[n]=S^{(N-m, k)}[n_{\rm red}]$. That is, the low-energy dynamics of this symmetry-broken descendent of SL$^{(N, k)}$ is effectively identical to that of SL$^{(N-m, k)}$ (see Fig. \ref{fig:cascade}). We remark that within the low-energy sector, the remaining continuous symmetry is $I^{(N-m)}$, which is generally smaller than $I^{(N, m)}$. Physically, this is because some elements of $I^{(N, m)}$ only act on the frozen DOFs of this symmetry-broken descendent, but not on its low-energy DOFs, as discussed above. Also notice that the discrete $\mc{R}$ and $\mc{T}$ symmetries defined in Eqs. \eqref{eq: canonical CRT actions} and \eqref{eq: CRT for N=5} are preserved for all $m\leqslant N-5$, and the $\mc{C}$ symmetry defined in Eq. \eqref{eq: canonical CRT actions} is preserved for $m<N-5$ and broken when $m=N-5$. 
    
Therefore, SL$^{(N, k)}$ for each $k$ form a cascade of theories, where the ones with a smaller $N$ can be obtained from the ones with a larger $N$ by appropriately perturbing the latter and focusing on the resulting low-energy sector.
    
If $m=N-4$ in the above, then all entries of $n$ are fixed, the remaining continuous symmetry is $(SO(4)\times SO(N-4))/Z_2$ for even $N$ and $SO(4)\times SO(N-4)$ for odd $N$, there is no longer any low-energy DOF in the system, and the resulting state is no longer a SL. In particular, the WZW term is now completely removed. Note that the $\mc{R}$ and $\mc{T}$ symmetries in Eq. \eqref{eq: canonical CRT actions} are still preserved, but the $\mc{C}$ symmetry defined there is broken. However, the following $\mc{C}'$ symmetry, which is a combination of that $\mc{C}$ and an element in $I^{(N)}$, is preserved:
\beq
\mc{C}': n_{ji}\rightarrow (-1)^{f'_{ji}}n_{ji}
\eeq
where $f'_{ji}=1$ if $(j=1\ \&\  i>1)$ or $(j>1\ \&\ i=1)$, and $f'_{ji}=0$ otherwise.

Since the WZW term is expected to capture the quantum anomalies of the SL, from the above cascade structure we see that the anomalies of a SL with a smaller $N$ should be contained in the anomalies of a SL with a larger $N$. Furthermore, by {\em explicitly} breaking the symmetries of SL$^{(N, k)}$ via tuning up $m$ from 0 to $N-4$ as above, its anomalies are gradually removed, and also localized to the low-energy sector defined by $n_{\rm red}$. For example, if the above procedure of symmetry breaking is applied to SL$^{(N)}$ with $m=N-5$, the anomaly of the system is reduced to that of SL$^{(5)}$, given by Eq. \eqref{eq: DQCP complete anomaly}. If this procedure is applied with $m=N-4$ instead, the anomaly of the system is completely removed. As another application, this cascade structure also implies that SL$^{(N)}$ with any $N\geqslant 6$ have symmetry-enforced gaplessness, similar to that of SL$^{(5)}$, which is reviewed in Sec. \ref{sec: review and summary}.
    
The cascade structure among the SLs remarked here will be repeatedly used later, and it plays an important role in simplifying the discussion of the anomaly.

\subsection{Possible fixed points at strong coupling} \label{subsec: IR fate}

Now we argue that for sufficiently large $N$ and small $k$, with certain mild physical assumptions, the action $S^{(N, k)}$ in Eq. \eqref{eq: level-k generalization} can flow to a conformally invariant fixed point under RG, so that these SLs can be a critical quantum liquid.

We will explore $S^{(N, k)}$ piece by piece, and we start by completely ignoring the WZW term in Eq. \eqref{eq: level-k generalization} and only considering $S_0$. When $N$ is suffiently large, $S_0$ is expected to be able to describe an order-disorder transition of the matrix $n$. 

One way to see this is to look at the saddle point corresponding to the $S_0$. The constraint $n^Tn=I_{N-4}$ can be eliminated by introducing an $(N-4)$-by-$(N-4)$-matrix-valued Lagrangian multiplier $\alpha$, such that the (Euclidean) path integral becomes
\beq
\begin{split}
\mc{Z}_0
=&\int\mc{D}n\mc{D}\alpha\exp\Big\{\int d^3x\left(-\frac{1}{2g}\right)\cdot\big[\partial_\mu n_{ji}\partial^\mu n_{ji}\\
&\qquad\qquad\qquad\quad
-i\alpha_{ij}(n_{kj}n_{ki}-\delta_{ij})\big]\Big\}\\
=&\int\mc{D}n\mc{D}\alpha\exp\left\{\int\frac{d^3x}{2g}\left[n_{kj}\left(\delta_{ij}\square+i\alpha_{ij}\right)n_{ki}\right]\right\}\\
&
\qquad\qquad\qquad
\cdot\exp\left(-i\int\frac{d^3x}{2g}\Tr(\alpha)\right)\\
=&\int\mc{D}\alpha\exp\left[-\frac{N}{2}\tr\ln\left(\delta_{ij}\square+i\alpha_{ij}\right)\right]\\
&
\qquad\quad
\cdot\exp\left(-i\int d^3x\frac{\Tr(\alpha)}{2g}\right)
\end{split}
\eeq
where $\square\equiv\partial_\mu\partial^\mu$. The Lagrangian multiplier $\alpha$ is introduced in the first step. In the second step we integrate by parts. In the last step we integrate out $n$ and ignore a constant multiplicative factor to the path integral, where the factor of $N$ appears because the index $k$ in the middle line runs from 1 to $N$, and trace is taken over both the matrix indices and the spacetime coordinates.

When $N\gg 1$, we expect that the remaining path integral will be dominated by the configuration of $\alpha$ that satisfies the saddle-point equation. Assuming an ansatz to the saddle-point equation with $\alpha_{ij}=i\Delta^2\delta_{ij}$, where $\Delta$ is a number (not a matrix), the saddle-point equation becomes
\beq
\frac{1}{2g}
=\frac{N}{2}\int\frac{d^3k}{(2\pi)^3}\frac{1}{k^2+\Delta^2}
\eeq
Taking the limit $N\rightarrow\infty$ while $gN$ is fixed, we find that the above equation has a solution with nonzero $\Delta$ if $g$ is larger than certain critical value, $g_0\sim\mc{O}(1/(N\Lambda))$, where $\Lambda$ is a UV cutoff. In this case, since $\Delta\neq 0$, the matrix $n$ acquires a gap, and the system is in the disordered phase. On the other hand, when $g<g_0$, the system is in the ordered phase. Note that the structure of this saddle-point equation is the same as the usual $O(N)$ vector NLSM \cite{Peskin1995}, despite that $n$ is a matrix.

The above results suggest that the beta function for $S_0$ at sufficiently large $N$ is
\beq \label{eq: beta function}
\beta(\tilde g)=-\tilde g+\beta_0(\tilde g)
\eeq
where $\tilde g\equiv g\Lambda$ is the dimensionless coupling. The first term comes from the engineering dimension of $g$, and the second term, $\beta_0(\tilde g)$, represents loop corrections. The precise form of $\beta_0(\tilde g)$ is hard to obtain even at large $N$. In fact, as $N\to\infty$ this model approaches the $SO(N)$ sigma model, for which even the leading correction to the beta function requires summing over all planar diagrams (see, for example, Ref.~\cite{Wegner1989} for the  $3$-loop result). For us, what is important is that $\beta(\tilde g)=0$ at $\tilde g=\tilde g_0\equiv g_0\Lambda\sim\mc{O}(1/N)$.

Physically, the above discussion suggests that for the NLSM defined by $S_0$, there is an attractive fixed point at $\tilde g=0$, corresponding to the ordered phase where the symmetry is broken spontaneously. There is also a repulsive fixed point at $\tilde g=\tilde g_0\sim\mc{O}(1/N)$, corresponding to the order-disorder transition.

Next, we include the WZW term and consider the full $S^{(N, k)}$. We will view the WZW term as a perturbation to $S_0$, and it is expected to contribute a term of the following form at the leading order to the beta function {\footnote{There is a class of ``leading-order contributions"
at large $N$, each of the form $-C_Vk^V\tilde g^{2V+1}$, with $V\geqslant 2$ an integer and $C_V\sim\mc{O}(1)$. Notice in the limit considered in the next paragraph, \ie $N\rightarrow\infty$, $k\rightarrow\infty$ and $k/N^2$ fixed, all these contributions are at $\mc{O}(1/N)$ if $\tilde g\sim\mc{O}(1/N)$.}}:
\beq
\label{eq:WZWbetafunction}
\delta\beta(\tilde g)=-C k^2\tilde g^{5}
\eeq
where $C$ is of order $1$. It is expected that $C>0$, because the WZW term yields a phase factor in the path integral and induces destructive interference of the paths, which tends to prevent $\tilde g$ from flowing to a large value, just like the effect of Haldane's phase in spin chains \cite{Haldane1983, Haldane1983a, Affleck1987}. To further understand the effect of this contribution to the beta function, let us first consider cases with $k\sim O(1)$, which are the main focus of this paper. At large $N$, the term Eq.~\eqref{eq:WZWbetafunction} is negligible when $\tilde{g}\sim 1/N$, so it does not affect the order-disorder transition significantly. The WZW could affect the nature of the disordered state. For example, for odd $k$ we know that the disordered state cannot be a trivially gapped phase because of the nontrivial 't Hooft anomaly -- it has to be either gapless or spontaneously break some other symmetry. It is hard to tell exactly what happens in the disordered regime since we no longer have analytic control. 

\begin{figure}
    \centering
    \includegraphics[width=0.5\textwidth]{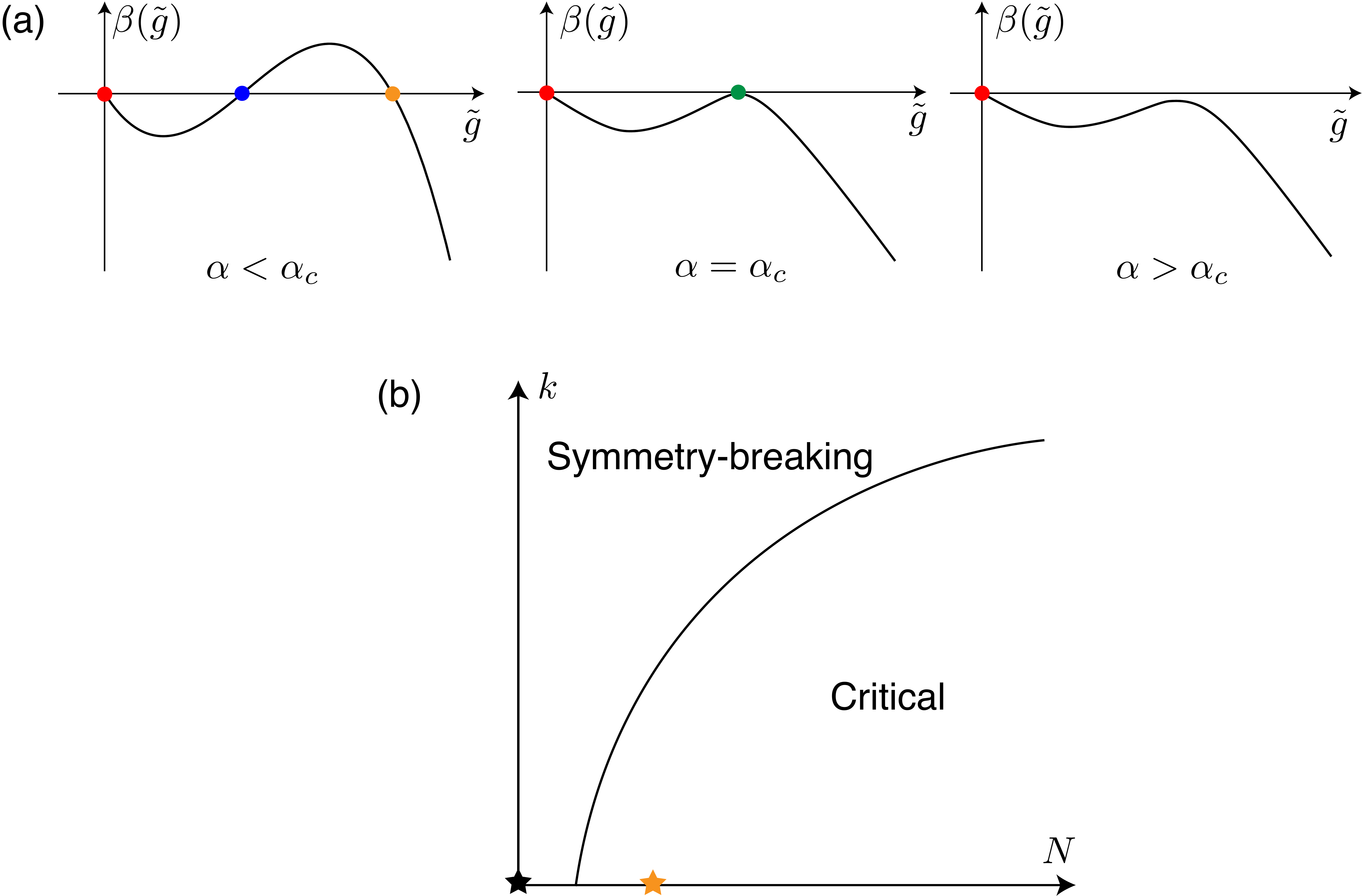}
    \caption{Fixed point structure and schematic phase diagram of SL$^{(N, k)}$. In (a), different fixed point structures for different values of $\alpha$ are shown. There is always an attractive fixed point, represented by the red circle, which corresponds to the symmetry-broken state. The structure of the other fixed points depends on the relation between $\alpha$ and $\alpha_c\sim\mc{O}(1)$. The precise value of $\alpha_c$ depends on $\beta_0$ and $C$. If $\alpha<\alpha_c$, there are two other fixed points, a repulsive one represented by the blue circle, corresponding to an order-disorder transition, and an attractive one, represented by the yellow circle, corresponding to a stable critical quantum liquid, \ie the SL. As $\alpha$ increases and approaches $\alpha_c$, the blue and yellow fixed points approach each other and collide when $\alpha=\alpha_c$. When $\alpha>\alpha_c$, the original blue and yellow fixed points disappear (become complex fixed points \cite{Wang2017, Gorbenko2018}). We would like to take $\alpha$ below $\alpha_c$. In (b), a schematic phase diagram of SL$^{(N, k)}$ is shown. For the $(N, k)$ in the critical regime, it is possible to tune the parameters of the system to yield a critical quantum liquid, while in the symmetry-breaking regime this is not possible. The yellow star represents $(N, k)=(6, 1)$, \ie the $U(1)$ DSL, and the black star represents $(N, k)=(5, 1)$, the DQCP. The precise  boundary between the critical and symmetry-breaking regimes is currently undetermined.}
    \label{fig:fixedpoint}
\end{figure}

It then turns out to be useful to consider a different limit with $N\rightarrow\infty$, $k\rightarrow\infty$ and $k/N^{2}=\alpha$ with $\alpha\sim O(1)$ fixed {\footnote{Instead of looking at this specific limit, one may also think in the following more general and heuristic way. When $k\sim\mc{O}(1)$, there are the attractive fixed point corresponding to the Goldstone phase, and the repulsive fixed point corresponding to the order-disorder transition. When $k$ is very large (for a fixed $N$), there should only be a single fixed point, \ie the attractive one corresponding to the Goldstone phase. As $k$ increases from $\mc{O}(1)$ to a large value, for the repulsive fixed point to disappear, it should collide with another fixed point and annihilate, schematically as shown in Fig. \ref{fig:fixedpoint} (a). This observation implies the existence of another interacting attractive fixed point for small enough $k$. According to the forms of Eqs. \eqref{eq: beta function} and \eqref{eq:WZWbetafunction}, the value of $k$ below which the interacting attractive fixed point exists scales with $N$ as $N^2$, and the coupling $\tilde g$ for this fixed point is $\tilde g\sim\mc{O}(1/N)$. This additional interacting attractive fixed point corresponds to the critical Stiefel liquid.}}.
Fig.~\ref{fig:fixedpoint} (a) illustrates several different scenarios. If $\alpha$ is smaller than some critical value $\alpha_c$, then the repulsive fixed point at $\tilde g=\tilde g_0$ will be shifted to a larger value, and another {\em attractive} fixed point at $\tilde g\sim\mc{O}(1/N)$ emerges\footnote{One may wonder at this new fixed point if there will be relevant perturbations not controlled by $g$. However, if the order-disorder transition has only one relevant singlet operator (\ie the tuning of $g$), as naturally expected, this new fixed point should have no relevant perturbation once the full symmetry of $S^{(N, k)}$ is preserved, \ie this fixed point is indeed attractive. Otherwise, there will be additional fixed points besides the ones in Fig. \ref{fig:fixedpoint} (a), which appears to be unnatural.}. Because this attractive fixed point is still weakly coupled at large $N$, we expect it to describe a symmetry-preserving critical state, rather than a gapped or Goldstone state. The repulsive fixed point then describes the transition from the critical phase to the symmetry breaking phase. On the other hand, if $\alpha>\alpha_c$, the attractive and repulsive fixed points will collide and annihilate with each other, and the only fixed point left will be the weakly-coupled one with spontaneously broken symmetry. In other words, with a smaller $k$, there is a stronger tendency for the WZW-NLSM to have an attractive fixed point at finite coupling. Although this conclusion is drawn in the $k/N^{2}\sim O(1)$ regime, it seems natural to \textit{assume} that this trend is qualitatively true even for $k\sim O(1)$. 
Namely, we expect that $S^{(N, k)}$ can describe a critical quantum liquid even for $k=1$, if $N\gg 1$.\footnote{Note that this nicely corroborates our results that $S^{(5, k)}$ and $S^{(6, k)}$ have dual descriptions based on $USp(2k)$ and $U(k)$ gauge theories, since these gauge theories are generally expected to have stronger tendency to be critical (unstable) for small (large) $k$. } More generally, we propose that for each $k\neq 0$, there exists an integer $N_c(k)$ that increases as $k$ increases, such that when $N\geqslant N_c(k)$, $S^{(N, k)}$ can flow to a conformally invariant attractive fixed point, corresponding to SL$^{(N, k)}$. The precise form of $N_c(k)$ is unknown at this stage (see Fig. \ref{fig:fixedpoint} (b)).

We can gain more confidence about the above arguments, or conjectures, from WZW sigma models on Grassmannian manifolds such as $U(2N)/U(N)\times U(N)$. Our arguments can be equally applied in those cases and we conclude that a strong-coupling attractive fixed point should occur for sufficiently large $N$ (see also Ref.~\cite{Bi2016}). Unlike the Stiefel WZW models, however, the Grassmannian WZW models are naturally related to various non-Abelian gauge theories (QCD) in $(2+1)$-d \cite{Komargodski2018}, which we review in Appendix~\ref{app: GrassmannianWZW}. The rank $N$ in the WZW model corresponds to flavor number in those QCD theories, and it is well known that a large-flavor QCD does flow to a nontrivial fixed point in $(2+1)$-d. This serves as a nontrivial check of our arguments on the existence of strong coupling fixed points.

Also notice that if $\alpha$ happens to be barely above $\alpha_c$ (as in Fig.~\ref{fig:fixedpoint}), the RG flow will be slow near the fixed-points-collision region. This gives a mechanism for the ``walking'' of the coupling constant \cite{Kaplan2009} and pseudo-critical behavior \cite{Wang2017,Gorbenko2018} of the system.

Recall that SL$^{(5, 1)}$ is just the DQCP. Later we will argue that SL$^{(6, 1)}$ is in fact the $U(1)$ DSL. As reviewed in Sec. \ref{sec: review and summary}, these two states are likely pseudo-critical and critical, respectively. This implies that for $k=1$, $5<N_c<6$, so we propose that $S^{(N, k)}$ for all $N\geqslant 6$ and $k=1$ can still flow to a conformally invariant fixed point and describe a critical state{\footnote{Note that this proposal is consistent with the symmetry-enforced gaplessness of these theories, as required by their anomalies. See Sec. \ref{subsec: cascade structure} for more details.}}. 

To summarize this section, we have proposed the existence of SLs, whose effective field theories can be directly formulated in terms of local DOF, given in Eqs. \eqref{eq: total action} and \eqref{eq: level-k generalization}. These SLs have an interesting symmetry structure discussed in Sec. \ref{subsec: symmetries}, and they are interrelated with a special cascade structure discussed in Sec. \ref{subsec: cascade structure}. We argue that SL$^{(N)}$ with $N\geqslant 6$ can flow to a CFT fixed point under RG. Furthermore, we conjecture that SL$^{(N)}$ with $N>6$ are all non-Lagrangian. Putting these compactly, SLs are cascades of extraordinary critical quantum liquids.

\section{$N=6$: Dirac spin liquids} \label{sec: N=6: DSL}

According to the previous discussion, it is readily seen that the SL$^{(N=5)}$ is special in many ways among all SLs. For example, its continuous symmetry $I^{(5)}=SO(5)$, which is qualitatively different from $I^{(N)}$ with $N\geqslant 6$, in the sense that the former has only one connected component, while the latter has two. In this section, we focus on the more nontrivial case, \ie SL$^{(N=6, k)}$, and we argue that SL$^{(N=6, k)}$ has a dual description in terms of $U(k)$ DSLs, \ie 4 flavors of gapless Dirac fermions coupled to a $U(k)$ gauge theory.

\subsection{Deriving $k=1$ WZW model from QED$_3$}
\label{subsec: k=1fromQED}

We begin with deriving the Stiefel WZW sigma model, $S^{(N=6)}$ in Eq. \eqref{eq: total action}, from QED$_3$ with $N_f=4$. The derivation is similar in spirit to that in Ref.~\cite{Senthil2005}, although somewhat more complicated in detail.

We start from the QED$_3$ Lagrangian:
\be
\sum_{\alpha=1}^{4}\bar{\psi}_\alpha i\slashed{D}_a\psi_\alpha-\frac{1}{2\pi}A_{\rm top}da
\ee
where $a$ is the dynamical $U(1)$ gauge field and $\psi$ is the Dirac fermion. Compared to Eq. \eqref{eq: DSL standard model}, here the Maxwell term of $a$ is suppressed, because it is unimportant for the current discussion. Also, the last term, $-\frac{1}{2\pi}adA_{\rm top}\equiv -\frac{1}{2\pi}\epsilon_{\mu\nu\lambda}A^\mu_{\rm top}\partial^\nu a^\lambda$, is introduced to keep track of the conserved flux of $a$ by introducing the probe $U(1)_{\rm top}$ gauge field $A_{\rm top}$. The subscript ``top" is due to the fact that the conservation of the current corresponding to this $U(1)$ symmetry, $j_\mu=\epsilon_{\mu\nu\lambda}\partial^\nu a^\lambda/(2\pi)$, does not rely on the equations of motion of the theory. For later convenience, here we introduce the following notation of the generators of the $SU(4)$ flavor symmetry of this theory: $\sigma_{ab}\equiv\frac{1}{2}\sigma_a\otimes\sigma_b$, with $a,\ b=0,1,2,3$ but $a$ and $b$ not simultaneously zero. Here $\sigma_0=I_2$ and $\sigma_{1,2,3}$ are the standard Pauli matrices.

We now consider dynamically breaking the $SU(4)$ flavor symmetry down to $(SU(2)\times SU(2)\times U(1))/Z_2$, which is believed to be the most likely symmetry breaking pattern for this theory \cite{Pisarski1984, VAFA1984, Vafa1984a, Polychronakos1988, Pisarski1991}. This introduces an order parameter $\mathcal{P}$, defined in the complex Grassmannian manifold $U(4)/\left(U(2)\times U(2)\right)$, that couples to the Dirac fermions as an $SU(4)$-adjoint mass:
\be
m\mathcal{P}_{\alpha \beta}\bar{\psi}_{\alpha}\psi_{\beta},
\ee
where $m$ is the coupling strength, which physically means the magnitude of the mass. Now we can formally integrate out the Dirac fermions and obtain an effective theory in terms of the $\mathcal{P}$ and $a$ fields. One can expand in $1/m$ and obtain a $U(4)/\left(U(2)\times U(2)\right)$ sigma model for the $\mathcal{P}$ fields. As shown in Ref.~\cite{Jian2018}, this sigma model comes with a WZW term with coefficient $k=1$. Note that this WZW term is well defined because the target Grassmannian manifold has $\pi_4=\mathbb{Z}$ and $\pi_3=0$.

However, this Grassmannian WZW theory is not the end of the story: the $U(1)$ gauge field is still present and we expect nontrivial couplings between the gauge field and the $\mathcal{P}$ field. The most important coupling is
\be
\label{eq: Skyrmiongaugecoupling}
a\cdot j^{Sk},
\ee
where $j^{Sk}$ is the Skyrmion current of the $\mathcal{P}$ field. The Skyrmion current is well defined because the target Grassmannian manifold has $\pi_2=\mathbb{Z}$. The existence of this coupling is due to that the elementary Skyrmion is a fermion that carries unit gauge charge, which is nothing but the original Dirac fermion, $\psi$. To see this, let us build up a simple type of Skyrmion by first considering a mass term that only couples to the first two flavors of Dirac fermions, $\psi_{\alpha=1,2}$, \ie a mass of the form $\bar{\psi}N_i\sigma_i\otimes(I_2+\sigma_3)\psi$, where $i=1,2,3$ and hence $\vec{N}\in S^2$. This $\vec N$ field can then form a standard Skyrmion configuration in space. Now this mass can be extended to an allowed configuration for the $\mathcal{P}$ field, by adding a constant mass for the other two Dirac fermions, say, $\bar{\psi}\sigma_3\otimes(I_2-\sigma_3)\psi$, which would not change the topological properties of the skyrmion. However, it is well known that this $\vec N$ field has a Hopf term with $\theta=\pi$ in its effective theory, and the Skyrmion is a fermion with gauge charge $1$ \cite{Abanov2000, Abanov2001}. Since this is the minimum gauge charge of the theory, we conclude that the elementary Skyrmion in $\mathcal{P}$ field also has gauge charge $1$ and is a fermion, \ie it is the original Dirac fermion $\psi$. This justifies the coupling in Eq.~\eqref{eq: Skyrmiongaugecoupling}. 

So the sigma model should be written as
\beq \label{eq: transforming from QED3 to WZW}
\begin{split}
S
=S_{G0}[P]&+S_{\rm G-WZW}[\mc{P}]\\
&+\int d^3x\left(a\cdot j^{Sk}-\frac{1}{2\pi}A_{\rm top}da\right)
\end{split}
\eeq
where $S_{G0}[P]$ is the NLSM on the Grassmannian manifold without any topological term, and $S_{\rm G-WZW}[\mc{P}]$ is the WZW action on this Grassmannian manifold. The precise expressions of $S_{G0}[P]$ and $S_{\rm G-WZW}[P]$ are unimportant for our purpose.

Next, notice that the complex Grassmannian $U(4)/(U(2)\times U(2))$ is equivalent to the real Grassmannian $SO(6)/(SO(4)\times SO(2))$, since $SU(4)\sim SO(6)$, $SU(2)\times SU(2)\sim SO(4)$ and $U(n)\sim U(1)\times SU(n)$ (all up to some discrete quotients, which do not change the conclusion here). The real Grassmannian $SO(6)/(SO(4)\times SO(2))$ can be rewritten as follows: introduce two orthonormal $SO(6)$ vectors $n_1$ and $n_2$, and rewrite the matrix field $\mc{P}=-2i\left(n_1^TT_{ab}n_2\right)\cdot\sigma_{ab}$, where $T_{ab}$ is the $SO(6)$ generator that corresponds to $\sigma_{ab}$ (see Appendix \ref{app: homomorphism}). The fact that $T_{ab}^T=-T_{ab}$ means that this rewriting introduces an $SO(2)$ gauge redundancy that rotates between $n_1$ and $n_2$, \ie there is a dynamical $SO(2)$ gauge field, denoted by $b$, that couples to $n_1$ and $n_2$. Notice that $n=(n_1,n_2)$ lives on nothing but the Stiefel manifold $SO(6)/SO(4)$. We have therefore rewritten the Grassmannian sigma model, $S_{G0}[\mc{P}]$, in terms of an $SO(2)=U(1)$ gauge field--$b$, coupled to an order parameter that lives on the Stiefel manifold $SO(6)/SO(4)$, \ie $S_0[n, b]$, the NLSM in Eq. \eqref{eq: bare NLSM} with $n$ minimally coupled to $b$. So what do the Grassmannian WZW term and the Skyrmion current become in this representation? Mathematically, our rewriting of the Grassmannian in terms of an $SO(2)$ gauge theory coupled to a Stiefel manifold corresponds to the fibration 
\beq
SO(2)\to \frac{SO(6)}{SO(4)}\to \frac{SO(6)}{SO(4)\times SO(2)}. \nonumber
\eeq 
The long exact sequence from this fibration leads to two isomorphisms:
\beq
p: & & \hspace{5pt}\pi_4\left(\frac{SO(6)}{SO(4)\times SO(2)}\right)\to \pi_4\left(\frac{SO(6)}{SO(4)}\right), \nn
\beta: & & \hspace{5pt}\pi_2\left(\frac{SO(6)}{SO(4)\times SO(2)}\right)\to\pi_1(SO(2)).
\eeq
This means that any nontrivial winding in $\pi_4$ and $\pi_2$ of the Grassmannian should be fully encoded in the corresponding homotopy groups of the Stiefel and $SO(2)$, respectively.
Since the Grassmannian WZW term comes from $\pi_4\left(SO(6)/\left(SO(4)\times SO(2)\right)\right)$, it should simply become the WZW term of the Stiefel manifold (which comes from $\pi_4(SO(6)/SO(4))$), given by Eq. \eqref{eq: WZW begins}. The Grassmannian Skyrmion current comes from $\pi_2\left(SO(6)/\left(SO(4)\times SO(2)\right)\right)$, so it should simply become the flux current of the $SO(2)$ gauge theory (which comes from $\pi_1(SO(2))$):
\be
j^{Sk}_{\mu}=\frac{1}{2\pi}\epsilon_{\mu\nu\lambda}\partial_{\nu}b_{\lambda}.
\ee

The complete theory in Eq. \eqref{eq: transforming from QED3 to WZW} can now be written as
\beq
S=&&S_0[n, b]+S^{(6)}_{\rm WZW}[n,b] \nn
&&+\int d^3x\left(\frac{1}{2\pi}adb-\frac{1}{2\pi}A_{\rm top}da\right).
\eeq
Now integrating out the $a$ gauge field, the $b$ gauge field will be set to $b=A$. The only IR degrees of freedom left is the Stiefel field $n$ with the action of WZW model at $k=1$. The $n$ fields couple to the probe gauge field $A$ as charge-$1$ fields, which leads to the interpretation that they correspond to the monopole operators in the original QED$_3$. This completes our derivation.

In passing, we mention that in Appendix \ref{app: DSL explicit} we also explicitly derive some properties regarding the quantum anomalies of the $U(1)$ DSL and show that they match with that of SL$^{(6)}$. This provides further evidence for the equivalence between the $U(1)$ DSL and SL$^{(6)}$.

\subsection{General $k$: $U(k)$ QCD with $N_f=4$}

The above derivation can be generalized quite readily to $U(k)=\left(U(1)\times SU(k)\right)/\mathbb{Z}_k$ gauge theories with $N_f=4$ fundamental Dirac fermions. Denote the $U(k)$ gauge field that is minimally coupled to the Dirac fermions by $\mathbf{a}=a+\tilde a\mathbf{1}$, where $a$ is an $SU(k)$ gauge field and $\tilde a$ is a $U(1)$ gauge field. Note that now the minimal local monopole carries $2\pi$ flux of $\Tr(\mathbf{a})=k\tilde a$, so the coupling of the theory to $A_{\rm top}$ takes the form $-\frac{1}{2\pi}A_{\rm top}d\Tr(\mathbf{a})$. Also notice that the Dirac fermion carries charge $1/k$ under $\Tr(\mb{a})$.

We can now proceed with the same arguments as in the QED case. First we introduce a mass operator on Grassmannian $SO(6)/(SO(4)\times SO(2))$, which is a color singlet and $SU(4)$ adjoint. Then we integrate out all Dirac fermions. This gives a Grassmannian sigma model with a WZW term, which is at level $k$ because of the color multiplicity of the Dirac fermions. There is also a Skyrmion term like Eq.~\eqref{eq: Skyrmiongaugecoupling} from each color, but with $a$ therein replaced by $\Tr(\mathbf{a})$, and with a coefficient $1/k$, since the Dirac fermions carry gauge charge $1/k$ under $\Tr(\mathbf{a})$. Summing over all colors gives precisely $\Tr(\mathbf{a})\cdot j^{Sk}$. The remaining $U(k)$ gauge field splits into an $SU(k)$ and $U(1)$ part. The $SU(k)$ gauge field now does not couple to any IR degrees of freedom, so we expect it to flow to strong coupling and eventually confine (and be gapped). The $U(1)$ part can be analyzed in the same way as we did for QED. At the end of the day we again obtain a Stiefel WZW sigma model, now at level $k$.

Therefore, we propose that SL$^{(6, k)}$ and a $U(k)$ 
DSL are dual. In passing, we note that a $U(k)$ DSL is proposed to arise in a spin-$k/2$ system \cite{Calvera2020}.

\section{Quantum anomaly of SL$^{(N, k)}$}
\label{Sec: Anomaly Analysis}

The above derivation of the effective theory of SL$^{(6, k)}$ from Dirac spin liquids should really be viewed as at the kinematic level, \ie this derivation does not guarantee that these two theories have identical IR dynamics. In fact, rigorously showing that two theories have identical IR dynamics is generally formidably challenging.

Now we investigate the kinematic aspects of the SLs in greater detail, by analyzing the quantum anomaly of SL$^{(N)}$ for general $N\geqslant 5$. The results for SL$^{(N,k)}$ can be readily obtained from those for SL$^{(N)}$ by viewing the former as $k$ copies of the latter. The anomaly takes the form of an invertible topological response term in $(3+1)$-d spacetime, characterizing an SPT phase whose boundary can host our physical SL$^{(N)}$. In principle, the anomaly can be calculated directly from the action of $S^{(N)}$, but in practice this appears to be quite complicated. Instead, we will show in this section that the anomaly of SL$^{(N)}$ is essentially fixed by the phase diagram, or more precisely, by the cascade structure discussed in Sec. \ref{subsec: cascade structure}.

Below we will first treat the continuous symmetry of SL$^{(N)}$ to be $SO(N)\times SO(N-4)$, and derive the topological response function corresponding to the full anomaly associated with the entire symmetry of the theory, \ie including both this continuous and the discrete symmetries. For $N$ odd, this treatment is faithful and complete. For $N$ even, the symmetry $SO(N)\times SO(N-4)$ is larger than the faithful $I^{(N)}=(SO(N)\times SO(N-4))/Z_2$ symmetry possessed by SL$^{(N)}$. Physically, one can think of this symmetry enlargement as from some trivially gapped local DOFs that, \eg transform as an $SO(N)$ vector and $SO(N-4)$ singlet. The risk of working with the larger $SO(N)\times SO(N-4)$ symmetry is that we may miss some anomalies that are nontrivial only for the original $I^{(N)}$ group.
In the later part of this section, by analyzing the properties of the monopoles of the $I^{(N)}$ symmetry, we will derive the anomaly associated with the faithful $I^{(N)}$ symmetry for all $N\geqslant 5$. However, there we will not explicitly derive the anomaly associated with the discrete symmetries, which is left for future work.

The final result with the continuous symmetry taken to be the enlarged $SO(N)\times SO(N-4)$ is given by Eq.~\eqref{eq: Anomaly11}. Some of the physical implications of this anomaly can be read off from Table \ref{tab: SO(N) and SO(N-4) monopoles for odd N}. If the continuous symmetry is taken to be the faithful $I^{(N)}$, for even $N$ the monopole corresponding to the $I^{(N)}$ symmetry of SL$^{(N)}$ has the structure given by root 3 of Eq. \eqref{eq: 4k+2 monopoles main}. One important implication of this improved characterization of the anomaly is that for even $N$, SL$^{(N, 2)}$ is still anomalous, and SL$^{(N, 4)}$ has no $I^{(N)}$-anomaly. This is to be contrasted from Eq. \eqref{eq: Anomaly11}, which implies that SL$^{(N, 2)}$ is anomaly-free for any $N$.

To derive Eq. \eqref{eq: Anomaly11}, we take the following three steps. We first put the system on an orientable manifold and also ignore the $\mc{C}$ symmetry. Next, we include the $\mc{C}$ symmetry, but still stay on an orientable manifold. This restriction to orientable manifolds means that we are not considering the full anomaly associated with the orientation-reversal (time-reversal and reflection) symmetries. Finally, we put the theory on a possibly unorientable manifold (still with $\mc{C}$ taken into account), in order to fully characterize the anomaly associated with all symmetries. Recall that from the discussion in Sec. \ref{subsec: symmetries}, it is sufficient to consider $\mc{C}$ and $\mc{T}$, and the results will already capture anomalies associated with $\mc{R}$.

\subsection{$SO(N)\times SO(N-4)$ on orientable manifolds} \label{subsec: anomaly continuous}

We first consider orientable 4-manifolds $X_4$, with vanishing first Stiefel-Whitney (SW) class $w_1^{TM}=0$ (mod $2$). We shall also neglect charge conjugation symmetry for now. A general response term in $4d$ takes the form
\beq
S_{\rm bulk}=&&i\pi\int_{X_4}(a_1w_4^{SO(N)}+a_2w_4^{SO(N-4)} \nn
&&+a_3w_2^{SO(N)}w_2^{SO(N-4)} +a_4w_2^{SO(N)}w_2^{SO(N)} \nn
&&+a_5w_2^{SO(N-4)}w_2^{SO(N-4)} +a_6 w_2^{TM}w_2^{TM}),
\eeq
where $a_{1,2,3,4,5,6}\in\{0,1\}$ are unknowns, $w_2\in H^2(X_4,\mathbb{Z}_2)$ and $w_4\in H^4(X_4,\mathbb{Z}_2)$ are the second and fourth SW classes of the corresponding bundles ($SO(N)$, $SO(N-4)$ and tangent bundles), respectively. The products among the SW classes here and below all refer to the cup product. The physical meanings of various of these topological response terms are given in Table \ref{tab: SO(N) and SO(N-4) monopoles for odd N}.

We now try to fix the unknown coefficients in the above expression. We do not attempt to directly gauge the SL$^{(N)}$ and compute the anomaly. Instead, we shall use two simple facts due to the cascade structure among the SLs discussed in Sec. \ref{subsec: cascade structure}:
\begin{enumerate}
    \item If the $SO(N)\times SO(N-4)$ symmetry is broken to $SO(5)\subset SO(N)$, the anomaly becomes simply $i\pi\int w_4^{SO(5)}$. Namely, if we set $w_2^{SO(N-4)}$ and $w_4^{SO(N-4)}$ to trivial, $w_2^{SO(N)}=w_2^{SO(5)}$ and $w_4^{SO(N)}=w_4^{SO(5)}$, the anomaly term should become just become $w_4^{SO(5)}$. This comes from the fact (as reviewed in Sec.~\ref{subsec: review of DQCP}) that SL$^{(5)}$ corresponds to the DQCP and has the simple $w_4$ anomaly.
    
    \item If the $SO(N)\times SO(N-4)$ symmetry is broken to $SO(4)\times SO(N-4)'$, where $SO(4)\subset SO(N)$ and $SO(N-4)'$ is a combination of $SO(N-4)\subset SO(N)$ and the original $SO(N-4)$, then there is no anomaly left since the theory admits a simple ordered state (see also Sec.~\ref{subsec: SL begins} for more details). This means that if we set $w_2^{SO(N)}=w_2^{SO(4)}+w_2^{SO(N-4)'}$ and $w_2^{SO(N-4)}=w_2^{SO(N-4)'}$, the anomaly should vanish.
\end{enumerate}

It turns out that the above two conditions unambiguously fix $S_{\rm bulk}$ to be
\beq
\label{eq: Anomaly0}
S_{\rm bulk}=&&i\pi\int_{X_4}(w_4^{SO(N)}+w_4^{SO(N-4)}+w_2^{SO(N)} w_2^{SO(N-4)} \nn &&+w_2^{SO(N-4)} w_2^{SO(N-4)}).
\eeq
To show this, we use the facts that (a) if $SO(N)$ symmetry is broken to $SO(N-m)\times SO(m)$, then $w_4^{SO(N)}=w_4^{SO(N-m)}+w_4^{SO(m)}+w_2^{SO(N-m)}w_2^{SO(m)}$, and (b) $w_4^{SO(n)}=0$ for $n\leqslant 4$.

\subsection{Charge conjugation}

We now consider the $\mc{C}$ symmetry, which is a $\mathbb{Z}_2$ symmetry that is improper (orientation-reversing) in both the $SO(N)$ and $SO(N-4)$ spaces. Upon including this improper $\mathbb{Z}_2$ symmetry, the probe $SO(N)\times SO(N-4)$ gauge field will be enhanced to an $O(N)\times O(N-4)$ bundle, with the restriction
\be
\label{eq: w1restriction}
w_1^{O(N)}=w_1^{O(N-4)} \hspace{5pt} ({\rm{mod}} \hspace{2pt} 2),
\ee
where $w_1$ is the first SW class of the corresponding bundle. This equation states that an improper rotation in the $O(N)$ space is also improper in $O(N-4)$. 

To study the anomaly, we utilize the simple fact that $O(N)\subset SO(N+1)$. So the anomaly of the $O(N)\times O(N-4)$ bundle in the $S^{(N)}$ theory is completely fixed by the known anomaly of the $SO(N+1)\times SO(N-3)$ bundle in the $S^{(N+1)}$ theory.

To be concrete, let us start from the $S^{(N+1)}$ theory, and condense the component $\la n_{11}\ra=1$ -- as discussed in Sec.~\ref{subsec: SL begins}, this leads to the $S^{(N)}$ theory. The condensate breaks the $SO(N+1)\times SO(N-3)$ symmetry down to $(SO(N)\times SO(N-4))\rtimes \mathbb{Z}_2^{\mathcal{C}}$, which is a subgroup of $O(N)\times O(N-4)$. Now starting from the $SO(N+1)\times SO(N-3)$ anomaly in Eq.~\eqref{eq: Anomaly0}, we can obtain the anomaly associated with the $O(N)\times O(N-4)$ bundle as follows. First we split the bundles $SO(N+1)\to O(1)^A\times O(N)$ and $SO(N-3)\to O(1)^B\times O(N-4)$ (remember $O(1)\sim \mathbb{Z}_2$), with the condition that
\be
w_1^{O(1)^A}=w_1^{O(N)}=w_1^{O(1)^B}=w_1^{O(N-4)}  \hspace{5pt} ({\rm{mod}} \hspace{2pt} 2).
\ee
The first and last equal signs come from the $SO(N+1)\times SO(N-3)$ ``parent" group, and the second equal sign comes from the fact that the condensate $\la n_{11}\ra$ forces the identification of $O(1)^A$ and $O(1)^B$. This gives rise to the restriction Eq.~\eqref{eq: w1restriction}. In the following we shall denote the common $w_1$ of these bundles to be simply $w_1$. Now take the terms in Eq.~\eqref{eq: Anomaly0} and apply Whitney product formula:
\beq
w_4^{SO(N+1)}&\to& w_4^{O(N)}+w_1w_3^{O(N)}, \nn
w_4^{SO(N-3)}&\to& w_4^{O(N-4)}+w_1w_3^{O(N-4)}, \nn
w_2^{SO(N+1)}&\to& w_2^{O(N)}+w_1^2, \nn
w_2^{SO(N-3)}&\to& w_2^{O(N-4)}+w_1^2.
\eeq
From the Wu formula we have $\int w_1^{O(N)}w_3^{O(N)}=\int\sq^1(w_3^{(O(N))})=\int w_1^{TM}w_3^{O(N)}=0$ (mod $2$) on orientable manifolds, where $\sq^1$ is the Steenrod square operation. After some algebra we obtain the following anomaly:
\beq
\label{eq: Anomaly0C}
S_{\rm bulk}&=&i\pi\int_{X_4}(w_4^{O(N)}+w_4^{O(N-4)}+w_2^{O(N)} w_2^{O(N-4)} \nn &+&(w_2^{O(N-4)})^2
+w_1^2(w_2^{O(N)}+w_2^{O(N-4)})).
\eeq

In particular, for $N=6$, the above anomaly agrees with an explicit computation for the $U(1)$ Dirac spin liquid in Ref.~\cite{Calvera2021}. This further strengthens the support for the equivalence between SL$^{(6)}$ and the $U(1)$ DSL.

\subsection{Unorientable manifolds}
\label{subsection: unorientable}

We now consider the anomaly on possibly unorientable manifolds. As discussed in Sec.~\ref{subsec: SL begins}, an orientation-reversing symmetry (such as time-reversal) should also be orientation-reversing in either the $SO(N)$ or $SO(N-4)$ space (but not both). This means that we should again consider an $O(N)\times O(N-4)$ bundle as we did for charge conjugation, but now the restriction Eq.~\eqref{eq: w1restriction} is modified:
\be
\label{eq: w_1restrictionunorientable}
w_1^{O(N)}+w_1^{O(N-4)}+w_1^{TM}=0  \hspace{5pt} ({\rm{mod}} \hspace{2pt} 2).
\ee
So it is now meaningful to ask which $w_1$'s participate in the anomaly terms like Eq.~\eqref{eq: Anomaly0C}  -- from Eq.~\eqref{eq: w_1restrictionunorientable} there are two linearly independent ones.

We now again take advantage of two facts due to the cascade structure among the SLs, as discussed in Sec.~\ref{subsec: SL begins}:

\begin{enumerate}
    \item If we reduce the theory to $S^{(5)}$ through a set of condensation, so that the $O(N-4)$ gauge symmetry is completely broken and $O(N)$ is broken to $O(5)$, the resulting theory is known to have the anomaly $i\pi\int w_4^{O(5)}$, with the restriction $w_1^{O(5)}=w_1^{TM}$ (mod $2$). 
    
    \item We can enter a completely ordered phase by condensing the first $N-4$ rows of the order parameter. This leaves behind an $O(4)\times O(N-4)'$ bundle ($O(N-4)'$ being a combination of an $O(N-4)\subset O(N)$ and the original $O(N-4)$) with the restriction $w_1^{O(4)}+w_1^{TM}=0$ (mod $2$). The anomaly should completely vanish for this bundle.
    
\end{enumerate}

One can check that there is only a single anomaly that satisfies the above two conditions, and reduces to Eq.~\eqref{eq: Anomaly0C} on orientable manifolds{\footnote{On orientable manifolds, $\int (w_1^{O(N-4)})^4=\int \sq^1(w_1^{O(N-4)})^3=\int w_1^{TM}(w_1^{O(N-4)})^3=0$, so the orientable anomaly Eq.~\eqref{eq: Anomaly0C} is reproduced.}}:
\beq
\label{eq: Anomaly11}
S_{\rm bulk}=&&i\pi\int_{X_4}(w_4^{O(N)}+w_4^{O(N-4)}+w_2^{O(N)} w_2^{O(N-4)} \nn &&+(w_2^{O(N-4)})^2+(w_1^{O(N-4)})^4 \nn
&&+(w_1^{O(N-4)})^2(w_2^{O(N)}+w_2^{O(N-4)})).
\eeq
It is relatively easy to see that this anomaly satisfies condition (1). To verify condition (2), the derivation goes as follows. We split the $O(N)$ bundle to $O(4)\times O(N-4)$ and identify the $O(N-4)\subset O(N)$ with the original $O(N-4)$. The SW classes of $O(N)$ split according to Whitney product formula. This leads to the following anomaly for the $O(4)\times O(N-4)$ bundle:
\beq
\label{eq: remnantanomaly}
&&i\pi \int_{X_4}(w_1^{O(4)}w_3^{O(N-4)}+w_1^{O(N-4)}w_3^{O(4)} \nn &&+w_1^{O(4)}w_1^{O(N-4)}w_2^{O(N-4)}+(w_1^{O(N-4)})^4 \nn &&+(w_1^{O(N-4)})^2w_2^{O(4)}+(w_1^{O(N-4)})^3w_1^{O(4)}),
\eeq
with the restriction $w_1^{O(4)}=w_1^{TM}$ (mod $2$). We now notice that these SW classes are not completely independent. There are several useful (mod $2$) relations, valid when integrated over $X_4$:
\beq
0&=&\sq^1(w_1^{O(N-4)}w_2^{O(4)})+w_1^{TM}w_1^{O(N-4)}w_2^{O(4)}  \nn
&=&(w_1^{O(N-4)})^2w_2^{O(4)}+w_1^{O(N-4)}w_1^{O(4)}w_2^{O(4)} \nn & &+w_1^{O(N-4)}w_3^{O(4)}+w_1^{TM}w_1^{O(N-4)}w_2^{O(4)} \nn 
&=&(w_1^{O(N-4)})^2w_2^{O(4)}+w_1^{O(N-4)}w_3^{O(4)}; \nn
0&=&\sq^1\cdot \sq^1(w_2^{O(N-4)}) \nn
&=&\sq^1(w_3^{O(N-4)}+w_1^{O(N-4)}w_2^{O(N-4)}) \nn
&=&w_1^{O(4)}w_3^{O(N-4)}+w_1^{O(4)}w_1^{O(N-4)}w_2^{O(N-4)}; \nn
0&=&(w_1^{O(N-4)})^4+\sq^1[(w_1^{O(N-4)})^3] \nn
&=& (w_1^{O(N-4)})^4+w_1^{O(4)}(w_1^{O(N-4)})^3.
\eeq
These relations come from several properties of the Steenrod square $\sq^1$: $\sq^1 x=w_1^{TM}x$ for $x\in H^3(X_4,\mathbb{Z}_2)$, $\sq^1 w_1^{O(n)}=(w_1^{O(n)})^2$, $\sq^1 w_2^{O(n)}=w_1^{O(n)}w_2^{O(n)}+w_3^{O(n)}$, $\sq^1(x\cup y)=(\sq^1x)\cup y+x\cup \sq^1y$, $\sq^1\cdot\sq^1=0$ as well as the (mod $2$) restriction $w_1^{O(4)}+w_1^{TM}=0$. The remnant anomaly Eq.~\eqref{eq: remnantanomaly} vanishes upon plugging in these relations, as promised. Furthermore, one can check that Eq. \eqref{eq: Anomaly11} is the unique anomaly that satisfies the above properties due to the cascade structure and reduces to Eq. \eqref{eq: Anomaly0C} on orientable manifolds.

We therefore conclude that Eq.~\eqref{eq: Anomaly11}, together with the restriction Eq.~\eqref{eq: w_1restrictionunorientable}, forms the complete anomaly of our theory.

\subsection{Anomaly for the faithful $I^{(N)}$ symmetry from monopole characteristics} \label{subsec: monopoles}

In the above we have derived the anomaly of the SLs by taking its continuous symmetry to be $SO(N)\times SO(N-4)$. As alluded before, this treatment is complete for odd $N$. For even $N$, this symmetry is larger than the faithful $I^{(N)}$ symmetry, and we may miss some anomalies by just looking at the enlarged symmetry. In this subsection, we will derive the anomaly associated with the faithful $I^{(N)}$ symmetry for even $N$. We will see that the $I^{(N)}$ anomaly of the SLs can still be unambiguously pinned down from the cascade structure. Interestingly, although the analysis in the previous subsections indicates that SL$^{(N, 2)}$ is anomaly-free, here we find that for even $N$, if the faithful $I^{(N)}$ symmetry is properly taken into account, SL$^{(N, 4)}$ is anomaly-free, but SL$^{(N, 2)}$ is still anomalous. In the following discussion we will mainly focus on anomalies that involve the continuous symmetries, and we leave the full anomaly (for example, on unorientable manifolds) to future works.

Our approach is to consider the $(3+1)$-d SPT whose boundary can host the SL, gauge the $I^{(N)}$ symmetry of this SPT, and use the properties of the $I^{(N)}$ monopoles as a characterization of the SPT. The bulk-boundary correspondence due to anomaly-inflow indicates that this is also a characterization of anomaly of the SL. This approach is a generalization of the one used in the study of symmetry-enriched $U(1)$ quantum spin liquids \cite{Wang2013, Wang2015, Zou2017, Zou2017a}. Note that since the properties of the monopoles capture the properties of the 't Hooft lines of the corresponding $I^{(N)}$ gauge theory, the discussion here can be equivalently phrased in terms of the 't Hooft lines. However, we will use the language of the monopoles. Here we will focus on the case with an even $N$, and in Appendix \ref{appsub: odd N} we apply this approach to odd $N$ to reproduce the results obtained before. 

To start, let us ask what is the fundamental monopole of an $I^{(N)}$ gauge theory, where by ``fundamental" we mean that all dyonic excitations can be viewed as a bound state of certain numbers of such a fundamental monopole and the pure gauge charge. Naively, one might expect that there are two types of fundamental monopoles: the $SO(N)$ monopole and the $SO(N-4)$ monopole. However, due to the locking of the $Z_2$ centers of the $SO(N)$ and $SO(N-4)$ symmetries, those are not the fundamental monopole. Instead, the fundamental monopole can be viewed as a bound state of half of an $SO(N)$ monopole and half of an $SO(N-4)$ monopole. More explicitly, denote the field configuration of a {\em unit} $U(1)$ monopole by $A_{U(1)}$, whose precise expression is unimportant, and a particular realization is given in Ref. \cite{Wu1975}. Write the $SO(N)$ and $SO(N-4)$ gauge fields as $A^{SO(N)}=A_a^LT_a^L$ and $A^{SO(N-4)}=A_a^RT_a^R$, with $\{T_a^{L}\}$ and $\{T_a^R\}$ the generators of $SO(N)$ and $SO(N-4)$, respectively. For example, $T_{12}^L$ generates the $SO(N)$ rotations in the $(1, 2)$-plane, $T_{34}^R$ generates the $SO(N-4)$ rotations in the $(3, 4)$-plane, etc. A fundamental $I^{(N)}$ monopole can be realized by the following field configuration:
\beq \label{eq: fundamental monopole main}
\begin{split}
&A_{12}^{L}=A_{34}^{L}=A_{56}^{L}=\cdots A_{N-1, N}^{L}\\
=&A_{12}^{R}=A_{34}^{R}=A_{56}^{R}=\cdots A_{N-5, N-4}^{R}=\frac{A_{U(1)}}{2}
\end{split}
\eeq
That is, this $I^{(N)}$ monopole is obtained by embedding {\it half}-$U(1)$ monopoles into the maximal Abelian subgroup of $I^{(N)}$. This configuration breaks the continuous $I^{(N)}$ symmetry to $(SO(2)^{N-2})/Z_2$. So it is convenient to denote a general excitation in this $I^{(N)}$ gauge theory by the following {\em excitation matrix}:
\begin{widetext}
\beq \label{eq: general excitation for even N}
\left(
\begin{array}{c}
\vec q\\
\vec m
\end{array}
\right)_s=
\left(
\begin{array}{cccc|cccc}
q_{12}^L & q_{34}^L & \cdots & q_{N-1,N}^L & q_{12}^R & q_{34}^R & \cdots & q_{N-5,N-4}^R\\
m_{12}^L & m_{34}^L & \cdots & m_{N-1,N}^L & m_{12}^R & m_{34}^R & \cdots & m_{N-5,N-4}^R
\end{array}
\right)_s
\eeq
\end{widetext}
where the first (second) row represents the electric (magnetic) charges of this excitation under $A_{ij}^{L,R}$, $s=0\ ({\rm mod\ }2)$ ($s=1\ ({\rm mod\ }2)$) represents that this excitation is a boson (fermion), and the vertical line separates the charges related to the original $SO(N)$ and $SO(N-4)$ subgroups of $I^{(N)}$. The fundamental monopole has $\vec m=(\half, \half, \cdots, \half)$, and its $\vec q$ and $s$ will characterize the corresponding SPT.

There are multiple constraints on the possible excitation matrices that a consistent theory should satisfy. For instance, a pure gauge charge should have an excitation matrix such that $\vec m=0$, and all entries of $\vec q$ are integers that sum up to an even integer, such as $\vec q=(1, 0, 0, \cdots, 0, 1)$. Another important constraint is the Dirac quantization condition, which in this case states that two excitations with excitation matrices $\left(\begin{array}{c}\vec q_1\\ \vec m_1\end{array}\right)_{s_1}$ and $\left(\begin{array}{c}\vec q_2\\ \vec m_2\end{array}\right)_{s_2}$ should satisfy $\vec q_1\cdot\vec m_2-\vec q_2\cdot\vec m_1\in\mathbb{Z}$. As a sanity check, consider a fundamental monopole with $\vec m=(\half, \half, \cdots, \half)$ as above, and an elementary pure gauge charge with $\vec m=0$ and $\vec q=(1, 0, 0, \cdots, 0, 1)$, we see that the Dirac quantization condition is indeed satisfied. This actually explains why the above configuration of the fundamental monopole is valid in this theory. Other constraints come from the $\mc{C}$, $\mc{R}$ and $\mc{T}$ symmetries, as well as the fact that the original theory has an $I^{(N)}$ gauge structure (see Appendix \ref{app: N=6 anomaly physical} for more details).

Taking all these constraints into account, as shown in Appendix \ref{app: N=6 anomaly physical}, there are only very few classes of distinct types of SPTs. In particular, if $N=2\ ({\rm mod\ }4)$, the structures of fundamental monopoles can be classified as $\mathbb{Z}_2\times\mathbb{Z}_2\times\mathbb{Z}_4$, where the three generators, or ``roots", are given by
\beq \label{eq: 4k+2 monopoles main}
\begin{split}
    &{\rm root\ }1:
    \left(
    \begin{array}{cccc|cccc}
    0 & 0 & \cdots & 0 & 0 & 0 & \cdots & 0\\
    \half & \half & \cdots & \half & \half & \half & \cdots & \half
    \end{array}
    \right)_f\\
    &{\rm root \ }2: \left(
    \begin{array}{cccc|cccc}
    0 & 0 & \cdots & 0 & 0 & 0 & \cdots & 1\\
    \half & \half & \cdots & \half & \half & \half & \cdots & \half
    \end{array}
    \right)_b\\
    &{\rm root \ }3: \left(
    \begin{array}{cccc|cccc}
    \frac{1}{4} & \frac{1}{4} & \cdots & \frac{1}{4} & -\frac{1}{4} & -\frac{1}{4} & \cdots & -\frac{1}{4}\\
    \half & \half & \cdots & \half & \half & \half & \cdots & \half
    \end{array}
    \right)_b
\end{split}
\eeq
For $N=0\ ({\rm mod\ }4)$, the structures of the fundamental monopoles can be classified as $\mathbb{Z}_2\times\mathbb{Z}_2\times\mathbb{Z}_4\times\mathbb{Z}_2$, \ie it has one more $\mathbb{Z}_2$ factor compared to the case with $N=2\ ({\rm mod\ }4)$. The fundamental monopoles in Eq. \eqref{eq: 4k+2 monopoles main} are still the roots for the first $\mathbb{Z}_2\times\mathbb{Z}_2\times\mathbb{Z}_4$ factor, and the root for the additional $\mathbb{Z}_2$ factor has the following fundamental monopole:
\beq
{\rm root\ }4:
\left(
\begin{array}{ccc|cccc}
0 & \cdots & 0 & \half & \half & \half & \half\\
\half & \cdots & \half & \half & \half & \cdots & \half
\end{array}
\right)_b
\eeq

It is useful to derive the properties of the $SO(N)$ and $SO(N-4)$ monopoles from these fundamental $I^{(N)}$ monopoles. The results are listed in Table \ref{tab: SO(N) and SO(N-4) monopoles main}, and the details of the derivation can be found in Appendix \ref{app: N=6 anomaly physical}. It is interesting to notice that the results for $N=2\ ({\rm mod\ }4)$, $N=0\ ({\rm mod\ }8)$ and $N=4\ ({\rm mod\ }8)$ are all different. It is known that the spinor representations of $SO(N)$ in these three classes are complex, real, and pseudoreal, respectively \cite{Zee2016}, which may be related to the difference of the monopoles in SLs with different $N$.

\begin{widetext}

\begin{table*}[h]
    \setlength{\tabcolsep}{0.2cm}
    \renewcommand{\arraystretch}{1.4}
    \centering
    \begin{tabular}{c|cc} 
        \hline \hline
        & $SO(N)$ monopole & $SO(N-4)$ monopole\\
        \hline
        root 1 & (singlet, singlet, boson) & (singlet, singlet, boson) \\
        root 2 & (singlet, singlet, fermion) & (singlet, singlet, fermion)\\
        root 3 & (spinor, spinor, boson) & (spinor, spinor, fermion)\\
        root 4 with $N=0\ ({\rm mod\ }8)$ & (singlet, singlet, fermion) & (singlet, vector, boson)\\
        root 4 with $N=4\ ({\rm mod\ }8)$ & (singlet, singlet, boson) & (singlet, vector, fermion)\\
        \hline \hline
        \end{tabular}
        \caption{Properties of the $SO(N)$ and $SO(N-4)$ monopoles of the root states for even $N$. The first three roots apply to all even $N$, and root 4 only applies to the case with $N$ an integral multiple of 4. The $SO(N)$ monopole breaks the $I^{(N)}$ symmetry to $(SO(2)\times SO(N-2)\times SO(N-4))/Z_2$, and it always has no charge under the $SO(2)$. Its three corresponding entries represent its representation under the $SO(N-2)$, its representation under the $SO(N-4)$, and its statistics, respectively. The $SO(N-4)$ monopole breaks the $I^{(N)}$ symmetry to $(SO(N)\times SO(N-6)\times SO(2))/Z_2$, and it always has no charge under the $SO(2)$. Its three corresponding entries represent its representation under the $SO(N)$, its representation under the $SO(N-6)$, and its statistics, respectively. For the case with $N=6$, the second entry does not exist for its $SO(N-4)$ monopole. Notice these properties are determined up to attaching pure gauge charges.}
    \label{tab: SO(N) and SO(N-4) monopoles main}
\end{table*}

\end{widetext}

The above discussion implies the existence of various $I^{(N)}$-SPTs, and thus also of the $I^{(N)}$-anomalies. Which of the anomalies are compatible with the cascade structure of the SLs, in particular, the two conditions in Sec. \ref{subsection: unorientable}? It is relatively easy to examine the first condition. Note that the $SO(N)$ monopole breaks the $SO(N)$ symmetry to $SO(N-2)$. To satisfy the first constraint, this $SO(N)$ monopole should carry a spinor representation of the remaining $SO(N-2)$, which means that root 3 or its inverse must be involved in the anomaly of the SL. It is a bit more complicated to examine the second condition, and we leave the details to Appendix \ref{app: N=6 anomaly physical}. The result is that only root 3 or its inverse can satisfy both constraints. Therefore, we conclude that for even $N$ the $I^{(N)}$-anomalies of SL$^{(N, \pm 1)}$ are those of root 3 and its inverse, respectively.

In passing, we mention that the monopole properties of the $U(1)$ DSL are explicitly derived in Appendix \ref{app: DSL explicit}, which agree with that of SL$^{(6)}$. This match provides further evidence for our statement in Sec. \ref{sec: N=6: DSL} that the $U(1)$ DSL and SL$^{(6)}$ are actually equivalent.

\subsection{Semion topological order from time-reversal breaking} \label{subsec: semion TO}

It is well known \cite{Vishwanath2012} that non-perturbative anomalies (such as those in this work) can sometimes be satisfied by gapped topological orders in dimension $d\geqslant (2+1)$. However, it is also known that for the $N=5$ theory (the deconfined criticality) the $w_4^{SO(5)}$ anomaly cannot be matched by a gapped topological order if time-reversal symmetry is not broken \cite{Wang2017}. This statement can be easily generalized to arbitrary $N\geqslant 5$ using similar arguments as in Ref. \cite{Wang2017}: consider an $SO(N)$ monopole, represented as a unit $SO(2)\subset SO(N)$ monopole in the first two components. The $w_4^{SO(N)}$ anomaly requires the monopole to carry spinor representation for the remaining $SO(N-2)$. For a gapped topologically ordered state, this condition can be satisfied only by attaching a gapped anyon excitation to the monopole, with the anyon carrying spinor representation under $SO(N-2)$. But an anyon should in general carry irrep under the entire $SO(N)$, which means that the $SO(N-2)$ spinor anyon should also carry $SO(2)$ charge $q=1/2$. This leads to a nontrivial Hall conductance for the $SO(2)$, which necessarily breaks time-reversal symmetry. For $N\geqslant 9$ the same argument applies to the $SO(N-4)$ symmetry since there is also a $w_4^{SO(N-4)}$ anomaly. 

If time-reversal is broken, either explicitly or spontaneously, then a gapped topological order becomes possible. For DQCP ($N=5$) and $U(1)$ Dirac spin liquid ($N=6$), it is known that the simplest topological order that satisfies the anomaly is a semion (or anti-semion) topological order, with only one nontrivial abelian anyon $s$ with exchange statistical phase $e^{i\pi/2}$ (or $e^{-i\pi/2}$ for anti-semion $\bar{s}$) -- basically each semion sees another semion as a $\pi$-flux. We now argue that for any $N\geqslant 5$, the anomaly can be matched by a semion topological order, in which the semion $s$ carries spinor representation under both $SO(N)$ and $SO(N-4)$ (for $N=6$ ``spinor rep'' for $SO(2)$ here means charge $1/2$). For simplicity we shall only consider the $SO(N)\times SO(N-4)$ symmetry below, neglecting the charge conjugation symmetry.

Consider an $SO(2)\subset SO(N)$ monopole. Since the semion carries charge $\pm1/2$ under this $SO(2)$, a semion sees a ``bare'' monopole as a $\pi$-flux. To make the monopole local, one has to attach a semion to the bare monopole to neutralize its mutual statistics with other semions. Since a semion carries spinor rep under both $SO(N)$ and $SO(N-4)$, the monopole now also carries spinor rep under $SO(N-2)\subset SO(N)$ and $SO(N-4)$. Using the same reasoning, an $SO(N-4)$ monopole will also carry spinor rep under $SO(N)$ and $SO(N-6)$ (the latter only if $N\geqslant 9$). These features match exactly with the general anomaly (without time-reversal and charge conjugation):
\be
\label{eq: anomalywithoutT}
i\pi\int_{X_4}(w_4^{SO(N)}+w_4^{SO(N-4)}+w_2^{SO(N)}\cup w_2^{SO(N-4)}).
\ee
Notice that compared to Eq.~\eqref{eq: Anomaly0}, the $(w_2^{SO(N-4)})^2$ term is missing from the above anomaly. This is because the $(w_2)^2$ term is equivalent to the standard $\Theta$-term for the $SO(N-4)$ gauge field. In the absence of time-reversal symmetry the $\Theta$ angle can be continuously tuned to zero and does not count as a nontrivial anomaly.

\section{Possible lattice realizations for $N>6$} \label{sec: N>6}

In the above we have conjectured, based on various evidence, that Stiefel liquids with $N>6$ and $k=1$ exist as an exotic type of critical quantum field theories. As quantum field theories they are interesting because of the possibility that they may be non-Lagrangian, which means that they cannot be UV completed by any weakly-coupled renormalizable continuum Lagrangian. We now discuss their relevance to condensed matter physics. Realizing a SL with $N>6$ in a condensed matter system will be particularly interesting, because it may represent a critical quantum state that has no ``mean-field'' description, not even one with partons, at any scale. In contrast, most correlated states theoretically constructed so far, at least for non-disordered phases at equilibrium, do admit some ``mean-field'' description at some energy scale (typically in the UV). Therefore, our Stiefel liquid states, if realized, will be an example beyond existing paradigms of quantum phases.

To be concrete, we shall discuss possible realizations of SL$^{(N>6)}$ in two dimensional lattice spin systems. In Sec.~\ref{subsec: generalstrategy} we discuss necessary conditions for a SL to be ``emergible'', that is, realizable in some local Hamiltonian systems. The two important conditions to be discussed are (1) anomaly matching and (2) dynamical stability. Subsequently we will discuss some concrete examples relevant to SL$^{(7)}$. We propose that on a triangular lattice, SL$^{(7)}$ can naturally arise as a competition (or intertwinement) between a tetrahedral magnetic order and the 12-site VBS order, and on a kagome lattice, SL$^{(7)}$ can naturally arise as a competition (or intertwinement) between a cuboctahedral magnetic order and a VBS order.

\subsection{General strategy}
\label{subsec: generalstrategy}

\subsubsection{Anomaly matching}

Given an effective IR field theory (such as our Stiefel liquids), an important question for condensed matter physicists is whether it can be realized as the low-energy theory of some lattice local Hamiltonian system. In general, it is very hard to definitively answer such questions, since the space of all local Hamiltonians has infinite dimensions (corresponding to infinitely many tuning parameters), and the vast majority of those Hamiltonians are not analytically solvable. 

A new approach to this type of questions has emerged in recent years based on Lieb-Schultz-Mattis (LSM) type of theorems and 't Hooft anomaly matching \cite{Lieb1961, Oshikawa1999, Hastings2003, Po2017, Cheng2015, Jian2017, Cho2017, Huang2017}. It has been well known, since Lieb-Schultz-Mattis, that certain structures of the lattice Hilbert space can forbid a trivial (symmetric and short-range entangled) ground state. For example, if a lattice spin system has an odd number of $S=1/2$ moments per unit cell, then as long as the $SO(3)$ spin rotation and lattice translation symmetries are unbroken, the ground state must either be gapless or topologically ordered. More recently, it has been appreciated that such LSM constraints are equivalent to certain 't Hooft anomaly matching conditions{\footnote{In this paper we assume that the constraints on the IR physics from the UV information are not associated with filling factors relevant for systems with $U(1)$ and translation symmetries, otherwise there can be further subtleties. See, for example, Ref. \cite{Song2019, Else2020} for more details.}}. Again we use the example with a spin-$1/2$ per unit cell, and now focus on $(2+1)$-d. If we try to couple the system to background gauge fields of the $SO(3)$ spin rotation symmetry and the $T_x$, $T_y$ translation symmetries (each with a group structure $\mathbb{Z}$) \cite{Thorngren2016}, the coupling should be anomalous, with an anomaly term in one higher dimension:
\be
\label{eq: LSManomaly}
i\pi\int_{X_4}w_2^{SO(3)}xy,
\ee
where $x,y\in H^1(X_4,\mathbb{Z})$ are the integer-valued gauge fields corresponding to the two translation symmetries. There may also be other anomalies involving lattice rotations, reflections and time-reversal, depending on the type of the lattice (we will discuss a concrete example in Sec.~\ref{subsec: triangleanomaly}). The LSM-like constraints state that the IR theories that emerge out of such lattice systems should also match the above anomalies, since anomalies are invariant under RG flow. This immediately rules out short-range entangled symmetric ground states, since there would be no IR degrees of freedom to match the anomaly. This also rules out conventional, Landau-Ginzburg-Wilson-Fisher type of theories since those theories do not carry any anomaly. The Stiefel liquids studied in this work do carry nontrivial anomalies, and it is natural to expect that they can match the LSM anomalies and emerge in certain situations. 

When a critical field theory emerges out of a lattice system in the IR limit, the local operators in the IR field theory can all be viewed as some coarse-grained versions of lattice operators. One way to characterize this coarse-graining is to utilize the fact that operators with low scaling dimensions are characterized by their symmetry properties. For example, in the Ising model the lattice spin, $S_z$ coarse-grains to the continuum real scalar field $\phi$ in the Wilson-Fisher theory, because both operators are odd under the global $\mathbb{Z}_2$ symmetry. More systematically, this coarse-graining is described by an embedding of the symmetries at the lattice scale, $G_{UV}$, to the symmetries of the IR theory, $G_{IR}$\footnote{Here we assume that $G_{IR}$ contains only $0$-form symmetries, which is likely the case for our Stiefel liquids. When $G_{IR}$ contains higher-form symmetries (such as in topological orders), the argument can be readily generalized. In this paper, we also assume that the many-body Hilbert space of the lattice system can be viewed as a tensor product of local Hilbert spaces on different lattice sites. If this is not the case, we expect that our approach still applies, as long as the appropriate LSM constraints are used (see Ref. \cite{Kobayashi2018} for some of such examples of lattice systems and LSM constraints).}. Typically $G_{UV}$ includes on-site symmetries like spin-rotation and time-reversal, as well as lattice symmetries like translations and rotations. For the Stiefel liquids, $G_{IR}$ symmetries include $SO(N)$, $SO(N-4)$, $\mathcal{C,R,T}$ as well as the emergent Poincar\'e symmetry. More formally, this embedding is characterized by a group homomorphism
\be
\varphi: G_{UV} \to G_{IR}.
\ee
As a simple example, when the $\mathbb{Z}_2$ Wilson-Fisher theory is realized from the Ising model near criticality, the lattice Ising $\mathbb{Z}_2$ symmetry is mapped under $\varphi$ to the $\mathbb{Z}_2$ symmetry of the Wilson-Fisher theory. If a Stiefel liquid is realized out of a spin system, both $G_{UV}$ and $G_{IR}$ will be more complicated than the Ising-Wilson-Fisher theory, and in general there are multiple nontrivial group homomorphisms $\varphi$ between $G_{UV}$ and $G_{IR}$. The natural question is: which  $\varphi$, if any, is physically legitimate? The LSM anomaly-matching conditions provide a strong constraint: the IR theory contains an anomaly $w[G_{IR}]$, as described in detail in Sec.~\ref{Sec: Anomaly Analysis}. We can now pullback the IR anomaly using $\varphi$, and obtain the corresponding anomaly for the UV symmetry:
\be
w[G_{UV}]=\varphi^*w[G_{IR}].
\ee
The requirement on $\varphi$ is that for the IR anomaly discussed in Sec.~\ref{Sec: Anomaly Analysis}, such as Eq. \eqref{eq: Anomaly11}, the pullback yields exactly the LSM anomalies, such as Eq.~\eqref{eq: LSManomaly} and its various generalizations.

Anomaly-matching has thus been established as a necessary condition for a low-energy theory to be emergible. Here we shall go one step further and conjecture that it is also sufficient. This conjecture can be phrased as

\fbox{
  \parbox{0.42\textwidth}{
  {\textbf {Hypothesis of emergibility}}: a low-energy theory is emergible out of a lattice system if and only if its anomaly matches with that from the lattice LSM-like theorems. 
  }
}\\

Although there is no proof to this statement, there is also no known counter example  {\footnote{There are systems with ``SPT-LSM constraints" that allow a symmetric short-ranged entangled ground state, but these symmetric short-range entangled ground state must be a nontrivial SPT in certain sense \cite{Lu2017, Yang2017, Else2019, Jiang2019}. We believe that these systems also satisfy the hypothesis, as long as both the UV and IR anomalies are properly accounted for. Furthermore, the low-energy theories we will consider below are all gapless, and stacking it with an SPT is not expected to change its low-energy dynamics (at least in the bulk). So we will ignore the subtlety associated with the SPT-LSM constraints in this paper and leave it for future work.}}. An indirect piece of supporting evidence of this conjecture is  the existence of ``featureless Mott insulators'': in certain systems such as the half-filled honeycomb lattice, the ground state is guaranteed to be gapless within free-fermion band theory, but there is no LSM-like constraint, so one may expect that with strong interactions a trivial state can emerge. Indeed, trivial states have been theoretically constructed in various such systems \cite{Kimchi2012, Jian2015, Kim2015,Latimer2020}.

Once we make the above conjecture, the task of finding emergible Stiefel liquids on certain lattice systems becomes the task of finding appropriate homomorphism, $\varphi: G_{UV}\to G_{IR}$, that pulls back the IR anomaly to the LSM anomaly.

\subsubsection{Dynamical stability}

The above hypothesis of emergibility based on anomaly matching only concerns about whether the IR theory can emerge at all, but does not make any statement about the stability of this IR theory, even if it is emergible. In order for the IR theory to be stable, we require that all $G_{UV}$-symmetry-allowed local perturbations to this IR theory are RG irrelevant.

For Stiefel liquids, although we have argued in Sec. \ref{subsec: IR fate} that all $G_{IR}$-symmetry-allowed local perturbations are RG irrelevant, since $G_{UV}$ is much smaller than $G_{IR}$, in general there will be operators that are nontrivial under $G_{IR}$ but trivial under $G_{UV}$, which may be RG relevant. Therefore, to avoid discussing unstable states (which are practically hard to access), we should look for $\varphi$ that only allows a small number of relevant perturbations. However, we do not know the accurate scaling dimensions of various operators in Stiefel liquids, although some guesswork can be done based various trends at $N=5$ and $N=6$, which have been numerically measured for DQCP and DSL, respectively. In general, we expect operators in sufficiently high-rank representations of either $SO(N)$ or $SO(N-4)$ to be irrelevant, but the ``critical rank'' is hard to determine. In fact, even for the $U(1)$ DSL (SL$^{(6)}$), it is still not entirely clear whether various rank-$2$ operators are irrelevant or not. These are represented in the QED$_3$ theory as various fermion quartic interactions and $4\pi$ monopoles. We will therefore consider embeddings $\varphi$ that disallow low-rank operators (such as vectors) as much as possible. 

More specifically, in the examples to be discussed in this section, two types of operators will be symmetry-disallowed: (1) the $n$ operators, which are vectors under both $SO(N)$ and $SO(N-4)$ -- these are believed to be the most relevant operators based on experience with DQCP and DSL; and (2) the conserved currents of either $SO(N)$ or $SO(N-4)$, these operators have scaling dimension $2$ and will be relevant if symmetry-allowed.

In the following we will construct some illuminating examples of lattice realizations of Stiefel liquids, characterized by the embeddings $\varphi$. These are by no means the only ways to realize Stiefel liquids on lattice systems. Instead, our goal is to illustrate the possibility of realizing Stiefel liquids in some lattice systems. We also note that all these realizations have an $SO(3)$ spin rotational symmetry that is embeded into the $SO(N)$ subgroup of $G_{IR}$, so it suffices to use Eq. \eqref{eq: Anomaly11} to characterize the anomaly of a SL, for any $N$.

\subsection{List of LSM-like anomalies in (2+1)d}
\label{subsec: triangleanomaly}

Here we list LSM-like anomalies that can arise in a two dimensional lattice spin system, with on-site $SO(3)$ and time-reversal symmetries as well as lattice symmetries. For simplicity, the lattice symmetry we consider will only include discrete translations, rotations and reflections.

First, if there is a odd number of $S=1/2$ moments per unit cell, there is the aforementioned anomaly involving $SO(3)$ and translations:
\be
\label{eq: LSManomaly}
S_{tr-LSM}=i\pi\int_{X_4}w_2^{SO(3)}xy,
\ee
where $x,y\in H^1(X_4,\mathbb{Z})$ are the $T_{a_1},T_{a_2}$ translation gauge fields. Likewise, if each spin-$1/2$ moment is also a Kramers doublet ($\mathcal{T}^2=-1$), then there should be another anomaly
\be
\label{eq: TLSManomaly}
S_{\mathcal{T}-LSM}=i\pi\int_{X_4}t^2  x y,
\ee
where $t\in H^1(X_4,\mathbb{Z}_2)$ is the gauge field for time-reversal symmetry. 

Next, if the location of each spin-$1/2$ moment is also an inversion ($C_2$ rotation) center, there is another anomaly:
\be
\label{eq: ILSManomaly}
S_{I-LSM}=i\pi\int_{X_4}c^2 (w_2^{SO(3)}+t^2),
\ee
where $c\in H^1(X_4,\mathbb{Z}_2)$ is a $\mathbb{Z}_2$ gauge field associated with the $C_2$ rotation symmetry. The $\mathbb{Z}_2$ nature of the topology of $SO(3)$ and $\mc{T}$ guarantees that other types of rotations like $C_3$ will not contribute to anomaly.

Now consider the reflection symmetry $\mathcal{R}_y$, we should also examine the reflection axis (the line that is invariant under reflection): we view the reflection axis as a $(1+1)$-d system, with a translation symmetry $T_{a_1}$ that commutes with reflection and possibly a $C_2$ rotation that acts like one dimensional inversion on the axis. If decorated on the reflection axis is a spin-$1/2$ chain, it will have its own LSM-like anomalies. We will then have the following anomalies
\be
\label{eq: RLSManomaly}
i\pi\int_{X_4}r(w_2^{SO(3)}+t^2)(x+c),
\ee
where $r\in H^1(X_4,\mathbb{Z}_2)$ is the gauge field for reflection. The fact that both time-reversal and reflection changes the space-time orientation means that we also have the following restriction:
\be
r+t=w_1^{TM} \hspace{10pt} ({\rm{mod}}\hspace{2pt}2).
\ee

If the lattice system has all the above features, the LSM anomalies add up to:
\beq \label{eq: full LSM}
i\pi\int_{X^4}(w_2^{SO(3)}+t^2)[xy+c^2+r(x+c)].
\eeq
Notice that there are also other LSM constraints that are not explicitly described above, but are nevertheless contained in this anomaly formula. For example, there can be analogous LSM constraints associated with the reflection symmetry $\mc{R}_x$ (the reflection symmetry with invariant axis perpendicular to that of $\mc{R}_y$). In Appendix \ref{app: LSM}, we derive some of these other LSM constraints from Eq. \eqref{eq: full LSM}. We also stress that in a given lattice system, there may be multiple different, say, $C_2$ rotation symmetries. Some of them have rotation centers hosting an odd number of spin-1/2's, but others do not. When the above formula is applied to the former, the contributions from Eq. \eqref{eq: ILSManomaly} should be nontrivial, while if it is applied to the latter, Eq. \eqref{eq: ILSManomaly} should vanish. 
In general, the anomalies associated with all these different $C_2$ centers should be specified separately. But on $C_6$-symmetric lattices such as triangular and Kagome, one typically only need to specify one inversion center and the others will be determined by symmetries. For example, consider triangular lattice. There are four inversion centers per unit cell: the $C_6$ center (lattice site) and three $C_2$ centers (bond center) that are related to each other through $C_6$ rotations. The parity of $2S$ ($S$ being the spin moment) of the entire unit cell is given by the sum of the parity on each inversion center. This means that if we have  anomaly $aw_2^{SO(3)}xy$ and $bw_2^{SO(3)}c^2$, where $a,b\in\{0,1\}$ and $c$ probes the site-centered inversion, then the anomaly associated with the bond-centered inversion (probed by $c'$) will be given by $(a+b)w_2^{SO(3)}(c')^2$.  For this reason we will only focus on one inversion center in these lattices.

The above situations (summarized in Eq. \eqref{eq: full LSM}) happen for a variety of $2d$ lattice system, including square, triangular and Kagome lattices {\footnote{In Appendix \ref{app: LSM}, an alternative expression for the LSM anomaly on a square lattice is given. We have checked that all the following statements about states on a square lattice hold if either the expression here or the alternative one there is used.}}. These are common playgrounds for studying frustrated quantum magnetism, and we will see some examples next.  

\subsection{Warm-up: anomaly-matching for DQCP}

Given the variety of anomaly terms presented above, it is rather nontrivial for a theory to exactly match all the anomalies. Below we show how this works for the DQCP on square lattice. In appendix \ref{app: anomaly matching triangular DSL},  we show this for the slightly more complicated case of $U(1)$ DSL on triangular lattice. Since these states admit explicit parton mean-field constructions on the lattices, we expect them to be emergible and the anomaly-matching should go through. So these exercises serve as a benchmark for the anomaly-matching approach. 

\begin{figure}
    \centering
    \includegraphics[width=0.36\textwidth]{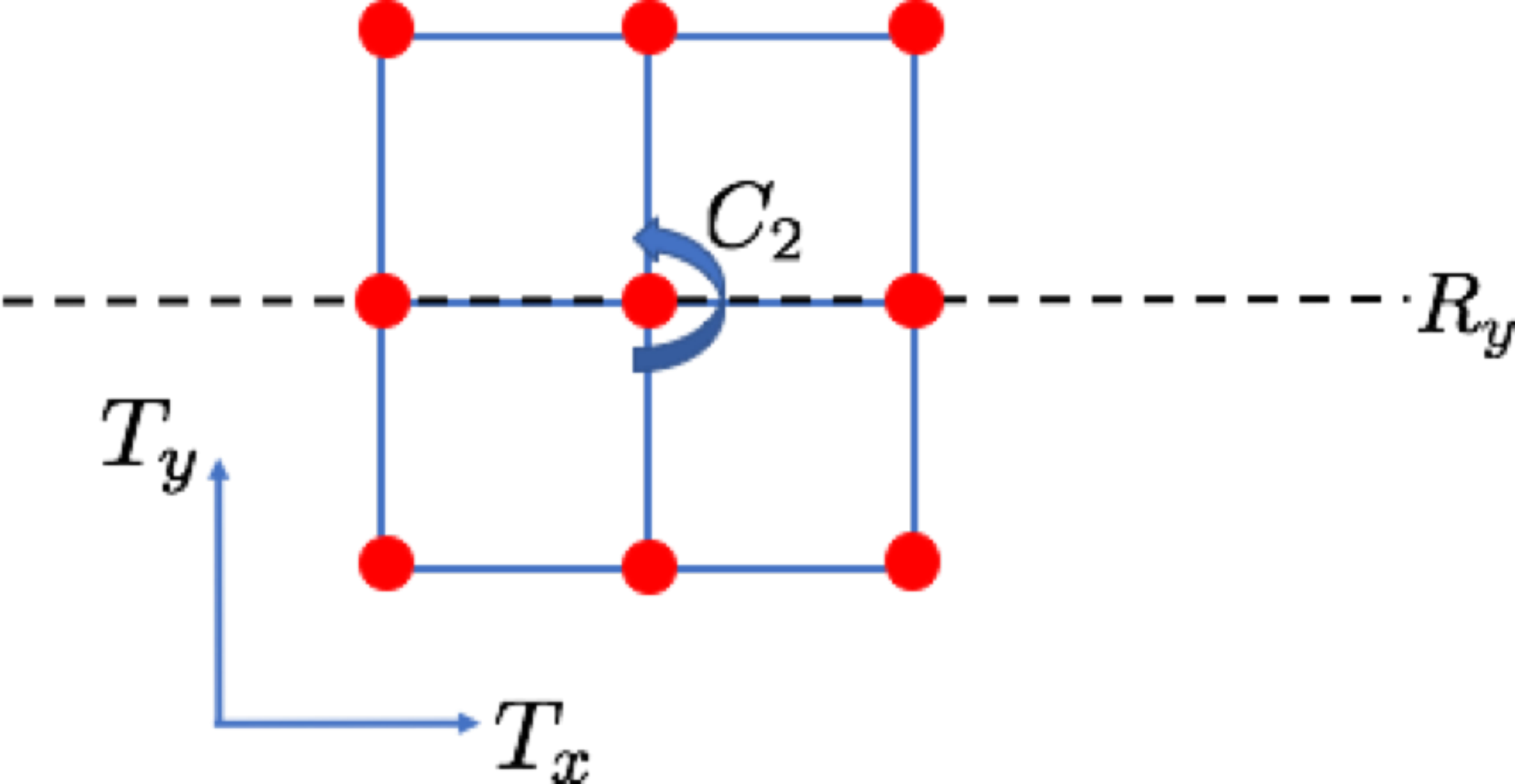}
    \caption{Square lattice and the relevant symmetries. Each filled red circle represents an odd number of spin-1/2's. The $C_2$ rotation is around a site that hosts the spins, and the dashed line is the reflection axis of $R_y$.}
    \label{fig:square}
\end{figure}

As we reviewed in Sec.~\ref{subsec: review of DQCP}, the DQCP corresponds to $S^{(5)}$, with IR anomaly
\be
i\pi\int_{X_4}w_4^{O(5)},
\ee
together with the restriction $w_1^{O(5)}=w_1^{TM}$.

We represent the microscopic symmetry implementations by their actions on the $n$ field, which for DQCP is simply a $5$-component vector. The $SO(3)$ spin rotation is implemented as
\be
n\to \left(\begin{array}{cc}
    SO^s(3) & 0  \\
     0 & I_{2}
\end{array} \right)n.
\ee
Time-reversal symmetry acts as
\be
n\to \left(\begin{array}{cc}
    -I_{3} & 0  \\
     0 & I_{2}
\end{array} \right)n, \hspace{10pt} i\to -i.
\ee

On square lattice the translation symmetries $T_x,T_y$ are implemented as (see Fig. \ref{fig:square})
\beq
T_x: && n\to \left(\begin{array}{ccccc}
    -1 & 0 & 0 & 0 & 0 \\
     0 & -1& 0 & 0 & 0 \\
     0 & 0 & -1 & 0 & 0 \\
      0 & 0 & 0 & -1 & 0 \\
       0 & 0 & 0 & 0 & 1
\end{array} \right)n, \nn
T_y: && n\to \left(\begin{array}{ccccc}
    -1 & 0 & 0 & 0 & 0 \\
     0 & -1& 0 & 0 & 0 \\
     0 & 0 & -1 & 0 & 0 \\
      0 & 0 & 0 & 1 & 0 \\
       0 & 0 & 0 & 0 & -1
\end{array} \right)n.
\eeq

As for the lattice rotation, since only the site-centered inversion ($C_2$) participates in the anomaly, we should just focus on it:
\be
C_2: n\to \left(\begin{array}{cc}
    I_{3} & 0  \\
     0 & -I_{2}
\end{array} \right)n.
\ee

Finally, for reflection $\mathcal{R}_y$:
\be
n\to \left(\begin{array}{ccccc}
    1 & 0 & 0 & 0 & 0 \\
     0 & 1& 0 & 0 & 0 \\
     0 & 0 & 1 & 0 & 0 \\
      0 & 0 & 0 & 1 & 0 \\
       0 & 0 & 0 & 0 & -1
\end{array} \right)n.
\ee

We can now pull back the $w_4^{O(5)}$ anomaly to the physical symmetries. The calculation proceeds as follows. First, since none of the microscopic symmetries mixes $n_{1,2,3}$ with $n_{4,5}$, we can decompose the $O(5)$ bundle into $O(3)\times O(2)$, where 
\beq
w_1^{O(3)}&=&t+x+y, \nn
w_1^{O(2)}&=&r+x+y, \nn
w_2^{O(3)}&=&w_2^{SO(3)}+t^2, \nn
w_2^{O(2)}&=&xy+c^2+xr+cr, \nn
w_3^{O(3)}&=&\sq^1(w_2^{O(3)})+w_1^{O(3)}w_2^{O(3)} \nn
&=&\sq^1(w_2^{SO(3)})+(t+x+y)(w_2^{SO(3)}+t^2),
\eeq
and all higher SW classes vanish. Notice that we have used the fact that $x,y$ live in $H^1(X_4,\mathbb{Z})$ so $x^2=y^2=0$. We also set $cx=cy=ry=0$, based on the physical understanding that non-commuting crystalline symmetries do not simultaneous contribute to the same anomaly term -- this can be seen, for example, from the dimension-reduction approach \cite{Song2017}. We can then write the $w_4^{O(5)}$ using Whitney product formula:
\beq
\begin{split}
w_4^{O(5)}
=&\sum_{i}w_i^{O(3)}w_{4-i}^{O(2)}\\
=&(w_2^{SO(3)}+t^2)(xy+c^2+xr+cr)\\ 
&+ r(\sq^1(w_2^{SO(3)})+tw_2^{SO(3)}+t^3)\\
&+ (x+y)[\sq^1(w_2^{SO(3)})+w_1^{TM}(w_2^{SO(3)}+t^2)]
\end{split}
\eeq
where we have used the constraint $r+t=w_1^{TM}$ to obtain the last line. The first line above is exactly what we expect from LSM constraints from Eqs.~\eqref{eq: LSManomaly}-\eqref{eq: RLSManomaly}, so our task now is to show that the last two lines vanish. This follows from the following relations:
\beq
r\sq^1(w_2^{SO(3)})&=&w_1^{TM}rw_2^{SO(3)}+r^2w_2^{SO(3)} \nn 
&=& trw_2^{SO(3)}, \nn
rt^3&=&w_1^{TM}t^3+t^4=0, \nn
w_1^{TM}(x+y)w_2^{SO(3)}&=&\sq^1[(x+y)w_2^{SO(3)}]\nn &=&(x+y)\sq^1(w_2^{SO(3)}), \nn
w_1^{TM}(x+y)t^2&=&\sq^1[(x+y)t^2]=0.
\eeq

\subsection{$N=7$: intertwining non-coplanar magnets with valence-bond solids} \label{subsec: N=7 emergibility}

In spin systems, as we reviewed in Sec.~\ref{sec: review and summary}, the DQCP (SL$^{(N=5)}$) naturally describes the competition (or intertwining) between collinear magnetic and valence-bond solid (VBS) orders, while the $U(1)$ Dirac spin liquid (SL$^{(N=6)}$) naturally describes the intertwining between coplanar magnetic and VBS orders. The natural extension to the intertwining between non-coplanar magnetic and VBS orders is the $N=7$ Stiefel liquid state. Below we discuss two lattice realizations of the SL$^{(7)}$ theory, one on triangular lattice and one on Kagome lattice.

\subsubsection{Triangular lattice}

We consider a triangular lattice with an odd number of half-integer spins per site, so the LSM anomalies are given by Eqs.~\eqref{eq: LSManomaly}, \eqref{eq: TLSManomaly}, \eqref{eq: ILSManomaly} and \eqref{eq: RLSManomaly}. These conditions impose strong constraints on the allowed lattice realizations of the $N=7$ SL theory. We now describe an embedding of the microscopic symmetries to the $N=7$ SL theory that matches the anomaly. We specify the embedding by the symmetry actions on the $SO(7)/SO(4)$ field $n_{ji}$, where the $SO(7)$ symmetry acts on the left and the $SO(3)$ symmetry acts on the right. 
\begin{figure}
    \centering
    \includegraphics[width=0.36\textwidth]{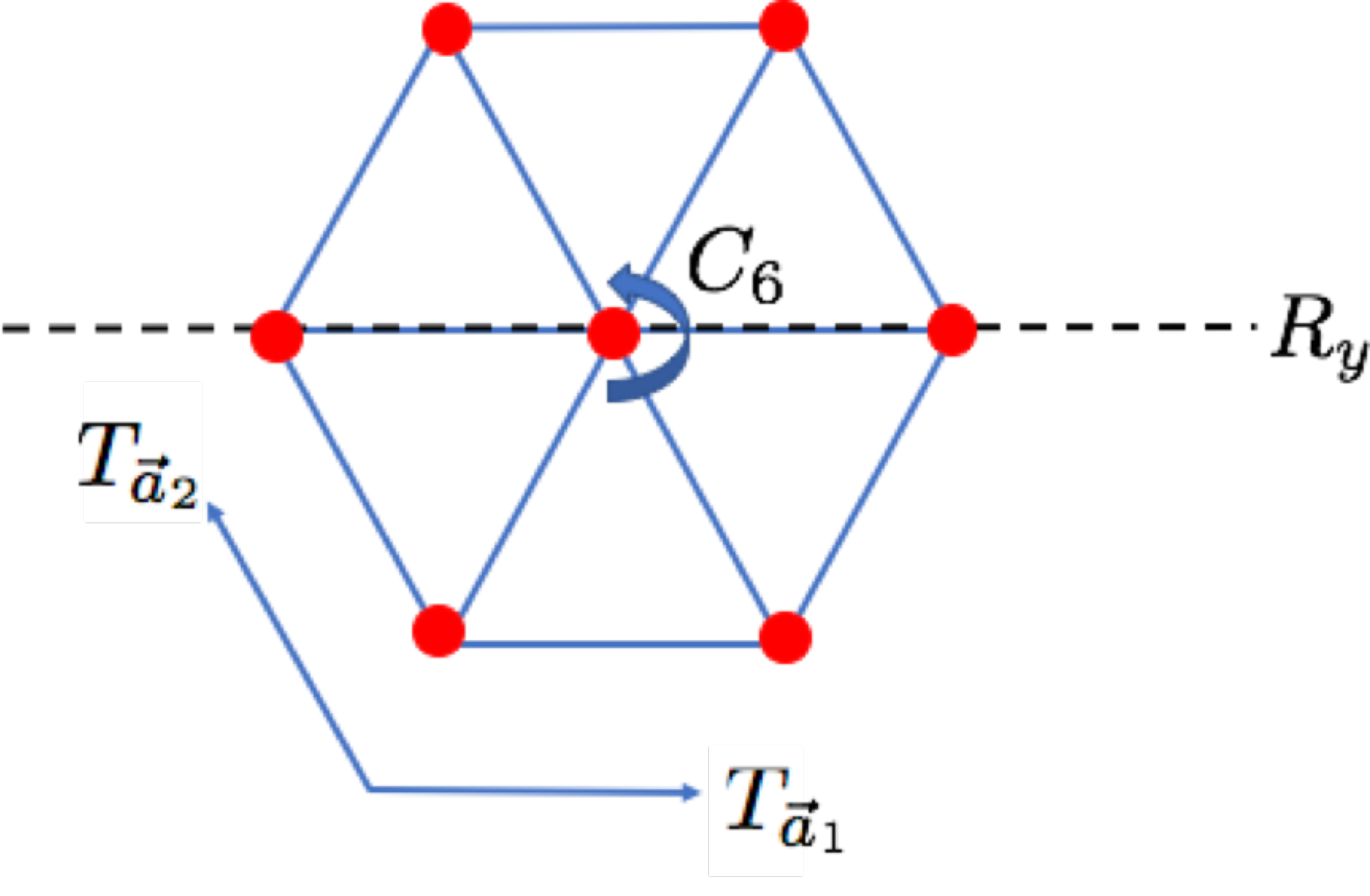}
    \caption{Triangular lattice and the relevant symmetries. Each filled red circle represents an odd number of spin-1/2's. The $C_6$ rotation is around a site that host the spins, and the dashed line is the reflection axis of $R_y$.}
    \label{fig:triangular}
\end{figure}

First, the on-site spin rotation $SO^s(3)$ symmetry acts as an $SO^s(3)$ subgroup of $SO(7)$:
\be
\label{eq: SO(3)embedding}
n\to \left(\begin{array}{cc}
    SO^s(3) & 0  \\
     0 & I_{4}
\end{array} \right)n.
\ee
Next we specify time-reversal symmetry as
\be
\label{eq: TRembedding}
n\to \left(\begin{array}{cc}
    -I_{3} & 0  \\
     0 & I_{4}
\end{array} \right)n, \hspace{10pt} i\to -i.
\ee

For translations along the three unit vectors $T_{\vec{a}_1}$, $T_{\vec{a}_2}$ and $T_{-\vec{a}_1-\vec{a}_2}$ (see Fig. \ref{fig:triangular}), we have (we shall use $\sigma_{\mu}^{i,j}$ to denote the $\mu$-th Pauli matrix acting on the $i,j$ indices)
\begin{widetext}
\beq
\label{eq: Translationaction}
T_{\vec{a}_1}: && n\to \left(\begin{array}{ccc}
    I_{3} & 0 & 0  \\
     0 & \exp\left(i\frac{2\pi}{3}\sigma_y^{4,5}\right) & 0 \\
     0 & 0 & \exp\left(-i\frac{2\pi}{3}\sigma_y^{6,7}\right)
\end{array}\right)n\left(\begin{array}{ccc}
    -1 & 0 & 0  \\
     0 & -1 & 0 \\
     0 & 0 & 1
\end{array}\right),  \nn
T_{\vec{a}_2}: && n\to \left(\begin{array}{ccc}
    I_{3} & 0 & 0  \\
     0 & \exp\left(i\frac{2\pi}{3}\sigma_y^{4,5}\right) & 0 \\
     0 & 0 & \exp\left(-i\frac{2\pi}{3}\sigma_y^{6,7}\right)
\end{array}\right)n\left(\begin{array}{ccc}
    -1 & 0 & 0  \\
     0 & 1 & 0 \\
     0 & 0 & -1
\end{array}\right),  \nn
T_{-\vec{a}_1-\vec{a}_2}: && n\to \left(\begin{array}{ccc}
    I_{3} & 0 & 0  \\
     0 & \exp\left(i\frac{2\pi}{3}\sigma_y^{4,5}\right) & 0 \\
     0 & 0 & \exp\left(-i\frac{2\pi}{3}\sigma_y^{6,7}\right)
\end{array}\right)n\left(\begin{array}{ccc}
    1 & 0 & 0  \\
     0 & -1 & 0 \\
     0 & 0 & -1
\end{array}\right).
\eeq
\end{widetext}
The $C_6=C_2\times C_3$ rotation is implemented as
\be
\label{eq: C6action}
C_6: n\to \left(\begin{array}{ccccc}
  I_{3} & 0 & 0 & 0 & 0 \\
  0 & 1 & 0 & 0 & 0 \\
  0 & 0 & -1 & 0 & 0 \\
  0 & 0 & 0 & 1 & 0 \\
  0 & 0 & 0 & 0 & -1 \end{array}\right) n \left(\begin{array}{ccc}
    0 & 1 & 0  \\
     0 & 0 & 1 \\
     1 & 0 & 0
\end{array}\right).
\ee
Finally, the reflection $\mathcal{R}_y$ (preserving $\vec{a}_1$ but exchanging $\vec{a}_2$ with $-\vec{a}_1-\vec{a}_2$) acts as
\be
\label{Raction}
\mathcal{R}_y: n\to \left(\begin{array}{ccccc}
  I_{3} & 0 & 0 & 0 & 0 \\
  0 & 1 & 0 & 0 & 0 \\
  0 & 0 & 1 & 0 & 0 \\
  0 & 0 & 0 & -1 & 0 \\
  0 & 0 & 0 & 0 & -1 \end{array}\right) n\left(\begin{array}{ccc}
  0 & 1 & 0  \\
  1 & 0 & 0  \\
  0 & 0 & 1  \\
 \end{array}\right).
\ee
The symmetry actions are chosen so that all components of the $n_{ji}$ field are nontrivial under some symmetry action. This makes the state somewhat stable since the most fundamental fields are not allowed by symmetry as perturbations. One can check that the conserved currents of the $SO(7)\times SO(3)$ symmetry are also forbidden, which is important since these operators have scaling dimension $2$ and are relevant. Whether the theory is actually a stable phase depends on the (yet unknown) relevance or irrelevance of composite operators like $n_{ji}n_{j'i'}$.

One can check that the symmetry actions are consistent with the group algebra and indeed gives a homomorphism. One can also check the anomaly-matching conditions as follows. First, we pull back the SW classes to the physical symmetries:
\beq
w_1^{O(7)}&=&t, \nn
w_2^{O(7)}&=&w_2^{SO(3)}+t^2+c^2+r^2+rc, \nn
w_4^{O(7)}&=&(w_2^{SO(3)}+t^2)(c^2+r^2+rc)+tcr(c+r), \nn
w_1^{O(3)}&=&r, \nn
w_2^{O(3)}&=&xy+xr.
\eeq
Notice that the restriction $w_1^{TM}=w_1^{O(7)}+w_1^{O(3)}=r+t$ is satisfied. Now plugging these into the IR anomaly Eq.~\eqref{eq: Anomaly11}, we obtain
\beq
\label{eq: anomalyofN=7embedding}
S_{\rm bulk}&=&i\pi\int_{X_4}((w_2^{SO(3)}+t^2)(xy+c^2+xr+rc) \nn & &+(c^2+r^2+rc)(r^2+xy+xr)+r^4 \nn
& &+tcr(c+r)+r^2(xy+xr)).
\eeq
We now examine the relations among these terms. Again, we set $ry=cx=cy=0$, since the crystalline symmetries involved in each product do not commute. So the second and third lines of the above anomaly become
\beq
& &c^2r^2+cr^3+tcr(c+r) \nn
&=&(r+t)cr(c+r) \nn
&=& w_1^{TM}cr(c+r) \nn
&=& \sq^1(c^2r+cr^2) \nn
&=& 0.
\eeq
So only the first line of Eq.~\eqref{eq: anomalyofN=7embedding} remains, which is exactly what is required as discussed in Sec.~\ref{subsec: triangleanomaly}.

How do we think of this $N=7$ SL theory on lattice? We can interpret the $n_{ji}$ field as a collection of fluctuating order parameters, whose nature is decided by their symmetry properties. The first three rows of the field $n_{ji}$ ($1\leqslant j\leqslant 3$), being a triplet in $SO^s(3)$, describes magnetic fluctuations at the three $\vec{M}$ points ($(\pi,0)$, $(0,\pi)$ and $(\pi,\pi)$) in the Brillouin zone. A possible pattern of symmetry-breaking order is
\be
\la n\ra\sim\left(\begin{array}{c}
  O_{3\times 3} \\
  0_{4\times 3}\end{array}\right),
\ee
where $O_{3\times 3}$ is a $3\times 3$ orthogonal matrix and $0_{4\times 3}$ is a zero-matrix. This describes a non-coplanar magnetic order, with the expectation of the spin operator $S_{i}$ ($i=1,2,3$) on the site $\vec{r}=n\vec{a}_1+m\vec{a}_2$ ($n,m\in\mathbb{Z}$)
\be
\la S_i\ra\sim O_{i1}\cos [(n+m)\pi]+O_{i2}\cos (n\pi)+O_{i3}\cos (m\pi),
\ee
where the three vectors $O_{i1}, O_{i2}, O_{i3}$ are by construction orthonormal. This non-coplanar magnetic order is also known as the tetrahedral order on triangular lattice. 

The theory can also form a VBS order by condensing $n_{ji}$ with $4\leqslant j\leqslant 7$. By Eq.~\eqref{eq: Translationaction} this VBS order has momentum $\vec{K}+\vec{M}$, which is the same as the commonly studied $12$-site VBS on triangular lattice. 

We therefore conclude that the SL$^{(7)}$ theory can naturally arise as a competition (or intertwinement) between the tetrahedral magnetic order and the $12$-site VBS order on triangular lattice. The tetrahedral order is known, numerically, to arise in the $J_1-J_2-J_{\chi}$ model \cite{Gong2017}, where $J_1$ and $J_2$ are the nearest and next-nearest neighbor Heisenberg couplings, respectively, and $J_{\chi}$ is the spin chirality $\vec{S}_{i}\cdot(\vec{S}_j\times \vec{S}_k)$. This model, however, explicitly breaks the time-reversal and reflection symmetries due to the chirality term. Since time-reversal breaking perturbations are relevant for Dirac spin liquids (SL$^{(6)}$), it may also be relevant for SL$^{(7)}$. Therefore, to search for SL$^{(7)}$, it may be useful to find a time-reversal invariant lattice Hamiltonian that realizes the tetrahedral order, and study the effect of various perturbations on top of it. We also note that the tetrahedral order may be realized in higher-spin systems with additional $(\vec{S}_i\cdot\vec{S}_j)^2$ coupling \cite{Yu2020}. So it is also interesting to explore whether SL physics can arise in those systems.
A smoking-gun signature for the SL$^{(7)}$ state is that both the non-coplanar magnetic and VBS order parameters (\ie the $21$ matrix elements of $n$) are critical with identical critical exponents, which is a consequence of the emergent $SO(7)\times SO(3)$ global symmetry.
Similar physics has been numerically confirmed for the SL$^{(5)}$ (\ie DQCP)~\cite{Nahum2015a}.

\subsubsection{Kagome lattice}

We now consider a Kagome lattice with spin-$1/2$ per site. The Kagome lattice has the same lattice symmetries as the triangular. There are three spin-$1/2$ moments in each unit cell, so the LSM anomaly involving translation symmetries is identical to the triangular case. The only essential difference with the triangular lattice is the lack of spin moment at the $C_6$ rotation center. So instead of Eq.~\eqref{eq: full LSM}, the full LSM anomaly on Kagome is
\be
\label{eq: KagomeLSM}
i\pi\int_{X^4}(w_2^{SO(3)}+t^2)(xy+rx).
\ee

\begin{figure}
    \centering
    \includegraphics[width=0.4\textwidth]{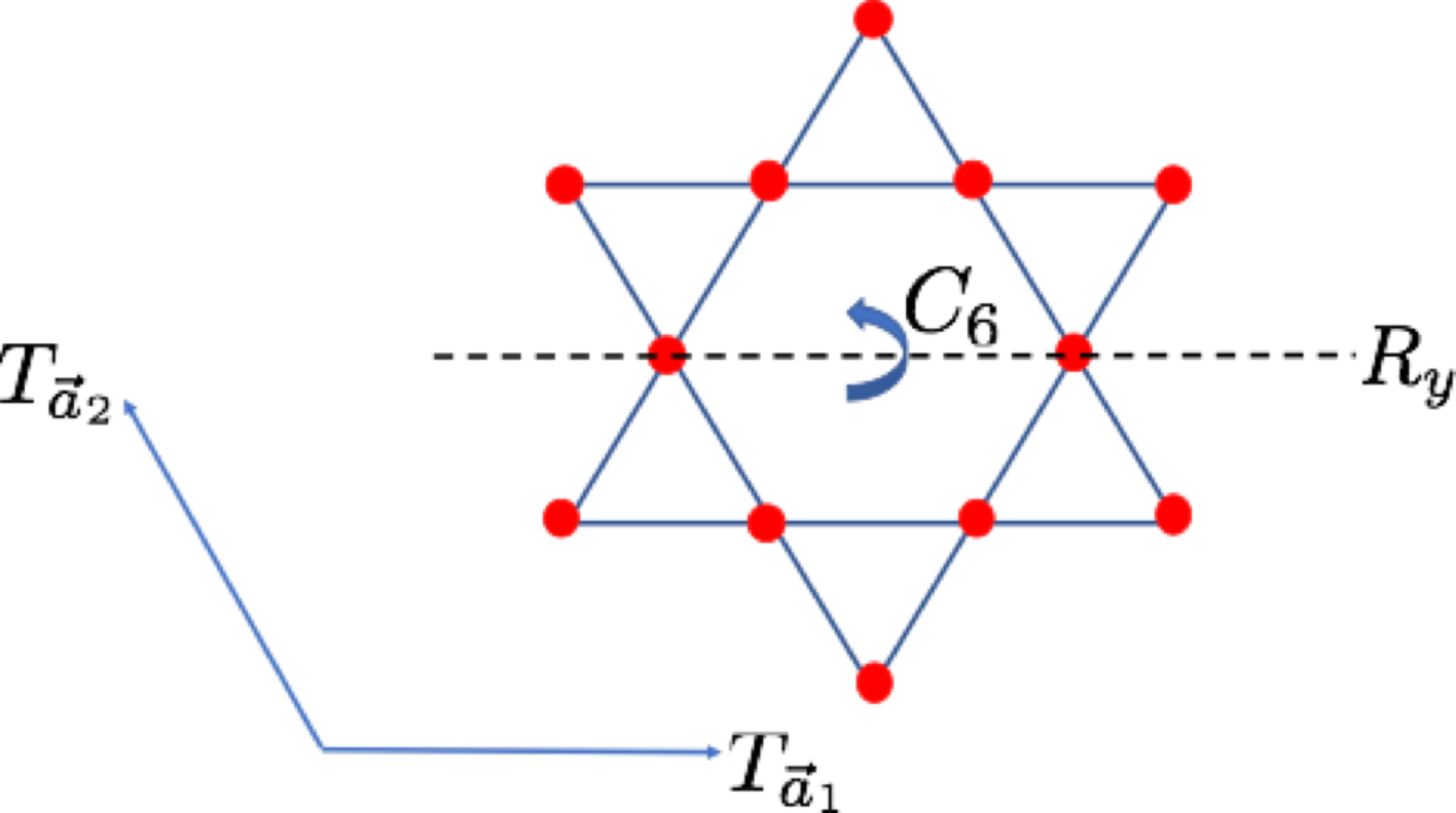}
    \caption{Kagome lattice and the relevant symmetries. Each filled red circle represents an odd number of spin-1/2's. In this case the rotation center of $C_6$ does not host any spin. Again, the dashed line is the reflection axis of $R_y$.}
    \label{fig:kagome}
\end{figure}

We now describe a symmetry embedding (a lattice realization) of the SL$^{(7)}$ theory on Kagome lattice. The spin rotation and time-reversal act on the first three rows of $n$, in the same way as the triangular realization following Eq.~\eqref{eq: SO(3)embedding} and \eqref{eq: TRembedding}. The translation symmetries act as (see Fig. \ref{fig:kagome})
\beq
\label{eq: KagomeTranslationaction}
T_{\vec{a}_1}: && n\to n\left(\begin{array}{ccc}
    -1 & 0 & 0  \\
     0 & -1 & 0 \\
     0 & 0 & 1
\end{array}\right),  \nn
T_{\vec{a}_2}: && n\to n\left(\begin{array}{ccc}
    -1 & 0 & 0  \\
     0 & 1 & 0 \\
     0 & 0 & -1
\end{array}\right),  \nn
T_{-\vec{a}_1-\vec{a}_2}: && n\to n\left(\begin{array}{ccc}
    1 & 0 & 0  \\
     0 & -1 & 0 \\
     0 & 0 & -1
\end{array}\right).
\eeq
The $C_6$ rotation acts as
\be
\label{eq: KagomeC6action}
C_6: n\to \left(\begin{array}{ccccc}
  -I_{3} & 0 & 0 & 0 & 0 \\
  0 & 1 & 0 & 0 & 0 \\
  0 & 0 & 1 & 0 & 0 \\
  0 & 0 & 0 & -1 & 0 \\
  0 & 0 & 0 & 0 & 1 \end{array}\right) n \left(\begin{array}{ccc}
    0 & 1 & 0  \\
     0 & 0 & 1 \\
     1 & 0 & 0
\end{array}\right).
\ee
Finally, the $\mathcal{R}_y$ reflection acts the same way as the triangular case:
\be
\mathcal{R}_y: n\to \left(\begin{array}{ccccc}
  I_{3} & 0 & 0 & 0 & 0 \\
  0 & 1 & 0 & 0 & 0 \\
  0 & 0 & 1 & 0 & 0 \\
  0 & 0 & 0 & -1 & 0 \\
  0 & 0 & 0 & 0 & -1 \end{array}\right) n\left(\begin{array}{ccc}
  0 & 1 & 0  \\
  1 & 0 & 0  \\
  0 & 0 & 1  \\
 \end{array}\right).
\ee

We can now go through the anomaly calculation in a similar way as our earlier examples. We shall omit the intermediate steps here and simply state that the final result is indeed the required anomaly in Eq.~\eqref{eq: KagomeLSM}. 

Similar to the triangular case, we can again interpret this SL$^{(7)}$ theory as a result of competition between non-coplanar magnetic and VBS orders, now both at momenta $\vec{M}$ from Eq.~\eqref{eq: KagomeTranslationaction}. The non-coplanar magnetic order also has a $C_6$ angular momentum $l=1$. This is known as the cuboctahedral order (more precisely, the cuboc1 order in the language of Ref. \cite{Messio2011}). This order has been numerically found for the $J_1-J_2-J_3$ Heisenberg model in certain regimes \cite{Gong2015}. Our results motivate further exploration of the phase diagram near the cuboctahedral order, and see if the SL$^{(7)}$ state could be realized. Again, just like the realization on the triangular lattice, a smoking-gun signature of the SL$^{(7)}$ state is that the non-coplanar magnetic and VBS order parameters are critical and have identical critical exponent, due to the emergent $SO(7)\times SO(3)$ symmetry.

\section{Discussion} \label{sec: discussion}

In this paper, based on a nonlinear sigma model defined on a Stiefel manifold, $SO(N)/SO(4)$, supplemented with a Wess-Zumino-Witten (WZW) term, we have put forward the theory of Stiefel liquids, which are a family of critical quantum liquids that have many extraordinary properties. For example, they have a large emergent symmetry, a cascade structure, and nontrivial quantum anomalies. Some of these Stiefel liquids are argued to be dual to the well known deconfined quantum critical point and $U(1)$ Dirac spin liquid, and others are conjectured to be non-Lagrangian, \ie its corresponding RG fixed point cannot be described by any weakly-coupled mean-field theory at any scale, which, in particular, means that these states are beyond parton (mean-field) construction widely used in the study of exotic quantum phases and phase transitions.

We make some comments on why the ``non-Lagrangian'' conjecture for the $N\geqslant7$ Stiefel liquids may be reasonable. The most obvious gauge theory candidates for such WZW fixed points are some kinds of QCD$_3$ with gapless Dirac fermions coupled to some gauge fields. However, as we review in Appendix~\ref{app: GrassmannianWZW}, typical QCD$_3$ correspond to WZW theories defined on Grassmannian manifolds $G(2N)/G(N)\times G(N)$, where $G$ can be $U, SU, USp, SO$. In fact, this is also why Stiefel liquids with $N=5,6$ do have gauge theory descriptions, since the corresponding Stiefel manifolds in these two cases also happen to be some kinds of Grassmannian: $SO(5)/SO(4)=USp(4)/(USp(2)\times USp(2))$ and $SO(6)/SO(4)=SU(4)/(SU(2)\times SU(2))$. For $SO(N\geqslant7)/SO(4)$, we do not have such identification, so the corresponding Stiefel liquids are not captured by some simple QCD$_3$. We also note the special role played by the $SO(N-4)$ symmetry in Stiefel liquids. For $N=6$, this $SO(2)$ symmetry is realized in the gauge theory as the flux conservation symmetry of the dynamical $U(1)$ gauge field. It is not clear how this $SO(2)$ flux conservation symmetry could be generalized to higher $SO(N-4)$ in different gauge theories. Besides these constraints from symmetries, the intricate anomaly structures of the Stiefel liquids discussed in Sec. \ref{Sec: Anomaly Analysis} also impose further nontrivial constraints on its possible renormalizable-Lagrangian description. One possibility is that the $N\geqslant 7$ Stiefel liquids can be realized by gauge theories with significantly lower symmetries in the UV Lagrangians, and the full IR symmetries emerge through some nontrivial dynamics. This scenario will be hard to rule out, and if true, it will likely shed new light on the dynamics of $(2+1)$-d gauge theories.

We mention that the fixed points of some quantum loop models were also proposed to be non-Lagrangian \cite{Freedman2004, Freedman2004a, Dai2019}. The nature of such loop quantum criticality appears to be very different from those studied in this paper. For example, they are not Lorentz invariant.

Note that although the most commonly used parton approach uses canonical bosonic/fermionic partons to construct a (non-interacting) mean field, there are also some parton approaches that use other types of partonic DOFs, in particular, ones that are subject to some constraints and are strongly fluctuating at all scales (see, \eg Ref. \cite{Xu2010}). One may wonder if the latter constrained-parton-based approach can lead to a construction to the non-Lagrangian Stiefel liquids. Although it is not ruled out, we believe this approach is difficult, and even if it can be achieved, novel ideas are still needed to make it work. This is because: i) As far as we know, all states constructed with constrained partons ultimately fall into the paradigm of mean field plus weak fluctuations, but these (conjectured) non-Lagrangian states are beyond this paradigm. ii) More technically, in such an approach, one often (if not always) encounters Dirac fermions coupled to sigma fields (\ie bosonic fields subject to some constraints), but these sigma fields live in certain Grassmannian manifold, which is difficult to be converted into a Stiefel manifold relevant here, unless some nontrivial mathematical facts can be used, in a way similar to the case of Stiefel liquids with $N=6$.

In Sec. \ref{sec: N>6}, we have proposed an approach based on the hypothesis of emergibility, which is complementary to the conventional parton approach, to study quantum phases and phase transitions. This approach is benchmarked with some known examples, and then applied to predict that spin-1/2 triangular and Kagome lattices can host one of the non-Lagrangian Stiefel liquids. This approach can also predict some detailed properties of such lattice realizations of these Stiefel liquids, such as the quantum numbers of various critical order parameters with identical scaling exponents.

It may be useful to comment on the parton approach and compare it with our anomaly-based approach. The parton approach is explicit, concrete, and relatively easy to manipulate, and it has led to tremendous success and deep insights in the study of strongly-correlated quantum matter. However, this parton approach also has some drawbacks. More specifically, there are two common treatments of a parton construction, leading to a projected wave function and an effective gauge theory, respectively. The wave-function-based treatment starts with an enlarged Hilbert space of the partons, and performs a projection of a valid ground state of these partons, in order to return to the physical Hilbert space. Although the projected wave function is indeed in the physical Hilbert space, a priori, it may not describe any {\it ground state of a local Hamiltonian} in the physical Hilbert space, and it is unclear what universal properties it exhibits -- these have to be checked case by case, say, using numerical calculations. On the other hand, from the effective gauge theory, it is more analytically tractable to deduce what ground state it describes and what universal properties it possesses. However, a priori, it is unclear whether such an effective gauge theory can really emerge from the physical Hilbert space, although it is often {\it assumed} so without any rigorous analytically controlled justification{\footnote{We note that sometimes such an effective gauge theory is phrased in terms of a lattice gauge theory, where the partons hop on the lattice sites and are coupled to the gauge fields living on the lattice links. In such an interpretation, the physical Hilbert space is often manifestly just a gauge-fluctuation-free subspace of the Hilbert space of the lattice gauge theory, where certain {\it hard} gauge constraints are imposed.}}. Our anomaly-based approach may be more abstract by the contemporary standards, but it is closer to the intrinsic characterization of the universal many-body physics discussed in the introduction, and it directly hinges on the emergibility: states that fail to satisfy the anomaly-matching condition are  necessarily not emergible, although at this stage we cannot rigorously prove that states that do satisfy the anomaly-matching condition must be emergible. We also note that the parton approach is in fact also essentially an attempt to verify anomaly-matching between the IR theory and the microscopic setup, but through an explicit mean-field-like construction.

Let us also comment on the significance of non-Lagrangian, or {\it intrinsically} non-renormalizable, theories specifically in condensed matter physics. Since most condensed matter systems do have some natural UV cutoffs, the concept of renormalizability should not play a fundamental role in condensed matter physics. This means that we should have a theoretical framework that can handle both Lagrangian and non-Lagrangian theories -- there should be no intrinsic difference between the two. However, the fact is that not only we do not have many tools to analyze non-Lagrangian theories, we did not even have many serious examples of non-Lagrangian theory prior to this work. From this perspective, what is really surprising is how difficult it was to find such non-Lagrangian examples -- this is perhaps rooted in our heavy reliance on perturbative quantum field theories in the past. The examples found here may also require and inspire us to develop new tools to analyze strongly correlated systems in more intrinsic manners. One example is the problem of ``emergibility", which was the focus of the later half of our paper: the intrinsic non-renormalizability, or lack of mean-field construction, forced us to further develop the anomaly-matching approach which may become useful for future works on strongly correlated systems in general. Therefore, even though renormalizability \textit{per se} is not of fundamental importance in condensed matter physics, being able to go beyond renormalizable theories is. Our work on (likely) non-Lagrangian quantum criticality represents a step toward this ambitious goal.

We finish this paper with some interesting open questions that we leave for future work.

\begin{enumerate}

\item Although we have given a derivation of the effective theory of some of the Stiefel liquids based on the gauge theoretic descriptions of Dirac spin liquids, it is desirable to give a more explicit derivation. For example, it is desirable to explicitly derive the expression of Skyrmion current in Eq. \eqref{eq: Skyrmiongaugecoupling} in terms of the $\mc{P}$ field. Also, it is nice to show how the WZW term on the Grassmannian manifold reduces to that on the Stiefel manifold.

\item The quantum anomalies of the Stiefel liquids labeled with an even $N$ have not been fully pinned down. It is interesting to finish this anomaly analysis. For the purpose of studying Stiefel liquids, using their cascade structure may be sufficient to fully determine their anomalies, just as what we have done in this paper. However, we note that deriving the full quantum anomalies directly based on the WZW action is also an intriguing theoretical challenge.

\item Although we have argued that the Stiefel liquids labeled by $N\geqslant 6$ can flow to a conformally invariant fixed point under RG, we have not been able to establish this with a controlled analysis. It is well motivated to find ways to systematically study the IR dynamics of the Stiefel liquids. For example, it is natural to ask if the Stiefel liquid fixed points predicted in this work can be found in numerical conformal bootstrap. The large emergent symmetry group $SO(N)\times SO(N-4)$ and the likely irrelevance of singlet operators may provide some reasonable starting point for such investigation. Also, given that the Stiefel liquids are described by a matrix model, it is interesting to explore if they have any holographic dual. 

\item The conjecture that the Stiefel liquids with $N>6$ are non-Lagrangian has not been proved. It is important to prove or falsify it. We expect that the method to prove or falsify it will necessarily bring in useful general insights. Also, even if it is falsified, these Stiefel liquids are still interesting. Relatedly, although a wave function is not strictly necessary for the intrinsic characterization of the universal physics of a many-body system, it may be interesting and useful to find a wave function for these Stiefel liquids.

\item It is natural to ask whether similar WZW models on target manifolds other than the Stiefel can lead to interesting fixed points. The most natural manifolds are the Grassmannians, which have been studied~\cite{Bi2016} and related to various gauge theories~\cite{Komargodski2018} (see also Appendix~\ref{app: GrassmannianWZW}). It will be interesting to either better understand the Grassmannian theories, or to contemplate on theories based on other types of manifolds.

\item The hypothesis of emergibility has not been proved rigorously, and it is important to prove or falsify it. If it can be proved, or at least further justified, it can be applied to other cases to study other exotic quantum phases and phase transitions. We expect this approach to give rise to many more interesting results and novel insights in the future. If it will be falsified, it is still very useful to find the correct general rules that govern the emergibility of a given low-energy theory for a system.

\item Perhaps the most important question is whether some of the critical Stiefel liquids can be realized in real materials. We have suggested that the $N=7$ Stiefel liquid may arise near certain non-coplanar magnetic orders. Numerically those non-coplanar orders can arise in some relatively simple lattice spin models \cite{Gong2015,Yu2020,Gong2017}. It will be fascinating to explore further in the phase diagram of those systems and see if the Stiefel liquid can indeed be found. This may provide valuable guidance towards ultimate experimental realizations.

\item We have seen that the theories of the deconfined quantum critical point and the $U(1)$ Dirac spin liquid can be formulated in terms of local DOFs. It may be interesting to see if other exotic non-quasiparticle critical quantum liquids can also be formulated in a similar way, and such a new formulation may bring in new insights. For example, can the theory of a Fermi surface coupled to a $U(1)$ gauge field in $(2+1)$-d be formulated purely in terms of local DOFs? It is likely that such a formulation will explicitly involve the infinitely many collective excitations around the Fermi surface, and the ideas from Ref. \cite{Mross2011} may be useful.

\end{enumerate}

\begin{acknowledgments}

We thank Maissam Barkeshli, Zhen Bi, Vladimir Calvera, Davide Gaiotto, Meng Guo, Chao-Ming Jian, Theo Johnson-Freyd, Steve Kivelson, John McGreevy, Subir Sachdev, Cenke Xu, Weicheng Ye and Yi-Zhuang You for illuminating discussions. Research at Perimeter Institute is supported in part by the Government of Canada through the Department of Innovation, Science and Industry Canada and by the Province of Ontario through the Ministry of Colleges and Universities.

\end{acknowledgments}

\bibliography{Stiefel.bib}

\onecolumngrid
\appendix

\section{More on the proposed WZW action} \label{app: more on WZW}

In the main text a WZW action for the $(2+1)$-d system of our interest is proposed in Eq. \eqref{eq: WZW begins}. In this appendix we present more details on its mathematical aspects.

For any even integer $d\geqslant 0$, we can define a WZW term for a $d+1$ (spacetime) dimensional system on a Stiefel manifold $V_{N, N-(d+2)}\equiv SO(N)/SO(d+2)$, where $N\geqslant d+3$. An element on this Stiefel manifold can be parameterized by an $N$-by-$\left(N-(d+2)\right)$ matrix, $n$, such that its columns are orthonormal, \ie $n^Tn=I_{N-(d+2)}$. The corresponding WZW action is
\beq \label{eq: WZW general}
S_\WZW^{(N, d)}[n]=\frac{2\pi}{\Omega_{d+2}}\int_0^1 du\int d^{d+1}x\sum_{k_1,k_2,\cdots,k_{\frac{d+2}{2}}=1}^{N-(d+2)}\det(\tilde n_{(k_1, k_2, \cdots, k_{\frac{d+2}{2}})})
\eeq
where $\Omega_{d+2}$ is the volume of $S^{d+2}$ with unit radius, and the $N$-by-$N$ matrix $\tilde n_{(k_1, k_2, \cdots, k_{\frac{d+2}{2}})}$ is given by
\beq \label{eq: n tilde app}
\tilde n_{(k_1, k_2, \cdots, k_{\frac{d+2}{2}})}=(n, \partial_{x_1}n_{k_1}, \partial_{x_2}n_{k_1}, \partial_{x_3}n_{k_2}, \partial_{x_4}n_{k_2},\cdots, \partial_{x_{d+1}}n_{k_{\frac{d+2}{2}}}, \partial_un_{k_{\frac{d+2}{2}}})
\eeq
where $x_{1,2,\cdots, x_{d+1}}$ is the coordinate of the physical spacetime, $n_{k_i}$ is the $k_i$th column of $n$ (note that the repeated subscripts $k_i$'s are not summed over in the right hand side of \eqref{eq: n tilde app}). That is, the first $N-(d+2)$ columns of $\tilde n_{(k_1, k_2, \cdots, k_{\frac{d+2}{2}})}$ is just $n$, and its last $d+2$ columns are derivatives of the columns of $n$ arranged in the above way. More explicitly,
\beq
\begin{split}
\det&(\tilde n_{(k_1, k_2, \cdots, k_{\frac{d+2}{2}})})
=\frac{1}{(N-(d+2))!}\epsilon^{i_1i_2\cdots i_{N-(d+2)}}\epsilon^{j_1j_2\cdots j_N}n_{j_1i_1}n_{j_2i_2}\cdots n_{j_{N-(d+2)}i_{N-(d+2)}}\\
&
\cdot\partial_{x_1}n_{j_{N-(d+2)+1}k_1}\partial_{x_2}n_{j_{N-(d+2)+2}k_1}\partial_{x_3}n_{j_{N-(d+2)+3}k_2}\partial_{x_4}n_{j_{N-(d+2)+4}k_2}\cdots \partial_{x_{d+1}}n_{j_{N-1}k_{\frac{d+2}{2}}}\partial_{u}n_{j_Nk_{\frac{d+2}{2}}}
\end{split}
\eeq
where the $\epsilon$'s are the fully anti-symmetric symbols with rank $N-(d+2)$ and $N$, respectively. 

It is straightforward to see that the WZW term presented in the main text is precisely the special case of the above one with $d=2$, and in the main text we denote $V_{N, N-4}$ by $V_N$. Also, it is clear that such a term can be defined only if $d$ is even, and it is interesting to compare this observation with the fact that the form of the homotopy groups of the Stiefel manifold $V_{N, N-(d+2)}$ is qualitatively different for even $d$ and odd $d$. That is, the first nontrivial homotopy group of $V_{N, N-(d+2)}$ is
\beq
\pi_{d+2}V_{N, N-(d+2)}=
\left\{
\begin{array}{lr}
\mathbb{Z}, & d+2\ {\rm even\ or\ } N=d+3\\
\mathbb{Z}_2, & d+2\ {\rm odd\ and\ } N>d+3
\end{array}
\right.
\eeq

The validity of the above WZW term requires it to be the integral of the pullback of a closed $(d+2)$-form on $V_{N, N-(d+2)}$. The $(d+2)$-form on $V_{N, N-(d+2)}$ that this WZW term is associated with is
\beq
\begin{split}
\omega
&=\frac{1}{(N-(d+2))!}\epsilon^{i_1i_2\cdots i_{N-(d+2)}}\epsilon^{j_1j_2\cdots j_N}n_{j_1i_1}n_{j_2i_2}\cdots n_{j_{N-(d+2)}i_{N-(d+2)}}\\
&\quad
\cdot dn_{j_{N-(d+2)+1}k_1}\wedge dn_{j_{N-(d+2)+2}k_1}\wedge dn_{j_{N-(d+2)+3}k_2}\wedge dn_{j_{N-(d+2)+4}k_2}\cdots \wedge dn_{j_{N-1}k_{\frac{d+2}{2}}}\wedge dn_{j_Nk_{\frac{d+2}{2}}}
\end{split}
\eeq
where the repeated subscripts $k_i$'s are summed over. It can be shown that the above form is indeed closed {\footnote{We thank Vladimir Calvera for giving a mathematical proof to the closedness.}}.

It remains to fix the normalization factor in front of the WZW term. We start with two observations: 
\begin{enumerate}
\item For $N=d+3$ Eq.~\eqref{eq: WZW general} is the familiar WZW term on $S^{d+2}$ with the correct normalization factor. 
\item For $N>d+3$, if we fix the first column of $n$ to a constant, say $n_1=(1,0,0...)^T$, the proposed WZW term for $V_{N,N-(d+2)}$ becomes that for $V_{N-1,(N-1)-(d+2)}$.
\end{enumerate}
Mathematically, fixing the first column of $n$ describes an inclusion map $i: V_{N-1,N-d-3}\to V_{N,N-d-2}$:
\beq
n_{(N-1)\times (N-d-3)}\to \left(\begin{array}{cc}
1 & 0 \\
0 & n_{(N-1)\times (N-d-3)}\end{array}\right).
\eeq
This map then induces a homomorphism between the homotopy groups $\pi_{d+2}(V_{N-1,N-d-3})\to \pi_{d+2}(V_{N,N-d-2})$. Based on the two observations made above, the normalization factor in Eq.~\eqref{eq: WZW general} will be justified if the homomorphism $\pi_{d+2}(V_{N-1,N-d-3})\to \pi_{d+2}(V_{N,N-d-2})$ induced by $i$ is an isomorphism. The last statement can be proved using the long exact sequence of homotopy groups associated with the fibration
\beq
V_{N-1,N-d-3}\to V_{N,N-d-2}\to S^{N-1}.
\eeq

A pictorial consequence of the above argument is that a ``generator" of $\pi_{d+2}(V_{N, N-(d+2)})$ is given by fixing the entries of the first $N-(d+2)-1$ columns of $n$ to be $n_{ji}=\delta_{ji}$, and letting the last column, which now lives on $S^{d+2}$ with unit radius, wrap around the $S^{d+2}$ once.

Some properties of the WZW term for the case with $d=2$ are discussed in Sec. \ref{subsec: SL begins}, and with minor modifications many of them also apply appropriately to the case with a general even $d$.

\section{A gauge theory description of SL$^{(N=5, k)}$} \label{app: gauge theory of (5, k)}

In this appendix, we show that SL$^{(N=5, k)}$ has a natural gauge-theoretic description, \ie a QCD$_3$ theory with $N_f=2$ flavor of fermions interacting with a $USp(2k)$ gauge field. 
A special case of $k=1$, i.e., SL$^{(5)}$ or the DQCP, was already discussed in Ref.~\cite{Wang2017}.

The global symmetry of the $N_f=2$ $USp(2k)$ QCD$_3$ theory is $USp(2N_f)=USp(4)\cong SO(5)$, which is identical to SL$^{(5, k)}$.
The fermions are in the fundamental representation of the $USp(2k)$ gauge group and $USp(4)$ global symmetry (\ie spinor representation of $SO(5)$).
The gauge invariant operators with lowest scaling dimensions shall be the fermion mass terms, which are $SO(5)$ vector and $SO(5)$ singlet.
The $SO(5)$ vector mass can be identified as the order parameter field $n$ of the NLSM.
To see the relation more explicitly, we can couple the $SO(5)$ vector mass to a $SO(5)$ bosonic vector $n_i$, and then integrate out fermions. This will yield an $SO(5)$ NLSM, and the original $USp(2k)$ gauge field is expected to just confine since it does not couple to any low-energy degrees of freedom.
Moreover, due to the Abanov-Wiegmann mechanism \cite{Abanov2000}, integrating out fermions also yields a level-$k$ WZW term of the $SO(5)$ vector $n_i$.
The level-$k$ comes from the fact that there are $k$ copies of fermions as the gauge group is $USp(2k)$. 
Therefore, we have derived that the $N_f=2$ $USp(2k)$ QCD$_3$ theory is dual to SL$^{(N=5, k)}$.

From the gauge-theoretic description of SL$^{(N=5,k)}$, we can gain some intuition about the properties of SL with a general $(N, k)$:
\begin{enumerate}

\item Stability. It is clear that the larger $k$ is, it is more likely that the QCD$_3$ theory will confine.
Naturally, we expect that this feature will also hold for $N>5$: for a given $N$ there exists a critical $k_c$, such that the SL can flow into a critical phase when $k\leqslant k_c$.

\item Neighboring topological order. By turning on a time-reversal-breaking singlet mass, the SL$^{(N=5, k)}$ will become the $USp(2k)_{\mp 1}^s$~\footnote{The superscript $s$ refers to the fact that the gauge field is a spin gauge field.} TQFT, which is dual to the $USp(2)_{\pm k}=SU(2)_{\pm k}$ TQFT.
$SU(2)_1$ is just the semion topological order discussed in the main text.
In Sec. \ref{subsec: semion TO}, we have argued that SL$^{(N, 1)}$ with a general $N\geqslant 5$ can flow to the $SU(2)_{\pm 1}$ TQFT under a time-reversal-breaking deformation.
So it is possible that the SL$^{(N>5, k)}$ will also flow to the $SU(2)_{\pm k}$ TQFT under appropriate time-reversal-breaking perturbation.

\item Higgs descendent. As shown in Sec. IV B of Ref. \cite{Wang2017}, for the case with $k=1$, adding a flavor-singlet Higgs field can break the gauge structure to $U(k)$ with $k=1$, and the resulting theory is equivalent to a QED$_3$ coupled to 4 gapless Dirac fermions, with one of the six monopole operators added to the Lagrangian, \ie schematically we have $USp(2)+{\rm Higgs}=U(1)+{\rm monopole}$. This is how the DQCP is related to the $U(1)$ DSL. The argument there can actually be generalized to any $k$ to show that $USp(2k)+{\rm Higgs}=U(k)+{\rm monopole}$. This is nicely compatible with the cascade structure of the SLs, and our results that SL$^{(5, k)}$ and SL$^{(6, k)}$ can also be described by $USp(2k)$ gauge theory and $U(k)$ gauge theory, respectively.

\item Microscopic realization. For a spin-$k/2$ system, there is a natural parton construction for the SL$^{(N=5, k)}$.

We first fractionalize spin operators into partons,
\begin{equation}\label{eq: USpparton}
S^i = -\frac{1}{4} \textrm{Tr}(X^\dag X \sigma^i),
\end{equation}
with $X$ being a $2k\times 2$ matrix,
\begin{equation}
X=\left( \begin{matrix}
\psi_1^\dag  & \cdots &\psi^\dag_{k}  &\psi^\dag_{k+1} & \cdots &\psi^\dag_{2k} \\ 
\psi_{k+1}  & \cdots & \psi_{2k}  & -\psi_{1} & \cdots & -\psi_{k}
\end{matrix} \right)^T.
\end{equation}
Here $\psi_i^\dag$ is the fermion creation operator. 
$X$ satisfies the reality condition $ X^* = \Omega^c X \Omega^s$
with 
\begin{equation}
\Omega^c = \left( \begin{matrix}
0 & I_{k} \\
-I_{k} & 0
\end{matrix}
\right), \quad  \Omega^S = \left( \begin{matrix}
0 & 1 \\
-1 & 0
\end{matrix}
\right).
\end{equation}
Here $I_k$ is a $k\times k$ identity matrix.
So the spin operators can be written as $S^i = -\frac{1}{4} \textrm{Tr}(\Omega^s X^T \Omega^c X \sigma^i)$.
It is apparent that this parton decomposition has a $USp(2k)$ gauge invariance, namely, the spin operators are invariant under a $USp(2k)$ (left) rotation of $X$, $RX$, as $R^T \Omega^c R = \Omega^c$.
On the other hand, the $SO(3)$ spin rotation acts as the $USp(2)\cong SU(2)$ (right) rotation of $X$. 
The local contraint of the parton construction is,
\begin{equation}
X \Omega^s X^T = -\Omega^c,
\end{equation}
or equivalently, $\psi_i^\dag \psi_i = \psi_{i+k}^\dag \psi_{i+k}$ for $i=1,\cdots,k$.
One can further show that the spin operators defined in Eq.~\eqref{eq: USpparton} satisfy $[S^i, S^j]=i\varepsilon^{ijk}S^k$ and $\sum_i (S^i)^2=\frac{k}{2}\left(\frac{k}{2}+1\right)$.

Therefore, the above parton construction has an emergent $USp(2k)$ gauge structure, and the fermions $(\psi_1, \cdots, \psi_k)$ form a $USp(2k)$ fundamental.
Putting  $\psi_i$ fermions into a band structure with two Dirac cones, we will get the $N_f=2$ $USp(2k)$ QCD$_3$, or equivalently, SL$^{(5, k)}$. 

So it is natural to look for SL$^{(5, k)}$ in a spin-$k/2$ system. Furthermore, as proposed in Sec. \ref{sec: N=6: DSL}, SL$^{(6, k)}$ is dual to the $N_f=2$ $U(k)$ QCD$_3$, so it is also possible that SL$^{(6, k)}$ can emerge in a spin-$k/2$ system. This is discussed in detail recently \cite{Calvera2020}. These observations lead us to further conjecturing that it is also true SL$^{(N>6, k)}$ can also emerge in a spin-$k/2$ system. Indeed, in Sec. \ref{subsec: N=7 emergibility}, we have argued that SL$^{(7)}$ can emerge in a spin-1/2 system.

\end{enumerate}

\section{Full quantum anomaly of the DQCP} \label{app: anomaly of DQCP}

The quantum anomaly of the DQCP, or equivalently, SL$^{(5)}$, was partially analyzed in Ref. \cite{Wang2017}, and an anomaly associated with the $SO(5)$ symmetry was found, which is described by a $(3+1)$-d topological response function, $i\pi\int_Mw_4^{SO(5)}$, where $M$ is the closed manifold in which the $(3+1)$-d bulk corresponding to the DQCP lives, and $w_4^{SO(5)}$ is the fourth Stiefel-Whitney (SW) class of the $SO(5)$ gauge bundle that couples to this bulk. 

Besides the $SO(5)$ symmetry, the DQCP also enjoys a time reversal symmetry, $\mc{T}$, which also contributes to the anomaly. To understand the full anomaly associated with both the $SO(5)$ and $\mc{T}$ symmetries, it is useful to enlarge the $SO(5)$ symmetry to $O(5)$ by including the improper $Z_2$ rotation. Then the action of the time reversal symmetry is a combination of this improper $Z_2$ rotation and a flip of the time coordinate. In this appendix, we show that the full anomaly of the DQCP is described by a $(3+1)$-d topological response function
\beq \label{eq: anomaly DQCP app}
S=i\pi\int_Mw_4^{O(5)}
\eeq
with a constraint $w_1^{O(5)}=w_1^{TM}\ ({\rm mod\ }2)$, where $TM$ denotes the tangent bundle of $M$. This constraint simply indicates that the improper $Z_2$ rotation of the $O(5)$ symmetry is accompanied with a flip of the time coordinate. For notational brevity, in the following we will suppress the superscript ``$TM$" of a SW class of $TM$.

To obtain the above result, let us look at the most general form that the anomaly can take:
\beq \label{eq: general anomaly for DQCP app}
S=i\pi\int_M\left[a_1w_4^{O(5)}+a_2[w_2^{O(5)}]^2+a_3w_1^2w_2^{O(5)}+a_4w_1^4+a_5w_2^2\right]
\eeq
where we have used the (mod 2) relations $w_1^{O(5)}=w_1$, $w_1w_3=w_1^4+w_2^2+w_4=0$, $w_1w_2=0$, $[w_2^{O(5)}]^2=(w_2+w_1^2)w_2^{O(5)}$, $\sq^1(w_3^{O(5)})=w_1w_3^{O(5)}$, $\sq^1\cdot\sq^1(w_2^{O(5)})=0$, and $w_3^{O(5)}=\sq^1(w_2^{O(5)})+w_1^{O(5)}w_2^{O(5)}$ to remove some terms. 

The above topological response function must satisfy the following known properties of the DQCP \cite{Vishwanath2012, Kapustin2014,Wang2017, Bi2013}, which help us to deduce the values of the $a$'s unambiguously:

\begin{enumerate}
    
    \item As mentioned above, if only the $SO(5)$ symmetry is considered, the anomaly is described by $i\pi\int_Mw_4^{SO(5)}$. 
    
    In this case, $w_4^{O(5)}=w_4^{SO(5)}$, $w_1=0$, and $w_2^{O(5)}=w_2^{SO(5)}$. So Eq. \eqref{eq: general anomaly for DQCP app} becomes $S=i\pi\int_M\left[a_1w_4^{SO(5)}+a_2[w_2^{SO(5)}]^2+a_5w_2^2\right]$, which implies that $a_1=1$ and $a_2=a_5=0$.
    
    \item Ignoring the $SO(5)$ symmetry and implementing $\mc{T}$ by $n\rightarrow -n$, there is an anomaly associated with $\mc{T}$, described by $i\pi\int_Mw_1^4$. The corresponding bulk is known as $eTmT$ \cite{Wang2013}.
    
    In this case, $w_4^{O(5)}=w_1^4$ and $w_2^{O(5)}=0$. So Eq. \eqref{eq: general anomaly for DQCP app} becomes $S=i\pi\int_M(1+a_4)w_1^4$, which implies that $a_4=0$.
    
    \item Suppose the full symmetry is broken to $SO(2)\times\mc{T}$, where the $SO(2)$ rotates the first 2 components of $n$ and $\mc{T}$ flips its last 3 components, the anomaly is described by $S_3=i\pi\int_Mw_1^2w_2^{SO(2)}$. The corresponding bulk is known as $eCmT$ \cite{Wang2013,Metlitski2013}.
    
    In this case, $w_4^{O(5)}=w_1^2w_2^{SO(2)}$ and $w_2^{O(5)}=w_2^{SO(2)}+w_1^2$. So Eq. \eqref{eq: general anomaly for DQCP app} becomes $S=i\pi\int_M\left[(1+a_3)w_1^2w_2^{SO(2)}+a_3w_1^4\right]$, which implies that $a_3=0$.
    
\end{enumerate}

In summary, the above three conditions imply that $a_1=1$ and $a_2=a_3=a_4=a_5=0$. So the full anomaly of the DQCP is described by Eq. \eqref{eq: anomaly DQCP app}.

\section{WZW models on the Grassmannian manifold $\frac{G(2N)}{G(N)\times G(N)}$}
\label{app: GrassmannianWZW}

The Grassmannian manifolds $\frac{G(2N)}{G(N)\times G(N)}$ (with $G=U, SU, SO, USp$) also have $\pi_4(\frac{G(2N)}{G(N)\times G(N)})=\mathbb Z$ and $\pi_3(\frac{G(2N)}{G(N)\times G(N)})=0$, so one can define $2+1$-d WZW models on these Grassmannians.
Using the argument in Sec.~\ref{subsec: IR fate} we can obtain a similar phase diagram (at least for large $N$). Namely,
as one tunes the coupling constant of NLSM, there are three fixed points: 1) an attractive fixed point of a spontaneous-symmetry-breaking phase, with the ground state manifold being $\frac{G(2N)}{G(N)\times G(N)}$; 2) a repulsive fixed point of order-disorder transition; 3) an attractive fixed point of a critical quantum liquid.
The last attractive fixed point is the Grassmannian version of our proposed SLs.
Interestingly, the Grassmannian WZW models have simple candidates of renormalizable Lagrangian descriptions, \ie Dirac fermions coupled to non-Abelian gauge fields~\cite{Komargodski2018}. 
More concretely, we have
\begin{enumerate}

\item The QCD$_3$ theory with $N_f=2N$ Dirac fermions coupled to a $SU(k)$ gauge field is a UV completion of the $\frac{U(2N)}{U(N)\times U(N)}$ NLSM model with a level $k$ WZW term.

\item The QCD$_3$ theory with $N_f=2N$ Majorana fermions coupled to a $SO(k)$ gauge field is a UV completion of the $\frac{SO(2N)}{SO(N)\times SO(N)}$ NLSM model with a level $k$ WZW term.

\item The QCD$_3$ theory with $N_f=2N$ Dirac fermions coupled to a $USp(2k)$ gauge field is a UV completion of the $\frac{USp(4N)}{USp(2N)\times USp(2N)}$ NLSM model with a level $k$ WZW term.

\item The QCD$_3$ theory with $N_f=2N$ Dirac fermions coupled to a $U(k)$ gauge field is a UV completion of the $\frac{SU(2N)}{SU(N)\times SU(N)}$ NLSM model with a level $k$ WZW term.\footnote{This one is relatively new and will be discussed more carefully elsewhere.}
\end{enumerate}

One may expect to see this correspondence by using the trick that appears several times in the paper.
We first couple the color-singlet fermion mass of the QCD$_3$ theory to a bosonic field that lives on the Grassmannian  $\frac{G(2N)}{G(N)\times G(N)}$, and then integrate out fermions.
This will give a NLSM of the bosonic field and will also generate a WZW term~\cite{Abanov2000}.
At last, the gauge field will confine by itself without doing anything to the Grassmannian $\frac{G(2N)}{G(N)\times G(N)}$ WZW models.
The level $k$ (instead of 1) comes from the color multiplicity of the gauge field.
It is worth mentioning if one couples the fermion mass to a field living on $\frac{G(2N)}{G(2N-M)\times G(M)}$ with $M\neq N$, integrating out Dirac fermions will generate a Chern-Simons term for the gauge field as well. 
In this case the gauge field will not confine, and one ends up with the $\frac{G(2N)}{G(2N-M)\times G(M)}$ WZW model coupled to a Chern-Simons gauge field.

We can also compare the global symmetry of the gauge theories and the Grassmannian WZW models. 
The simplest one is the last case, where both the gauge theory and the Grassmannian WZW models have an explicit $USp(4N)$ global symmetry.
For the first case, the $SU(k)$ gauge theory has an explicit $SU(2N)\times U(1)$ symmetry, and the $U(1)$ symmetry is carried by the baryon operator.
The $\frac{U(2N)}{U(N)\times U(N)}$ WZW model has an explicit $SU(2N)$ symmetry, which acts directly on the NLSM field.
The nontrivial part is the $U(1)$ symmetry, which comes from the topological property of the manifold $\pi_2(\frac{U(2N)}{U(N)\times U(N)})=\mathbb Z$.
The operator charged under this topological $U(1)$ symmetry is the Skyrmion creation operator, which is fermionic (or bosonic) if the level $k$ of the WZW term is odd (or even)~\cite{Komargodski2018}.
This nicely matches the statistics of the baryon operator of the $SU(k)$ gauge theory.
Similarly, for the second case one can also match the global symmetry by using $\pi_2(\frac{SO(2N)}{SO(N)\times SO(N)})=\mathbb Z_2$ for $N> 2$.

The identification of Grassmannian WZW models as the QCD$_3$ theories further corroborates the existence of SLs as critical quantum liquids, as discussed in Sec. \ref{subsec: IR fate}.

\section{Explicit homomorphism between the $su(4)$ and $so(6)$ generators} \label{app: homomorphism}

To be self-contained, in this appendix we present the explicit homomorphism between the $su(4)$ and $so(6)$ generators that is used in this paper.

Recall that the we write the $su(4)$ generators as $\sigma_{ab}\equiv\frac{1}{2}\sigma_a\otimes\sigma_b$, with $a,\ b=0,1,2,3$ but $a$ and $b$ not simultaneously zero. Here $\sigma_0=I_2$ and $\sigma_{1,2,3}$ are the standard Pauli matrices. The correspondence between the $su(4)$ and $so(6)$ generators are given as follows
\beq
\begin{split}
    &\sigma_{01}\leftrightarrow T_{16},\  \sigma_{02}\leftrightarrow T_{62},\  \sigma_{03}\leftrightarrow T_{12},\\
    \sigma_{10}\leftrightarrow T_{54},\  &\sigma_{11}\leftrightarrow T_{32},\  \sigma_{12}\leftrightarrow T_{31},\  \sigma_{13}\leftrightarrow T_{63},\\
    \sigma_{20}\leftrightarrow T_{53},\  &\sigma_{21}\leftrightarrow T_{24},\  \sigma_{22}\leftrightarrow T_{14},\  \sigma_{23}\leftrightarrow T_{46},\\
    \sigma_{30}\leftrightarrow T_{34},\  &\sigma_{31}\leftrightarrow T_{25},\  \sigma_{32}\leftrightarrow T_{51},\  \sigma_{33}\leftrightarrow T_{56}
\end{split}
\eeq
where $(T_{ij})_{kl}=i(\delta_{ik}\delta_{jl}-\delta_{il}\delta_{jk})$ is a 6-by-6 matrix, which generates rotations on the $(i, j)$-plane. One can explicitly check that the above correspondence is indeed a homomorphism between the $su(4)$ and $so(6)$ algebras.

\section{$I^{(N)}$ anomalies of SL$^{(N)}$} \label{app: N=6 anomaly physical}

In this appendix, we
present the details of the monopole-based approach to the anomalies associated with the $I^{(N)}$ symmetry, where $I^{(N)}=(SO(N)\times SO(N-4))/Z_2$ for even $N$ and $I^{(N)}=SO(N)\times SO(N-4)$ for odd $N$. Our strategy is to consider $(3+1)$-d bosonic $I^{(N)}$-SPTs that are also compatible with the discrete $\mc{C}$, $\mc{R}$ and $\mc{T}$ symmetries, gauge the $I^{(N)}$ symmetry, and use the statistics and quantum numbers of the fundamental $I^{(N)}$-monopoles of the resulting gauge theory to characterize the SPT we start with. This also gives us a characterization and classification of the $I^{(N)}$-anomalies of the $(2+1)$-d theories. Note that this is not a full classification of the anomalies associated with both $I^{(N)}$ and the discrete symmetries, and such a full classification is expected to be a more refined version of the one presented here.

The main results are as follows.
\begin{enumerate}
    
    \item If $N=2\ ({\rm mod\ }4)$, the SPTs or anomalies are classified into a $\mathbb{Z}_2^2\times\mathbb{Z}_4$ structure. The fundamental monopoles of the root states are given by \eqref{eq: 4k+2 roots}.
    
    \item If $N=0\ ({\rm mod\ }4)$, the SPTs or anomalies are classified into a $\mathbb{Z}_2^3\times\mathbb{Z}_4$ structure, \ie it has one more $\mathbb{Z}_2$ factor compared to the case with $N=2\ ({\rm mod\ }4)$.  In addition to fundamental monopoles of the types given in \eqref{eq: 4k+2 roots}, this additional $\mathbb{Z}_2$ factor corresponds to one more possible type of the fundamental monopole, given in \eqref{eq: root 4}.
    
    \item If $N$ is odd, the SPTs or anomalies are classified into a $Z_2^5$ structure. The fundamental monopoles of root states are listed in Table \ref{tab: SO(N) and SO(N-4) monopoles for odd N}.
    
    \item In all these cases, the statistics and quantum numbers of the $SO(N)$ and $SO(N-4)$ monopoles of the root states are derived, given by Table \ref{tab: SO(N) and SO(N-4) monopoles} for even $N$ and Table \ref{tab: SO(N) and SO(N-4) monopoles for odd N} for odd $N$.
    
    \item In all these cases, we identify anomalies corresponding to theories that are compatible with the cascade structure of the SLs, as discussed in Sec. \ref{subsec: cascade structure}. In particular, two conditions need to be satisfied:
    
    \begin{enumerate}
    
    \item If the symmetry is broken to $SO(5)$, we can consider the $SO(5)$ monopole of the resulting theory. An $SO(5)$ monopole breaks the $SO(5)$ symmetry to $SO(2)\times SO(3)$. For a SL, the $SO(5)$ monopole of the resulting theory should carry no charge under the $SO(2)$ but a spinor representation under $SO(3)$.
    
    \item If the symmetry of a SL is broken to $(SO(4)\times SO(N-4))/Z_2$, the resulting theory should have no anomaly.
    
    \end{enumerate}
    
    For even $N$, we find that only root 3 in \eqref{eq: 4k+2 roots} and its inverse satisfy both conditions. For odd $N$, there is a single anomaly class that satisfies both conditions, as discussed at the end of Appendix \ref{appsub: odd N}.
    
\end{enumerate}

\subsection{The case with an even $N$}

We start the discussion with the case with an even $N$, as in this case results not captured by Eq. \eqref{eq: Anomaly11} may arise.

As in the main text, we write the $SO(N)$ and $SO(N-4)$ gauge fields as $A^{SO(N)}=A_a^LT_a^L$ and $A^{SO(N-4)}=A_a^RT_a^R$, respectively, where $\{T^L\}$ and $\{T^L\}$ form the generators of $SO(N)$ and $SO(N-4)$, respectively. For even $N$, the field configuration of a fundamental monopole will be taken as
\beq \label{eq: fund monopole app}
A_{12}^{L}=A_{34}^{L}=A_{56}^{L}=\cdots A_{N-1, N}^{L}=A_{12}^{R}=A_{34}^{R}=A_{56}^{R}=\cdots A_{N-5, N-4}^{R}=\frac{A_{U(1)}}{2}
\eeq
where $A_{ij}^{L}$ ($A_{ij}^{R}$) is the gauge field corresponding to the generator associated with rotations on the $(i,j)$-plane of $SO(N)$ ($SO(N-4)$) symmetry, and $A_{U(1)}$ is the field configuration of a unit monopole in a $U(1)$ gauge theory, which can taken to be of the form in Ref. \cite{Wu1975}. Namely, this monopole is obtained by embedding many half-$U(1)$-monopoles into the maximal Abelian group of $I^{(N)}$. The configuration of such a monopole breaks the gauge symmetry from $I^{(N)}$ to $(SO(2)^{N-2})/Z_2$. So it is convenient to denote a general excitation in this $I^{(N)}$ gauge theory by the following {\em excitation matrix}:
\beq \label{eq: general excitation for even N}
\left(
\begin{array}{c}
\vec q\\
\vec m
\end{array}
\right)_s=
\left(
\begin{array}{cccc|cccc}
q_{12}^L & q_{34}^L & \cdots & q_{N-1,N}^L & q_{12}^R & q_{34}^R & \cdots & q_{N-5,N-4}^R\\
m_{12}^L & m_{34}^L & \cdots & m_{N-1,N}^L & m_{12}^R & m_{34}^R & \cdots & m_{N-5,N-4}^R
\end{array}
\right)_s
\eeq
where the first (second) row represents the electric (magnetic) charges of this excitation under $A_{ij}^{L,R}$, $s=0\ ({\rm mod\ }2)$ ($s=1\ ({\rm mod\ }2)$) represents that this excitation is a boson (fermion), and the vertical line separates the charges related to the original $SO(N)$ and $SO(N-4)$ subgroups of $I^{(N)}$. The above fundamental monopole has $\vec m=(\half, \half, \cdots, \half)$, and its $\vec q$ and $s$ will characterize the corresponding SPT. Because the statistics of any excitation can be unambiguously determined by its $\vec q$ and the statistics of the fundamental monopole, later we will sometimes suppress the subscript related to the statistics of this excitation. 

In such a theory, the structures of the possible excitations are constrained by the following conditions.
\begin{enumerate}
    
    \item The pure gauge charges are built up with bosons in the bifundamental representation of $I^{(N)}$. That is, if $\vec m=0$, then all entries of $\vec q$ are integers that add up to an even integer. An example of the elementary pure gauge charge has $\vec m=0$ and $\vec q=(1, 0, 0, \cdots, 0, 0, 1)$. 
    
    \item The Dirac quantization condition for two excitations $\left(\begin{array}{c}\vec q_1\\ \vec m_1\end{array}\right)$ and $\left(\begin{array}{c}\vec q_2\\ \vec m_2\end{array}\right)$: $\vec q_1\cdot\vec m_2-\vec q_2\cdot\vec m_1\in\mathbb{Z}$.
    
    \item If an excitation exists, its $\mc{C}$, $\mc{R}$ and $\mc{T}$ partners also exist. We take the actions of these discrete symmetries on the excitation given by \eqref{eq: general excitation for even N} to be 
    \beq
    \begin{split}
       &\mc{C}: \left(
        \begin{array}{c}
            \vec q\\
            \vec m
        \end{array}
        \right)_s\rightarrow
        \left(
            \begin{array}{cccc|cccc}
                -q_{12}^L & q_{34}^L & \cdots & q_{N-1,N}^L & -q_{12}^R & q_{34}^R & \cdots & q_{N-5,N-4}^R\\
                -m_{12}^L & m_{34}^L & \cdots & m_{N-1,N}^L & -m_{12}^R & m_{34}^R & \cdots & m_{N-5,N-4}^R
            \end{array}
        \right)_s\\
        &\mc{R}: \left(
        \begin{array}{c}
            \vec q\\
            \vec m
        \end{array}
        \right)_s\rightarrow
        \left(
            \begin{array}{cccc|cccc}
                -q_{12}^L & q_{34}^L & \cdots & q_{N-1,N}^L & q_{12}^R & q_{34}^R & \cdots & q_{N-5,N-4}^R\\
                m_{12}^L & -m_{34}^L & \cdots & -m_{N-1,N}^L & -m_{12}^R & -m_{34}^R & \cdots & -m_{N-5,N-4}^R
            \end{array}
        \right)_s\\
        &\mc{T}: \left(
        \begin{array}{c}
            \vec q\\
            \vec m
        \end{array}
        \right)_s\rightarrow
        \left(
            \begin{array}{cccc|cccc}
                q_{12}^L & q_{34}^L & \cdots & q_{N-1,N}^L & -q_{12}^R & q_{34}^R & \cdots & q_{N-5,N-4}^R\\
                -m_{12}^L & -m_{34}^L & \cdots & -m_{N-1,N}^L & m_{12}^R & -m_{34}^R & \cdots & -m_{N-5,N-4}^R
            \end{array}
        \right)_s
    \end{split}
    \eeq
    
    \item The remaining $(SO(2)^{N-2})/Z_2$ has a normalizer subgroup in $I^{(N)}$. If an excitation exists, its partners under the actions of the normalizer subgroup also exist.
    
\end{enumerate}
These are necessary conditions for a theory to be consistent, and we believe they are also sufficient.

The above conditions impose strong constraints on the possible $\vec q$ of a fundamental monopole. First, there is an element in the normalizer subgroup associated with the remaining $(SO(2)^{N-2})/Z_2$, whose action is to exchange the first two columns of the excitation matrix. Applying this operation to the fundamental monopole yields a normalizer-partner of it. The bound state of this normalizer-partner and the anti-particle of the fundamental monopole has all entries in the excitation matrix being $0$, except that the first two entries in the first row are $\pm(q_{12}^L-q_{34}^L)$. Because this is a pure gauge charge, $q_{12}^L=q_{34}^L\ ({\rm mod\ }1)$. Similarly, it is easy to see that the first $N/2$ entries in $\vec q$ of the fundamental monopoles are all equal mod 1, and the last $(N-4)/2$ entries in $\vec q$ are also all equal mod 1. Second, it is always possible to attach the fundamental monopole with some pure gauge charge, such that all its $\vec q$-entries are in the interval $(-1,1]$. These two observations imply that we can always write a fundamental monopole as $\left(
\begin{array}{cccc|cccc}
q^L & q^L & \cdots & q^L & q^R & q^R & \cdots & q^R\\
\half & \half & \cdots & \half & \half & \half & \cdots & \half
\end{array}
\right)_s$ if $N=2\ ({\rm mod\ }4)$, while if $N=0\ ({\rm mod\ } 4)$, besides this possibility, there is one more possible type: $\left(
\begin{array}{cccc|cccc}
q^L & q^L & \cdots & q^L & q^R & q^R & \cdots & q^R+1\\
\half & \half & \cdots & \half & \half & \half & \cdots & \half
\end{array}
\right)_s$. 

\subsubsection{The case with $N=2\ ({\rm mod\ }4)$}

Now let us focus on the case with $N=2\ ({\rm mod\ } 4)$. The case with $N=6$ needs some special treatment, so we defer the discussion on it for a moment. If $N>6$, the 4-particle bound state of the fundamental monopole and its $\mc{C}$, $\mc{R}$ and $\mc{T}$ parnters is a pure gauge charge with excitation matrix $\left(\begin{array}{ccccc|ccccc}0 & 4q^L & 4q^L & \cdots & 4q^L & 0 & 4q^R & 4q^R & \cdots & 4q^R\\0 & 0 & 0 & \cdots & 0 & 0 & 0 & 0 & \cdots & 0\end{array}\right)$.
This implies that $4q^{L,R}\in\mathbb{Z}$. The Dirac quantization condition on the fundamental monopole and its $\mc{T}$-partner implies $\frac{Nq^L+(N-8)q^R}{2}\in\mathbb{Z}$. So $q^L+q^R\in\mathbb{Z}$. Now it is straightforward to see that there are only two elementary possibilities of $(q^L, q^R)$, \ie $(q^L, q^R)=(0, 1)$ and $(q^L, q^R)=(\frac{1}{4}, -\frac{1}{4})$, and these possibilities are elementary in the sense that all other possibilities can be obtained from them by forming bound states of them and/or attaching pure gauge charges, \ie they can be taken as the monopoles of the root states. So far we have not considered the statistics of the fundamental monopole, and it can actually be either bosonic or fermionic. These results suggest a $\mathbb{Z}_2^2\times \mathbb{Z}_4$ classification of the fundamental monopoles, and the roots can be taken to be
\beq \label{eq: 4k+2 roots}
\begin{split}
    &{\rm root\ }1:
    \left(
    \begin{array}{cccc|cccc}
    0 & 0 & \cdots & 0 & 0 & 0 & \cdots & 0\\
    \half & \half & \cdots & \half & \half & \half & \cdots & \half
    \end{array}
    \right)_f\\
    &{\rm root \ }2: \left(
    \begin{array}{cccc|cccc}
    0 & 0 & \cdots & 0 & 0 & 0 & \cdots & 1\\
    \half & \half & \cdots & \half & \half & \half & \cdots & \half
    \end{array}
    \right)_b\\
    &{\rm root \ }3: \left(
    \begin{array}{cccc|cccc}
    \frac{1}{4} & \frac{1}{4} & \cdots & \frac{1}{4} & -\frac{1}{4} & -\frac{1}{4} & \cdots & -\frac{1}{4}\\
    \half & \half & \cdots & \half & \half & \half & \cdots & \half
    \end{array}
    \right)_b
\end{split}
\eeq
Notice that in writing root 2, we have attached pure gauge charge to it. One can check that these roots satisfy all 4 conditions listed at the beginning of this subsection.

The above results also apply to the case with $N=6$, but the argument needs to be slightly modified. For $N=6$, these 4 excitations exist: $\left(\begin{array}{ccc|c}-q^L & -q^L & -q^L & q^R\\ \half & \half& \half & -\half\end{array}\right)_s$, $\left(\begin{array}{ccc|c}q^L & -q^L & q^L & q^R\\-\half & \half& -\half & -\half\end{array}\right)_s$, $\left(\begin{array}{ccc|c}-q^L & -q^L & q^L & q^R\\-\half &-\half& \half & \half\end{array}\right)_s$, and $\left(\begin{array}{ccc|c}q^L & -q^L & -q^L & q^R\\ \half & -\half& -\half & \half\end{array}\right)$. The 4-particle bound state of these 4 excitations is $\left(\begin{array}{ccc|c}0 & -4q^L & 0 & 4q^R\\ 0 & 0& 0 & 0\end{array}\right)_b$, which means in this case we also have $4q^{L, R}\in\mathbb{Z}$. Furthermore, the Dirac quantization condition for a fundamental monopole and its $\mc{T}$-partner gives $-3q^L+q^R\in\mathbb{Z}$. These conditions still suggest a $Z_2^2\times Z_4$ classification, and the roots can still be taken as the ones in \eqref{eq: 4k+2 roots}.

It is useful to derive the structure of the $SO(N)$ and $SO(N-4)$ monopoles from these fundamental monopoles. An $SO(N)$ monopole can be viewed as the 2-particle bound state of the fundamental monopole and its $\mc{R}$ partner: $\left(\begin{array}{cccc|ccc}0 & 2q^L & \cdots & 2q^L & 2q^R & \cdots & 2q^R\\ 1 & 0 &\cdots & 0 & 0&\cdots&0\end{array}\right)_{q_L+q_R}$. The $SO(N-4)$ monopole can be viewed as the 2-particle bound state of the fundamental monopole and its $\mc{T}$ partner: $\left(\begin{array}{ccc|cccc}2q^L & \cdots & 2q^L & 0 & 2q^R & \cdots & 2q^R\\ 0 &\cdots & 0 & 1 & 0&\cdots&0\end{array}\right)_{q_L-3q_R}$. In physical terms, the properties of these monopoles are summarized in Table \ref{tab: SO(N) and SO(N-4) monopoles}.

\begin{table}[h]
    \setlength{\tabcolsep}{0.2cm}
    \renewcommand{\arraystretch}{1.4}
    \centering
    \begin{tabular}{c|cc} 
        \hline \hline
        & $SO(N)$ monopole & $SO(N-4)$ monopole\\
        \hline
        root 1 & (singlet, singlet, boson) & (singlet, singlet, boson) \\
        root 2 & (singlet, singlet, fermion) & (singlet, singlet, fermion)\\
        root 3 & (spinor, spinor, boson) & (spinor, spinor, fermion)\\
        root 4 with $N=0\ ({\rm mod\ }8)$ & (singlet, singlet, fermion) & (singlet, vector, boson)\\
        root 4 with $N=4\ ({\rm mod\ }8)$ & (singlet, singlet, boson) & (singlet, vector, fermion)\\
        \hline \hline
        \end{tabular}
        \caption{Properties of the $SO(N)$ and $SO(N-4)$ monopoles of the root states for even $N$. The first three roots apply to all even $N$, and root 4 only applies to the case with $N$ an integral multiple of 4. The $SO(N)$ monopole breaks the $I^{(N)}$ symmetry to $(SO(2)\times SO(N-2)\times SO(N-4))/Z_2$, and it always has no charge under the $SO(2)$. Its three corresponding entries represent its representation under the $SO(N-2)$, its representation under the $SO(N-4)$, and its statistics, respectively. The $SO(N-4)$ monopole breaks the $I^{(N)}$ symmetry to $(SO(N)\times SO(N-6)\times SO(2))/Z_2$, and it always has no charge under the $SO(2)$. Its three corresponding entries represent its representation under the $SO(N)$, its representation under the $SO(N-6)$, and its statistics, respectively. For the case with $N=6$, the second entry does not exist for its $SO(N-4)$ monopole. Notice these properties are determined up to attaching pure gauge charges.}
    \label{tab: SO(N) and SO(N-4) monopoles}
\end{table}

Let us check which anomalies correspond to states that satisfy the two conditions listed at the beginning of this appendix. In order for the first condition to be satisfied, according to Table \ref{tab: SO(N) and SO(N-4) monopoles}, the state must contains the anomaly corresponding to root 3 or its inverse. 

It is a bit more complicated to check the second condition, but it actually suffices to check a weaker condition: if the remaining $SO(N-4)$ symmetry is further broken to $SO(2)^{\frac{N}{2}-2}$, the system is anomaly-free. To this end, let us condense $\left(\begin{array}{cccccc|cccc} 1 & 1 & \cdots & 1 & 0 & 0 & -1 & -1 & \cdots & -1 \\ 0 & 0 & \cdots & 0 & 0 & 0 & 0 & 0 & \cdots & 0\end{array}\right)_b$. This condensate breaks the $I^{(N)}$ symmetry into $(SO(2)^{\frac{N}{2}-2}\times SO(4))/Z_2$, and the gauge fields corresponding to the remaining $\frac{N}{2}-2$ $SO(2)$ symmetries can be taken as $A_{12}'=\frac{1}{2}(A_{12}^L+A_{12}^R), A_{34}'=\frac{1}{2}(A_{34}^L+A_{34}^R), \cdots, A_{N-5,N-4}'=\frac{1}{2}(A_{N-5, N-4}^L+A_{N-5, N-4}^R)$, and for the $SO(4)\backsimeq\frac{SU(2)\times SU(2)}{Z_2}$, we denote the gauge fields corresponding to these two $SU(2)$ subgroups by $B_{1,2}$, which together form the $SO(4)$ gauge field.

The fundamental monopole remains deconfined in this condensate and becomes the fundamental monopole of the resulting theory. In terms of the remaining symmetries, it should be written as $\left(\begin{array}{cccc|cc} q^L+q^R & q^L+q^R & \cdots & q^L+q^R & q^L & q^L\\ \half & \half & \cdots & \half & \half & \half\end{array}\right)_s$. By adding to the theory a proper theta-term of the $SO(4)$ gauge field, $\theta \epsilon_{\mu\nu\lambda\rho}\Tr(\partial^\mu B_1^{\nu}\partial^\nu B_1^{\rho}-\partial^\mu B_2^{\nu}\partial^\lambda B_2^{\rho})$, which preserves all remaining symmetries (including the remaining $\mc{C}$, $\mc{R}$ and $\mc{T}$ symmetries), this fundamental monopole can be converted to $\left(\begin{array}{cccc|cc} q^L+q^R & q^L+q^R & \cdots & q^L+q^R & 0 & 0\\ \half & \half & \cdots & \half & \half & \half\end{array}\right)_s$. In order for the theory to be anomaly-free, this monopole should be a boson with trivial projective quantum numbers. This means $s=0\ ({\rm mod\ }2)$ and $q^L+q^R=0\ ({\rm mod\ }2)$. Therefore, only root 3 and multiple copies of it satisfy this condition.

To satisfy both conditions, the only possibilities are root 3 and its inverse. Because the SLs with $(N, \pm 1)$ satisfy both conditions, we conclude that the anomalies of these SLs are precisely the same as root 3 and its inverse. Notice that from this analysis we cannot determine which of $(N, \pm 1)$ corresponds to root 3, and which corresponds to its inverse.

Now that we have identified the SLs with $(N, \pm 1)$ as states that realize the anomalies of root 3 and its inverse, it may also be worth mentioning which states realize the anomalies of the other 2 roots.

The $I^{(N)}$-anomaly of root 1 can be realized by a $Z_2$ topological order (TO). Denote the $Z_2$ charge and flux by $e$ and $m$, respectively, the symmetry actions on these topological sectors are given in Table \ref{tab: root 1 realization}.
        
    \begin{table}[h]
        \setlength{\tabcolsep}{0.2cm}
        \renewcommand{\arraystretch}{1.4}
        \centering
        \begin{tabular}{c|cc} 
        \hline \hline
        & $e$ & $m$\\
        \hline
        $I^{(N)}$ & (singlet, vector) & (singlet, vector) \\
        $\mc{C}$ & $e$ & $m$\\
        $\mc{R}$ & $e$ & $m$\\
        $\mc{T}$ & $e$ & $m$\\
        \hline \hline
        \end{tabular}
        \caption{Projective symmetry actions on the topological sectors of the $Z_2$ TO that realizes the $I^{(N)}$ anomaly of root 1.  In the row corresponding to $I^{(N)}$, the two entries represent the representations of this excitation under the $SO(N)$ and $SO(N-4)$ subgroups of $I^{(N)}$, respectively. When $N=6$, the vector representation of the $SO(2)\subset I^{(6)}$ means charge 1 under $SO(2)$. Their partners under $\mc{C}$, $\mc{R}$ and $\mc{T}$ are shown as above. Notice one can further specify data like $T^2$ for $e$ and $m$, but it is unnecessary for our purpose.}
        \label{tab: root 1 realization}
    \end{table}

    To see that this $Z_2$ TO realizes the $I^{(N)}$-anomaly of root 1, consider threading a fundamental monopole through the system, which leaves a flux. Because of the above symmetry assignment, when an $e$ or $m$ circles around this flux, it acquires a phase factor $-1$, no matter how far it is away from the flux. Since this monopole threading process is local for the $(2+1)$-d system, this $-1$ phase factor has to be cancelled by requiring that the flux also trap an anyon that has $-1$ mutual braiding with both $e$ and $m$. This anyon is $\epsilon$, the fermionic bound state of $e$ and $m$. Furthermore, time reversal symmetry ensures that no polarization charge is induced around this flux. Then the composite of this flux and $\epsilon$ is a fermion in the trivial representation of $I^{(N)}$. Therefore, the fundamental monopole of this $Z_2$ TO has precisely the structure as in root 1.
    
    This $Z_2$ TO can be explicitly constructed using the layer construction in Ref. \cite{Wang2013}, and it is an analog of $eCmC$, the surface state of a $(3+1)$-d bosonic topological insulator protected by $U(1)$ charge conservation and time reversal \cite{Vishwanath2012, Wang2013}. 
    
The $I^{(N)}$-anomaly of root 2 can be realized by a $Z_4\times Z_2$ TO. Denote the $Z_4$ charge and flux by $e_1$ and $m_1$, and the $Z_2$ charge and flux by $e_2$ and $m_2$, respectively. We take the convention such that the mutual statistics between $e_1$ and $m_1$ is $i$ if $N=6\ ({\rm mod\ }8)$, and $-i$ if $N=2\ ({\rm mod\ }8)$. The symmetry actions on these topological sectors are given in Table \ref{tab: root 2 realization}.
    
\begin{table}[h]
    \setlength{\tabcolsep}{0.2cm}
    \renewcommand{\arraystretch}{1.4}
    \centering
    \begin{tabular}{c|ccccc} 
    \hline \hline
    & $e_1$ & $m_1$ & $e_2$ & $m_2$ & $\epsilon_2\equiv e_2m_2$\\
    \hline
    $I^{(N)}$ & (singlet, fund.) & (singlet, anti-fund.) & (anti-fund., anti-fund.) & (fund., fund.) & (singlet, singlet)\\
    $\mc{C}$ & $e_1^{-1}$ & $m_1^{-1}$ & $e_2^{-1}$ & $m_2^{-1}$ & $\epsilon_2$\\
    $\mc{R}$ & $e_1$ & $m_1^{-1}e_1^{-2}\epsilon_2$ & $e_2^{-1}e_1^{-2}$ & $m_2^{-1}e_1^{2}$ & $\epsilon_2$\\
    $\mc{T}$ & $e_1^{-1}$ & $m_1e_1^2\epsilon_2$ & $e_2e_1^{2}$ & $m_2e_1^{-2}$ & $\epsilon_2$\\
    \hline \hline
    \end{tabular}
    \caption{Projective symmetry actions on the topological sectors of the $Z_4\times Z_2$ TO that realizes the $I^{(N)}$ anomaly of root 2. In the row corresponding to $I^{(N)}$, the two entries in each parenthesis represent the representations of this excitation under the associated {\em spin} group of the correponding $SO(N)$ and $SO(N-4)$ subgroups of $I^{(N)}$, where ``fund." (``anti-fund.") represents fundamental (anti-fundamental) representation of the relevant spin group (when $N=6$, the fund. (anti-fund.) in the second entry in the parenthesis means charge $1/2$ ($-1/2$) under $SO(2)\subset I^{(6)}$). Their partners under $\mc{C}$, $\mc{R}$ and $\mc{T}$ are shown as above. Notice one can further specify data like $T^2$ for various anyons, but it is unnecessary for our purpose.}
    \label{tab: root 2 realization}
\end{table}

To see that this theory realizes the $I^{(N)}$-anomaly of root 2, we can again consider threading a fundamental monopole through the system and use a similar argument as before. One can check that the flux left by the fundamental monopole will trap an anyon $e_1m_1^{-1}$, and this fundamental monopole is indeed a boson in the vector representation under $SO(N-4)$, so we conclude that this theory realizes the $I^{(6)}$-anomaly of root 2.
    
We believe this $Z_4\times Z_2$ TO with the above symmetry implementation is a consistent theory, although we do not have an explicit construction for it.
    
\subsubsection{The case with $N=0\ ({\rm mod\ }4)$}

Next we turn to the case with $N=0\ ({\rm mod\ }4)$. From similar analysis as before, now the constraints we obtain are $4q^{L, R}\in\mathbb{Z}$ and $2(q^L+q^R)\in\mathbb{Z}$. The three roots in \eqref{eq: 4k+2 roots} still satisfy these constraints, but we find one additional root {\footnote{This root is absent in the case with $N=2\ ({\rm mod\ }4)$ because $q_L+q_R\in\mathbb{Z}$ in that case, which is violated here.}}:
\beq \label{eq: root 4}
{\rm root\ }4:
\left(
\begin{array}{ccc|cccc}
0 & \cdots & 0 & \half & \half & \cdots & \half\\
\half & \cdots & \half & \half & \half & \cdots & \half
\end{array}
\right)_b
\eeq

Similar as before, we can also derive the properties of the $SO(N)$ and $SO(N-4)$ monopoles for this root, and the results are given in Table \ref{tab: SO(N) and SO(N-4) monopoles}. From this Table, we observe that there are not only differences in the anomalies for the cases with $N=2\ ({\rm mod\ }4)$ and $N=0\ ({\rm mod\ }4)$, but also differences in the anomalies for the cases with $N=0\ ({\rm mod\ }8)$ and $N=4\ ({\rm mod\ }8)$. It is known that the spinor representations of $SO(N)$ in these three cases are different, \ie they are complex, real, and pseudoreal for $N=2\ ({\rm mod\ }4)$, $N=0\ ({\rm mod\ }8)$ and $N=4\ ({\rm mod\ }8)$, respectively \cite{Zee2016}. Our analysis suggests a connection between these two results.

Lastly, analogous arguments as before indicate that only root 3 and its inverse satisfy both conditions discussed at the beginning of this appendix. Because SLs with $(N, \pm 1)$ also satisfy those two conditions, we conclude that these SLs have the same anomaly as root 3 and its inverse.

\subsection{The case with an odd $N$} \label{appsub: odd N}

Now we turn to the case with odd $N$, so $I^{(N)}=SO(N)\times SO(N-4)$. Notice this analysis also applies to the case with an even $N$ if we add to the system bosonic DOF in the vector representation of $SO(N)$.

In this case, the fundamental monopoles are simply the usual $SO(N)$ and $SO(N-4)$ monopoles, so these two monopoles will characterize the $I^{(N)}$ SPTs and anomalies.

An $SO(N)$ monopole breaks $I^{(N)}$ into $SO(2)\times SO(N-2)\times SO(N-4)$. This monopole can be denoted as $\left(
\begin{array}{cc|c}
q_N & r_N^L & r_N^R\\
1 & 0 & 0
\end{array}
\right)_{s_N}$, where $q_N$ represents the fractional charge under the remaining $SO(2)$, $r_N^L$ represents the projective quantum number under the remaining $SO(N-2)$, $r_N^R$ represents the projective quantum number under $SO(N-4)$, and $s_N$ is the statistics of this monopole. Similarly, an $SO(N-4)$ monopole can be denoted by $\left(
\begin{array}{c|cc}
r_{N-4}^L & q_{N-4} & r_{N-4}^R\\
0 & 1 & 0
\end{array}
\right)_{s_{N-4}}$. Just as in the usual $(3+1)$-d bosonic topological insulator, time reversal symmetry and the bosonic statistics of the pure gauge charge require that $q_N=q_{N-4}=0$ \cite{Vishwanath2012}. Furthermore, the Dirac quantization condition requires that whenever the $SO(N)$ monopole carries a spinor representation under $SO(N-4)$, the $SO(N-4)$ monopole must also carry a spinor representation under $SO(N)$, and vice versa. These are all the constraints, and we get a $\mathbb{Z}_2^5$ classification of the $I^{(N)}$ SPTs or anomalies, and the structures of the monopoles of the 5 root states are in Table \ref{tab: SO(N) and SO(N-4) monopoles for odd N}.

\begin{table}[h]
    \setlength{\tabcolsep}{0.2cm}
    \renewcommand{\arraystretch}{1.4}
    \centering
    \begin{tabular}{c|ccc} 
        \hline \hline
        & $SO(N)$ monopole & $SO(N-4)$ monopole & topological response function\\
        \hline
        root 1 & (singlet, singlet, fermion) & (singlet, singlet, boson) & $(w_2^{SO(N)})^2$\\
        root 2 & (singlet, singlet, boson) & (singlet, singlet, fermion) & $(w_2^{SO(N-4)})^2$\\
        root 3 & (spinor, singlet, boson) & (singlet, singlet, boson) & $w_4^{SO(N)}$\\
        root 4 & (singlet, singlet, boson) & (singlet, spinor, boson) & $w_4^{SO(N-4)}$\\
        root 5 & (singlet, spinor, boson) & (spinor, singlet, boson) & $w_2^{SO(N)}w_2^{SO(N-4)}$\\
        \hline \hline
        \end{tabular}
        \caption{Properties of the $SO(N)$ and $SO(N-4)$ monopoles of the root states for odd $N$. The $SO(N)$ monopole breaks the $I^{(N)}$ symmetry to $SO(2)\times SO(N-2)\times SO(N-4)$, and it always has no fractional charge under the $SO(2)$. Its three corresponding entries represent its representation under the $SO(N-2)$, its representation under the $SO(N-4)$, and its statistics, respectively. The $SO(N-4)$ monopole breaks the $I^{(N)}$ symmetry to $SO(N)\times SO(N-6)\times SO(2)$, and it always has no charge under the $SO(2)$. Its three corresponding entries represent its representation under the $SO(N)$, its representation under the $SO(N-6)$, and its statistics, respectively. The last column lists the topological response function corresponding to these root states.}
    \label{tab: SO(N) and SO(N-4) monopoles for odd N}
\end{table}

Using similar arguments as before, one can show that only one of the $2^5=32$ anomaly classes satisfies both conditions discussed at the beginning of this appendix, which is (root 2)$\otimes$(root 3)$\otimes$(root 4)$\otimes$(root 5), \ie a composed system made of root 2, root 3, root 4 and root 5. This anomaly class is identified with the one for the SLs, and this result agrees with Eq. \eqref{eq: Anomaly11}.

\section{Explicit calculations of the $U(1)$ DSL} \label{app: DSL explicit}

In this appendix, we explicitly derive the properties of the fundamental $I^{(6)}$-monopole in a $U(1)$ DSL, whose effective theory is given by Eq. \eqref{eq: DSL standard model}. We will see that this fundamental $I^{(6)}$-monopole has precisely the structure of root 3 in Eq. \eqref{eq: 4k+2 monopoles main}, which further strengthens our proposal that the $U(1)$ DSL and SL$^{(6)}$ are equivalent. Our results also agree with a more formal calculation in Ref.~\cite{Calvera2021}.

Recall that we can take the Dirac fermions in a DSL to be in either the fundamental or anti-fundamental representation of $SU(4)$, and take the monopole of $a$ to have either charge $1$ or $-1$ under $U(1)_{\rm top}$, so there are 4 different choices of the symmetry implementation. To be general, we will consider the 4 cases together by introducing parameters $\zeta$ and $\xi$, such that $\zeta=1$ ($\zeta=-1$) if the Dirac fermions are in the fundamental (anti-fundamental) representation of $SU(4)$, and $\xi=\pm 1$ if the monopole of $a$ carries charge $\pm 1$ under $U(1)_{\rm top}$\footnote{Since we can redefine the theory through a charge conjugation $\tilde{\psi}=\psi^{\dagger}$, $\tilde{a}=-a$, the two signs $\zeta$ and $\xi$ can be flipped simultaneously without physical effect. Only the product $\zeta\xi$ will eventually matter in the following discussions.}.

Next we will calculate the $q^L$, $q^R$ and $s$ of the fundamental $I^{(6)}$-monopole for a DSL. In particular, we will thread a flux corresponding to the fundamental $I^{(6)}$-monopole in Eq. \eqref{eq: fundamental monopole main}, which has a $\pi$-flux for $A_{12}$, $A_{34}$, $A_{56}$ and $A_{\rm top}$.
    
It is useful to denote the 4 flavors of Dirac fermions by $\psi_{\uparrow+}$, $\psi_{\uparrow-}$, $\psi_{\downarrow+}$, and $\psi_{\downarrow-}$, respectively. This notation is motivated by the lattice realizations of a DSL, where $\uparrow$ and $\downarrow$ represent two physical spins, and $+$ and $-$ represent the two valleys. According to the homomorphism between $su(4)$ and $so(6)$ in Appendix \ref{app: homomorphism}, the charges under $(A_{12}, A_{34}, A_{56})$ carried by $\psi_{\uparrow+}$, $\psi_{\uparrow-}$, $\psi_{\downarrow+}$ and $\psi_{\downarrow-}$ are respectively $\zeta(1/2, 1/2, 1/2)$, $\zeta(1/2, -1/2, -1/2)$, $\zeta(-1/2, 1/2, -1/2)$ and $\zeta(-1/2, -1/2, 1/2)$. Note that the Dirac fermions are neutral under $A_{\rm top}$. 
    
When the flux specified above is thread, $\psi_{\uparrow+}$ sees a total $3\zeta\pi/2$ flux, while the other 3 flavors of Dirac fermions see a total $-\zeta\pi/2$ flux. To construct a gauge invariant state corresponding to the local fundamental $I^{(6)}$-monopole, one can consider a state where the internal gauge field, $a$, has a flux of $\zeta\pi/2$, such that at the end $\psi_{\uparrow+}$ sees a $2\zeta\pi$ flux and contributes a zero mode in this flux background, and the other 3 flavors see no flux and contribute no zero mode. Because there is a single zero mode in the background of a flux with magnitude $2\pi$, no matter it is occupied or not, we get a bosonic state, so $s=b$ for the fundamental $I^{(6)}$ monopole in Eq. \eqref{eq: fundamental monopole main}. To determine $q^{L,R}$ for this fundamental monopole, we need to determine whether this zero mode is occupied or not.
    
The usual way to do this is to demand that the zero modes are half-filled. However, that works only if the theory has a symmetry that preserves the flux but flips the charge. In the present case, there is no such a symmetry, so a different approach should be taken. To proceed, we will regularize the DSL as follows. First, to preserve the $SU(4)$ flavor symmetry, for each flavor we add to the system a gapped Dirac fermion, which contributes to the effective action a term $-\pi\eta/2$ when combined with the original gapless Dirac fermion with the same flavor, where $\eta$ is the $\eta$-invariant of the Dirac operator corresponding to each flavor of gapless and gapped Dirac fermions \cite{Witten2015}. These $\eta$-invariants will generally break the $\mc{T}$ symmetry. To maintain the $\mc{T}$ symmetry, next we put the system on the boundary of a $(3+1)$-d bulk that contributes an appropriate theta-term to the bulk partition function, such that the combined partition function of the boundary and bulk is $\mc{T}$ invariant. This particular regularization of the theory should suffice to yield $q^{L, R}$ of the fundamental monopole in Eq. \eqref{eq: fundamental monopole main}, up to attaching local DOF.
    
With this regularization, the effective action\footnote{More precisely the partition function is $Z=|\det(\slashed{D})|\exp(iS_{\rm{eff}})$ in Euclidean signature \cite{Witten2015}.} is
\beq
S_\eff=&\sum_{i=1}^4\left[-\frac{\pi}{2}\eta(a_i)+\half\frac{1}{4\pi}\int d^3xa_ida_i\right]+\frac{\xi}{2\pi}\int d^3xA_{\rm top}da
\eeq
with
\beq
\begin{split}
&a_1=a+\frac{\zeta (A_{12}+A_{34}+A_{56})}{2}\\
&a_2=a+\frac{\zeta(A_{12}-A_{34}-A_{56})}{2}\\
&a_3=a+\frac{\zeta(-A_{12}+A_{34}-A_{56})}{2}\\
&a_4=a+\frac{\zeta(-A_{12}-A_{34}+A_{56})}{2}
\end{split}
\eeq
and $i=1,2,3,4$ correspond to contributions from $\uparrow+$, $\uparrow-$, $\downarrow+$, $\downarrow-$, respectively. Notice that the second term in the above effective action comes from reducing the theta-term of the $(3+1)$-d bulk to the boundary. Furthermore, the coefficient of the CS term of the dynamical gauge field $a$ is $1/(2\pi)$, which is well defined in $(2+1)$-d and implies that the $(3+1)$-d bulk does not really need a dynamical gauge field, consistent with the general expectation that a theory with a 't Hooft anomaly can live on the boundary of a short-range entangled bulk (see, \eg Ref. \cite{Ning2019} for examples of theories that have more severe anomalies than a 't Hooft anomaly and thus can only live on the boundary of a long-range entangled bulk). Also notice that all dependence on the metric of the spacetime manifold of the system is suppressed, which will not affect our following analysis.
    
This effective action can be used to read off the resulting charges under various gauge fields when the flux is thread:
\beq
\begin{split}
&Q_a=N_{\psi_{\uparrow+}}+\frac{B_{a_1}+B_{a_2}+B_{a_3}+B_{a_4}}{4\pi}+\frac{\xi B_{\rm top}}{2\pi}=N_{\psi_{\uparrow+}}+\frac{B_a}{\pi}+\frac{\xi B_{\rm top}}{2\pi}\\
&Q_{12}=\zeta\left(\frac{ N_{\psi_{\uparrow+}}}{2}+\frac{B_{a_1}+B_{a_2}-B_{a_3}-B_{a_4}}{8\pi}\right)=\zeta\cdot\frac{N_{\psi_{\uparrow+}}}{2}+\frac{B_{12}}{4\pi}\\
&Q_{34}=\zeta\left(\frac{ N_{\psi_{\uparrow+}}}{2}+\frac{B_{a_1}-B_{a_2}+B_{a_3}-B_{a_4}}{8\pi}\right)=\zeta\cdot\frac{N_{\psi_{\uparrow+}}}{2}+\frac{B_{34}}{4\pi}\\
&Q_{56}=\zeta\left(\frac{ N_{\psi_{\uparrow+}}}{2}+\frac{B_{a_1}-B_{a_2}-B_{a_3}+B_{a_4}}{8\pi}\right)=\zeta\cdot\frac{N_{\psi_{\uparrow+}}}{2}+\frac{B_{56}}{4\pi}\\
&Q_{\rm top}=\frac{\xi B_a}{2\pi}
\end{split}
\eeq
where $Q_{(\cdot)}$ and $B_{(\cdot)}$ represent the charge and flux under the corresponding gauge field, respectively, and $N_{\psi_{\uparrow+}}$ determines whether the zero mode contributed by $\psi_{\uparrow+}$ is occupied, \ie if the flux seen by the fermion is positive (negative), then $N_{\psi_{\uparrow+}}=0$ means it is occupied (unoccupied).
    
According to the previous discussion, now we have $B_{12}=B_{34}=B_{56}=B_{\rm top}=\pi$ and $B_a=\zeta\pi/2$. To be gauge invariant, $Q_a=0$, which means $N_{\psi_{\uparrow+}}=-(\zeta+\xi)/2$. Substituting this into the rest of the equations yields $Q_{12}=Q_{34}=Q_{56}=-\frac{\zeta\xi}{4}$ and $Q_{\rm top}=\frac{\zeta\xi}{4}$. That is, if $(\zeta, \xi)=(1, -1)$ or $(\zeta, \xi)=(-1, 1)$, $q^L=-q^R=1/4$, corresponding to root 3 in Eq. \eqref{eq: 4k+2 monopoles main}. If $(\zeta, \xi)=(1, 1)$ or $(\zeta, \xi)=(-1, -1)$, $q^L=-q^R=-1/4$, corresponding to the inverse of root 3 in Eq. \eqref{eq: 4k+2 monopoles main}. Therefore, the $I^{(N)}$ anomaly of the $U(1)$ DSL is indeed identical to that of the SL$^{(6, \pm 1)}$, which even further strengthens our proposal that they are dual.

\section{More on the LSM constraints} \label{app: LSM}

In the main text, we have used some physical arguments to propose that Eq. \eqref{eq: full LSM} describes the complete set of LSM constraints for various lattice systems. In this appendix, we extract some LSM contraints from Eq. \eqref{eq: full LSM} that were not used in deriving Eq. \eqref{eq: full LSM}. We will also give an alternative expression for the LSM anomaly on a square lattice.

For convenience, we copy Eq. \eqref{eq: full LSM}:
\beq \label{eq: full LSM app}
S_{\rm LSM}=i\pi\int_{X_4}(w_2^{SO(3)}+t^2)[xy+c^2+r(x+c)]
\eeq
where $w_2^{SO(3)}$ is the second SW class of the $SO(3)$ gauge field corresponding to the spin rotational symmetry, $t$ is the gauge field corresponding to the time reversal symmetry, $x$ and $y$ are the gauge field corresponding to translation along $T_1$ and $T_2$, respectively, $c$ is the gauge field corresponding to $C_2$ site-centered lattice rotation, and $r$ is the gauge field corresponding to the reflection symmetry $R_y$. There is a constraint $r+t=w_1^{TM}\ ({\rm mod\ }2)$. This anomaly polynomial should be viewed as a topological response function of the system under the various gauge fields.

Physically, we expect that there will be LSM anomalies associated with reflection symmetry $R_x$. To read them off, we need to design a configuration of the gauge field corresponding to $R_x$ in Eq. \eqref{eq: full LSM app}. Although the gauge field corresponding to $R_x$ does not explicitly appear in Eq. \eqref{eq: full LSM app}, because $R_x$ is a combination of $C_2$ and $R_y$, we can still have a gauge connection of $R_x$ by writing $c=c_0+r$. This means that whenever there is a gauge connection corresponding to $R_y$, a gauge connection corresponding to $C_2$ is also induced. Therefore, now $r$ actually represents a gauge connection corresponding to $R_x$, and $c_0$ is the gauge connection for pure $C_2$ rotation. Substituting $c=c_0+r$ into Eq. \eqref{eq: full LSM app} yields
\beq
S_{\rm LSM}=i\pi\int_{X_4}(w_2^{SO(3)}+t^2)[xy+c_0^2+r(x+c_0)]
\eeq

The physical meaning of this new anomaly polynomial can be understood by looking at various sub-symmetries of the system. For example, ignoring translation symmetries, \ie setting $x=y=0$, it becomes $S_{\rm LSM}=i\pi\int_{X_4}(w_2^{SO(3)}+t^2)(c_0^2+rc_0)$. The first term in the second parenthesis, $c_0^2$, physically means that there is an LSM anomaly if there is an odd number of spin-1/2's at the $C_2$ center, just as Eq. \eqref{eq: ILSManomaly}. The other term, $rc_0$, represents an LSM anomaly associated with $R_x$ and $C_2$, if there is a $C_2$ center at the $R_x$-invariant line and this $C_2$ center hosts an odd number of spin-1/2's. 

As another example, we can also ignore the $C_2$ symmetry by setting $c_0=0$, and consider translations. On a triangular lattice, the $R_x$-invariant line has a translation symmetry generated by $T_1T_2^2$. The gauge field corresponding to such a translation symmetry can be obtained by writing $y=y_0+2x$. Similar as above, now $x$ represents the gauge field corresponding to $T_1T_2^2$, and $y_0$ represents the gauge field corresponding to $T_2$. Substituting $c_0=0$ and $y=y_0+2x$ into Eq. \eqref{eq: full LSM app} yields $S_{\rm LSM}=i\pi\int_{X^4}(w_2^{SO(3)}+t^2)(xy_0+rx)$. Now the first term in the second parenthesis, $xy_0$, represents an LSM anomaly associated with having an odd number of spin-1/2's in each unit cell corresponding to the translations $T_2$ and $T_1T_2^2$. The second term, $rx$, represents an LSM anomaly associated with $R_x$ and $T_1T_2^2$, if there is an odd number of spin-1/2's in each unit cell of $T_1T_2^2$.

Similarly, we can also obtain the LSM anomaly associated with $R_x$ and $R_y$. To do so, we can set $x=y=0$ and $r=c+r_0$ in Eq. \eqref{eq: full LSM app}. The first of these two conditions amounts to ignoring translation symmetries, while the second condition means that now $c$ is really a gauge field corresponding to the $R_x$ symmetry, and $r_0$ is the gauge field corresponding to the $R_y$ symmetry. Substituting these conditions into Eq. \eqref{eq: full LSM app} yields $S_{\rm LSM}=i\pi\int_{X_4}(w_2^{SO(3)}+t^2)cr_0$. This represents an LSM anomaly if there is an odd number of spin-1/2's at the intersecting point of the reflection axes of $R_x$ and $R_y$. This LSM anomaly of course encodes the one associated with the $C_2$ center. To see it formally, one way is to further restrict $c=r_0$, which turns both $c$ and $r_0$ into the gauge field corresponding to the $C_2$ rotation. Then this anomaly polynomial becomes Eq. \eqref{eq: ILSManomaly}.

The above discussion motivates us to write the LSM anomaly on a square lattice in terms of gauge fields corresponds to $T_{1,2}$, $R_{x, y}$, $SO(3)$ and $\mc{T}$ symmetries. Denote the gauge fields corresponding to $R_{x,y}$ by $r_{x,y}$. Using an argument similar to that in the main text, the LSM anomaly on a square lattice can be written as
\beq \label{eq: LSM square}
S_{\rm LSM}=i\pi\int_{X_4}(w_2^{SO(3)}+t^2)(x+r_x)(y+r_y)
\eeq
with a constraint $t+r_x+r_y=w_1^{TM}\ ({\rm mod\ }2)$. Note that this expression is manifestly $C_4$ rotationally invariant. 

To reproduce the Eq. \eqref{eq: full LSM app}, we want to write this anomaly in terms of gauge fields of $R_y$, $C_2$, $T_{1,2}$, $SO(3)$ and $\mc{T}$. So we let $r_x=c$ and $r_y=c+r$, then $c$ is the gauge field for $C_2$ and $r$ is the gauge field for $R_y$. Then the above LSM anomaly precisely recover Eq. \eqref{eq: full LSM app}, after using that $cx=cy=0$. It is interesting to note that in order to derive Eq. \eqref{eq: LSM square} from Eq. \eqref{eq: full LSM app}, one needs to first replace the latter by
\beq
S_{\rm LSM}=i\pi\int_{X_4}(w_2^{SO(3)}+t^2)[xy+c^2+r(x+c)+cy]
\eeq
Because $cy=0$, this expression should be equivalent to Eq. \eqref{eq: full LSM app}. Then by setting $c=r_x$ and $r=r_y+r_x$ and using that $r_xx=r_yy=0$, this anomaly becomes Eq. \eqref{eq: LSM square}.

All the above results are consistent with the physical expectations. The method employed above can also be readily applied to other situations to extract other LSM anomalies.

\section{Anomaly matching of the $U(1)$ DSL on a triangular lattice} \label{app: anomaly matching triangular DSL}

In this appendix we show that the $U(1)$ DSL on a triangular lattice indeed has the correct LSM anomaly.

The symmetries we will focus on are $SO^s(3)$ spin rotation, time reversal $\mc{T}$, translations $T_{\vec a_{1}, \vec a_2}$, site-centered $C_2$ rotation, and reflection $R_y$ that keeps $\vec a_1$ invariant. 
Their actions on the $U(1)$ DSL on a triangular lattice are \cite{Song2018, Song2018a}:
\beq
\begin{split}
    &SO^s(3):\ n\rightarrow \left(
    \begin{array}{cc}
    I_3 & \\
    & SO^s(3)
    \end{array}
    \right)n,\\
    &\mc{T}:\ n\rightarrow
    \left(
    \begin{array}{cc}
        I_3 &  \\
         & -I_3
    \end{array}
    \right)n,\\
    & T_{\vec a_1}: n\rightarrow
    \left(
    \begin{array}{cccc}
    -1 & & & \\
    & 1 & & \\
    & & -1 &\\
    & & & I_3
    \end{array}
    \right)
    n\exp\left(i\frac{2\pi}{3}\sigma_y\right),\\
    &T_{\vec a_2}: n\rightarrow 
    \left(
    \begin{array}{cccc}
    1 & & & \\
    & -1 & & \\
    & & -1 &\\
    & & & I_3
    \end{array}
    \right)
    n\exp\left(i\frac{2\pi}{3}\sigma_y\right),\\
    & C_2: n\rightarrow \left(\begin{array}{cc}I_3 & \\ & -I_3\end{array}\right)n\sigma_z,\\
    & R_y: n\rightarrow
    \left(
    \begin{array}{cccc}
    & & -1 & \\
    & 1 & & \\
    -1 & & & \\
    & & & I_3
    \end{array}
    \right)n
\end{split}
\eeq

From these symmetry actions, we get
\beq
\begin{split}
    &w_1^{O(6)}=t+r+c,\ w_1^{O(2)}=c,\\
    &w_2^{O(6)}=xy+xr+rt+rc+c^2+w_2^{SO^s(3)}+t^2,\ w_2^{O(2)}=0,\\
    &w_4^{O(6)}=(w_2^{SO^s(3)}+t^2)(xy+xr+rc)+rc^2(t+c)
\end{split}
\eeq
with the meanings of these symbols identical as those in the main text. Substituting these expressions into Eq. \eqref{eq: Anomaly11} and performing some algebraic manipulations yield
\beq
S_{\rm bulk}=i\pi\int_M[xy+c^2+r(x+c)]\left(w_2^{SO^s(3)}+t^2\right)=S_{\rm LSM}
\eeq
which shows that the $U(1)$ DSL on a triangular lattice indeed has the correct LSM anomaly.

\end{document}